\documentclass[]{JHEP3}
\pdfoutput = 1

\usepackage{graphicx}
\usepackage{amsmath,amssymb,amscd,amsfonts,bm}
 
\newcommand{\ME}{M}

\newcommand{\as}{\alpha_{\mathrm{s}}}

\newcommand{\LA}{\mathrm{A}}
\newcommand{\LB}{\mathrm{B}}
\newcommand{\LE}{\mathrm{E}}
\newcommand{\LF}{\mathrm{F}}
\newcommand{\LI}{\mathrm{I}}
\newcommand{\LS}{\mathrm{S}}
\newcommand{\La}{\mathrm{a}}
\newcommand{\Lb}{\mathrm{b}}
\newcommand{\Lc}{\mathrm{c}}
\newcommand{\Ls}{\mathrm{s}}

\def\ket#1{\big|{#1}\big\rangle}
\def\bra#1{\big\langle{#1}\big|}
\def\brax#1{\big\langle{#1}}   

\def\sket#1{\big|{#1}\big)}
\def\sbra#1{\big({#1}\big|}
\def\sbrax#1{\big({#1}}        

\def\dualL{\raisebox{-5 pt}{$\scriptstyle D$}\!}   
\def\dualR{\!\raisebox{-5 pt}{$\scriptstyle D$}} 

\newbox\charbox
\newbox\slabox
\def\s#1{{      
        \setbox\charbox=\hbox{$#1$}
        \setbox\slabox=\hbox{$/$}
        \dimen\charbox=\ht\slabox
        \advance\dimen\charbox by -\dp\slabox
        \advance\dimen\charbox by -\ht\charbox
        \advance\dimen\charbox by \dp\charbox
        \divide\dimen\charbox by 2
        \raise-\dimen\charbox\hbox to \wd\charbox{\hss/\hss}
        \llap{$#1$}
}}

\title{Parton showers with quantum interference}

\author{Zolt\'an Nagy \\
Theory Division,
CERN\\
CH-1211 Geneva 23, Switzerland\\
E-mail: \email{Zoltan.Nagy@cern.ch}
}

\author{Davison E.\ Soper\\
Institute of Theoretical Science\\
University of Oregon\\
Eugene, OR  97403-5203, USA\\
E-mail: \email{soper@uoregon.edu}
}

\abstract{
We specify recursive equations that could be used to generate a lowest order parton shower for hard scattering in hadron-hadron collisions. The formalism is based on the factorization soft and collinear interactions from relatively harder interactions in QCD amplitudes. It incorporates quantum interference between different amplitudes in those cases in which the interference diagrams have leading soft or collinear singularities. It incorporates the color and spin information carried by partons emerging from a hard interaction. One motivation for this work is to have a method that can naturally cooperate with next-to-leading order calculations. 
}

\preprint{CERN-PH-TH/2007-082\\
24 September 2007}

\begin{document}


\section{Introduction}
\label{sec:introduction}

Parton shower Monte Carlo event generators, such as  \textsc{Herwig}  \cite{Herwig} and \textsc{Pythia} \cite{Pythia}, have proven to be enormously useful since the development of the main ideas in the 1980s \cite{EarlyPythia, Gottschalk, angleorder}. These computer programs perform calculations of cross sections according to an approximation to the standard model or some of its possible extensions. Because of the great success of these programs, it is worthwhile to investigate possible improvements. In this paper, we propose a theoretical structure for event generators that generalizes the structure of current programs and allows the elimination of certain approximations used currently.

Parton showers are mostly reflections of QCD interactions. In order to present a reasonably complete discussion of the QCD issues in a parton shower while keeping the length of this paper within reasonable bounds, we limit the presentation to QCD and omit any discussion of how electroweak and beyond-the-standard-model interactions are to be added to the QCD interactions to make a useful event generator.

What is a parton shower Monte Carlo event generator? Let us consider hadron-hadron collisions, which is the case relevant for the Tevatron and the Large Hadron Collider. An experiment will produce a large number of events $f$, where one can characterize an event as a list of the momenta and flavors of the final state particles produced. The experiment can measure a cross section $\sigma[F]$ corresponding to an observable\footnote{In order to be subject to reliable calculation in QCD perturbation theory, the function $F$ should have the property known as infrared safety. However, a parton shower event generator is also useful for observables that are not infrared safe.} that assigns to each event $f$ a number $F(f)$. The relation of the cross section and the function $F$ is
\begin{equation}
\sigma[F] \approx \frac{1}{{\cal L}}\sum_{n=1}^N
F(f_{\!n})\;\;,
\label{eq:experiment}
\end{equation}
where ${\cal L}$ is the integrated luminosity for an experimental run and the $f_n$ are the observed events. For example, the cross section to produce a Higgs boson and two jets having certain characteristics is specified by setting $F(f) = 1$ if $f$ contains a Higgs boson and two jets having these characteristics and $F(f) = 0$ otherwise.\footnote{The case in which $F(f)$ takes values 0 or 1 is the most common, but other possibilities are allowed. For instance, the energy-energy correlation function in electron-positron annihilation is of the more general variety.
} 

A parton shower Monte Carlo event generator calculates this cross section by producing a large number $N$ of simulated events $f_n$, each with an accompanying weight $w_n$. The calculated cross section is then
\begin{equation}
\sigma[F] \approx \frac{1}{N}\sum_{n=1}^N
w_n\,F(f_{\!n})\;\;.
\label{eq:MCaverage}
\end{equation}
Most typically, the weights are all equal, so that $1/w_n$ is the simulated luminosity per point ${\cal L}/N$. Our definition of the category of parton shower Monte Carlo event generator includes the possibility that the weights are complex numbers produced for each event. It is always possible to throw away the imaginary parts of the $w_n$ since we know in advance that the imaginary part of the sum in Eq.~(\ref{eq:MCaverage}) vanishes, so having complex weights is equivalent to having real weights that can be positive or negative. This situation occurs in typical event generators \cite{MCNLO, KMS} that are based on next-to-leading order perturbation theory.\footnote{The recent paper \cite{positive} provides an exception to this rule.}

In a typical parton shower event generator, the physics is modeled as a process in classical statistical mechanics. Some number of partons are produced in a hard interaction. Then each parton has a chance to split into two partons, with the probability to split determined from an approximation to the theory. Parton splitting continues in this probabilistic style until a complete parton shower has developed.

The parton splitting probability is biggest when the two daughter partons are almost massless with nearly collinear momenta or when one of their momenta is soft (near $p = 0$), or both. There is a simple underlying approximation used: the amplitude for producing $m+1$ partons when two of the momenta $p_i$ and $p_j$ are nearly collinear or one is soft factors into a splitting function times the matrix element for producing $m$ partons.

The underlying approximation is the factorization of amplitudes in the soft or collinear limits. However, further approximations are usually added:
\begin{enumerate}
\item The interference between a diagram in which a soft gluon is emitted from one hard parton and a diagram in which the same soft gluon is emitted from another hard parton is treated in an approximate way,  with the ``angular ordering'' approximation.
\item Color is treated in an approximate way, valid when $1/N_{\rm c}^2 \to 0$ where $N_{\rm c} = 3$ is the number of colors.
\item Parton spin is treated in an approximate way. According to the full quantum amplitudes, when a parton splits, the angular distribution of the daughter partons depends on the mother parton spin and even on the interference between different mother-parton spin states. This dependence is typically ignored.
\end{enumerate}
With the use of these further approximations, one can get to a formalism in which the shower develops according to classical statistical mechanics with a certain evolution operator.

Our purpose in this paper is to investigate whether one can have a formulation of parton showers based on the factorization of amplitudes in the soft or collinear limits in which one does not make the additional approximations enumerated above. For this, one would have to use quantum statistical mechanics instead of classical statistical mechanics. 

It might seem that doing the problem in quantum mechanics is hopelessly complicated. However, within the soft/collinear factorization approximation, the problem is fairly simple because it is almost classical. In fact, if partons did not have color or spin, the problem would be classical (as we discuss in Sec.~\ref{sec:notation}). Thus what we need is a fully quantum treatment of color and spin. We arrange for this by making use of the quantum density operator in color $\otimes$ spin space.

In the subsequent sections, we define evolution equations for the quantum density matrix within the soft/collinear factorization approximation. The matrix evolves in ``shower time'' from harder splittings to softer splittings. The iterative solution of these equations gives $\sigma[F]$ in the form of a sum of integrals. To give some idea of the structure, we omit any mention of hadronization and write the result in a notation that is quite abbreviated compared to the notation in the body of the paper,
\begin{equation}
\sigma[F] =
\int d P_{0}\ f_0
\sum_{N=1}^\infty \prod_{j=1}^N 
\left(
\int d\zeta_j\ f_j
\right)\
F
\;\;.
\end{equation}
There is, first of all, an integration (including sums, for discrete variables) over momenta, flavors, spins, and colors for initial partons that emerge from the hard matrix element and its complex conjugate. Here we call all of these variables collectively $P_0$, the initial partonic variables. There is a function $f_0$ that depends on $P_0$ and represents the hard matrix element at the start of the shower times its complex conjugate. Then there is a sum over how many splittings, $N$, there are.\footnote{We have formally iterated the evolution equation an infinite number of times, allowing any number of splittings. However, we imagine that there is a cutoff on splitting hardness, so that very large values of $N$ are seldom encountered. Some of our splittings are $1 \to 1$ self interactions rather than $1 \to 2$ splittings.} Next there is an integration over splitting variables $\zeta_j$ for the $j$th splitting. The splitting variables include the label telling which parton split and momentum variables, for which a dimensionless virtuality $y$, a momentum fraction $z$, and an azimuthal angle $\phi$ might be used. There are also discrete flavor, color, and spin variables. At each splitting, there is a set of starting partonic variables, $P_{j-1}$ and a set of new partonic variables $P_j$ that are determined by $P_{j-1}$ and the splitting parameters $\zeta_j$. For each splitting, there is a function $f_j$ that depends on $P_{j-1}$ and $\zeta_j$. We have integrations over the splitting parameters for splittings 1 through $N$. At the end, there is the measurement function $F$ that depends on the partonic variables $P_N$ reached after all of the splittings.

The structure of this representation is similar to that in conventional parton showers, with the functions $f_j$ made from splitting functions and Sudakov exponentials that express the probability for not splitting. There are, however, some important structural differences that result from not making the approximations 1, 2, and 3 above. Chief among them is the use of the spin and color variables.

What we develop in this paper is an evolution equation that results in a representation of $\sigma[F]$ as integrals of known functions. Of course, one will want to turn the integrals into numbers. How to do that is a question of numerical integration that we leave for future work. However, it may be useful to sketch how a numerical evaluation might work.

To evaluate $\sigma[F]$ numerically, one has to construct the functions involved as described in the body of this paper and then perform all of the integrations (and sums). In particular, Monte Carlo integration can be used for many of the integrations. In a numerical method that is very, very simple, one can choose random points $P_0$ first, according to a density $\rho_0$. Then one would choose the first splitting variables, $\zeta_1$, according to a density $\rho_1$ that is determined by the variables $P_0$. This would determine new partonic variables $P_1$. Continuing, we choose $\zeta_j$ according to a density $\rho_j$ that is determined by variables $P_{j-1}$ and we use $\zeta_j$ together with $P_{j-1}$ to determine $P_j$. At the end, we use $P_N$ as input to the measurement function $F$. This process constitutes a Markov chain that produces ``events'' with a final state $P_N$. The probability density for getting a final state $P_N$ with a given shower history is $\rho_0$ times the product of the $\rho_j$ for $j \ge 1$. We multiply $F(P_n)$ by a weight equal to $f_0/\rho_0$ times a product of the $f_j/\rho_j$.

It would be a design goal to choose the $\rho_j$ to be roughly proportional to the absolute values of $f_j$. This kind of importance sampling would produce weights that do not vary over a wide range.

We may note that in a conventional parton shower, the $f_j$ are everywhere positive and, for $j \ge 1$, integrate to 1. Thus one can choose the $\rho_0$ to be proportional to $f_0$ and $\rho_j = f_j$ for $j \ge 1$. Then the weight function is a constant. In our case, the factors in $f_j$ are not everywhere positive, so we expect to need weights, which could have either sign. 

What we have described above would generate a very conceptually simple numerical solution to the evolution equation. We expect that one could do much better, particularly by performing the spin sums not by numerical Monte Carlo summation but by exact summation. For this, one could adapt the method proposed by Collins \cite{JCCspin} and elaborated by Knowles \cite{KnowlesSpin} and by Richardson \cite{HerwigSpin}.

We leave issues of the numerical evaluation of the integrals, beyond this simple discussion, for future work.

Some features of the formalism presented here can best be understood by asking what would happen if we kept only the leading $1/N_\Lc^2 \to 0$ limit and averaged over spins everywhere, thus making approximations 2 and 3 above. We would then have a shower based on gluon emission from color dipoles. With such a picture, the imposition of a cut to enforce angular ordering (approximation 1) is not needed: interference between gluon emissions from both halves of a color dipole is already included. The resulting evolution equation would then be similar to what is implemented as $k_\perp$-showers in \textsc{Pythia}, as described in Ref.~\cite{SjostrandSkands}. One could also modify the formalism presented here change from $k^2$ as the evolution variable to the version of $k_\perp^2$ used in \textsc{Pythia}. Then there would be two main features that differed between the present formalism and \textsc{Pythia}. One difference is in the choice of splitting functions. The \textsc{Pythia} choice is the Altarelli-Parisi splitting functions, defined with a certain definition of the momentum fraction $z$. Our splitting functions are made from the Feynman diagrams for one off-shell parton producing two on-shell partons with only a minimal manipulation to separate this part of the diagram from the hard scattering to which it attaches. The definitions match in limit of collinear splittings, but differ away from this limit. The other difference is in the momentum mapping that connects the momentum space for two initial state partons and $m+1$ final state partons to that with $m$ final state partons. We have investigated \cite{Ringberg} the possibility of using the Catani-Seymour \cite{CataniSeymour} mapping and splitting functions, which are commonly used for next-to-leading order calculations. However, we have here adopted a mapping that avoids the use of designated ``spectator'' partons that share some of their momenta.\footnote{One motivation for avoiding a special role for designated ``spectator'' partons is that if one wants to go to a next-to-leading order splitting kernel or to subtractions for a NNLO perturbative calculation, problems can arise from a third parton becoming collinear with the designated spectator parton \cite{TrocsanyiPrivateComm}.} \textsc{Pythia} uses the Catani-Seymour momentum mapping for final state splittings and something more complicated for initial state splittings. We note here the recent paper \cite{Vincia}, which explores other possibilities for both momentum mapping and splitting functions. We also note that S.~Schumann and F.~Krauss and, separately, M.~Dinsdale, M.~Temick, and S.~Weinzierl have very recently implemented the Catani-Seymour dipole subtraction functions and momentum mappings as the basis for a parton shower. The first results look promising \cite{Schumann}.

One of our goals has been to have a formulation that can coexist easily with matching the probabilities generated by showering to known exact tree level matrix elements, as in Refs.~\cite{CKKWL} and with using next-to-leading order hard matrix elements as in Refs.~\cite{MCNLO, KMS, NagySoper}. However, we leave for future work the analysis of how one can match the showers to the exact tree level matrix elements or to next-to-leading order hard matrix elements.

One may wonder whether removing the approximations 1, 2, and 3 listed above is numerically important. We do not have a definitive answer. What we would like to do is to set up a formalism that does not make approximations beyond the basic soft/collinear factorization approximation, then (in future work) implement this formalism as a working algorithm and computer code. One could then make the further approximations separately or all together and see what difference they make.

We close this introduction with some comments on whether a parton shower Monte Carlo event generator ought to allow weights for generated events, and in particular negative weights. 

We first note that in a real experiment the relation between the measurement function $F$ and the measured cross section is a little more complicated than we indicated in Eq.~(\ref{eq:experiment}). Instead, we have 
\begin{equation}
\sigma[F] \approx \frac{1}{{\cal L}}\sum_{n=1}^N
\frac{1}{a_n}\,F(f_{\!n})\;\;,
\label{eq:experiment2}
\end{equation}
where ${\cal L}$ is the integrated luminosity for an experimental run in which $N$ total events $n$ are collected and $a_n$ is the acceptance for the event resulting from the way the detector is triggered. For example, if a fraction $10^{-3}$ of a certain kind of event is recorded, then for those events $a_n$ is $10^{-3}$. Thus the weight factors $w_n/N$ in Eq.~(\ref{eq:MCaverage}) are analogous to $1/(a_n{\cal L})$ in the analysis of real data.

We can also examine the effect of weights on the statistical error in Eq.~(\ref{eq:MCaverage}). The expected error ${\cal E}$ is given by
\begin{equation}
{\cal E}^2 \sim \frac{1}{N}
\left(\frac{1}{N}\sum_{n=1}^N\,
\big[w_n\,F(f_{\!n})- \langle wF\rangle \big]^2\right)
\;\;.
\end{equation}
Here $\langle wF\rangle$ indicates an average. If it is very expensive to use a large $N$, for instance because calculating $F(f_n)$ requires a full detector simulation, then one would like to make ${\cal E}^2$ as small as possible for a fixed $N$. That suggests not that $w_n$ should be constant, but that $w_n\,F(f_{\!n})$ should be approximately constant for the observables $F$ of most interest.\footnote{Thus, if our primary interest were in the high $P_T$ tail of a jet $P_T$ distribution, we would not want to use most of the available computer time to generate low $P_T$ events. Rather, we would want to generate few low $P_T$ events, giving each of them a high weight to compensate.} It is never possible to make $w_n\,F(f_{\!n})/\langle wF\rangle = 1$ for all events, but one does not want to have lots of events for which this ratio is much smaller than 1 nor any events for which the ratio is much bigger than 1. Having events for which $w_n\,F(f_{\!n})/\langle wF\rangle \sim - 1$ is not a good thing, but, since $|-1|$ is not much larger than 1, it is not really damaging from the point of view of avoiding large statistical errors.

Evidently having weights that are real numbers of either sign, or complex numbers, does not make it impossible to apply Eq.~(\ref{eq:MCaverage}). It does, however, make the analysis a little more complicated. However, we believe that that the added complication does not present a serious problem.

\paragraph{Preview.} 
Since this is a rather lengthy paper, some preview of what is in it may be helpful. Section \ref{sec:notation} contains an introduction to the notation we use. This notation is, we think, useful for thinking about a variety of formulations of the parton shower idea. We present it in the context of a simple scalar field theory that is free from a lot of the complications of quantum chromodynamics (QCD). We then turn to QCD, with its complications. We present in Sec.~\ref{sec:structure} the structure that we propose for a parton shower that contains quantum interference. In order to present this structure in just a few pages, we leave for later sections most of the detailed definitions. The first of these, the momentum and flavor mapping, is covered in Sec.~\ref{sec:mapping}. Then Sec.~\ref{sec:spinstates} covers spin. This provides enough background to present the splitting functions for the quantum amplitudes in Sec.~\ref{sec:QuantumSpltting}. The description of color, which is rather more complicated than that of spin, is presented in Sec.~\ref{sec:color}. We are then able to specify the shower evolution operator in Secs.~\ref{sec:ClassicalEvolution} and \ref{sec:HIoft}. We follow this with discussions of two interesting issues, the evolution of color in Sec.~\ref{sec:colorstructure} and soft gluon coherence in Sec.~\ref{sec:softcoherence}. We analyze the structure of the functions that appear in the Sudakov exponent in Sec.~\ref{sec:inclusiveH}. Eventually shower evolution stops and a hadronization model is inserted. We discuss this in Sec.~\ref{sec:ShowerEnd}. We present some concluding remarks in Sec.~\ref{sec:conclusions}. There are two appendices that deal with certain technical issues.

\section{A notation for parton showers}
\label{sec:notation}

Starting in the next section, we present a formulation for parton showers in QCD hard scattering events in hadron-hadron collisions, taking into account the complexities introduced by spin and color correlations and by soft, wide angle gluon emissions in addition to collinear splittings. In order to do this, we use a mathematical language that helps to organize the algorithm. Alas, the complexities of the real physical situation make the needed construction a bit subtle. Therefore, in this section we first introduce some of the needed language in a simpler situation. The notation introduced here will be used again for QCD in the subsequent sections.

Consider the process $e^+e^- \to {\it hadrons}$ in a world in which hadrons consist of just one kind of massless scalar particle, which has no color. The $e^+e^-$ annihilation produces a pair of virtual scalar particles through an interaction that we need not specify. In the evolution of the hadronic state, we can still have collinear singularities similar to those found in QCD if the theory consists of $\phi^3$ theory in six dimensional space-time. The cross section to measure an observable $F$ can be written as
\begin{equation}
\label{eq:sigmaFphi3}
\sigma[F] = \sum_{m}\frac{1}{m!}
\int\!\big[d\{p\}_{m}\big]\ 
|\ME(\{p\}_{m})|^2
F(\{p\}_{m})
\,\,.
\end{equation}
Here there is a sum over the number $m$ of produced particles, $\{p\}_{m} = \{p_1,\dots,p_m\}$, and
\begin{equation}
\label{eq:pmeasure}
\begin{split}
\int \big[d\{p\}_{m}\big] \equiv{}& 
\prod_{i=1}^m
\left\{\int  \frac{d^6p_i}{(2\pi)^{6}}\,2\pi\delta_{+}(p_{i}^{2})\right\}
(2\pi)^{6}
\delta\bigg(P_0 -\sum_{i=1}^{m}p_{i}\bigg)
\;\;,
\end{split}
\end{equation}
with $P_0 = (\sqrt s,\vec 0)$. The function $\ME(\{p\}_{m})$ gives the matrix element to produce $m$ particles with momenta $\{p\}_{m}$, while $F(\{p\}_{m})$ describes the measurement to be done.

We want to describe this using an algorithm that approximates $|\ME(\{p\}_{m})|^2$ based on $|\ME(\{p\}_{m})|^2$ for $m=2$ and the subsequent generation of the rest of the particles based on a narrow angle approximation for one particle to split into two. This is to be done using a Monte Carlo simulation in which the system evolves from 2 particles to many particles as a simulation time $t$ progresses from $t=0$ to a large value, at which the simulation is terminated. There are various possibilities for the physical meaning of the time $t$. We will take it that a splitting $l \to i + j$ occurs at time $t = \log(Q_0^2/(2p_i  \cdot p_j))$, where $Q_0^2$ is the hardness scale of the hard interaction with which we start.

At each stage of this simulation, let the cross section to have $m$ particles with momenta $\{p\}_{m}$ be $\rho(\{p\}_{m},t)$. Summing over the number of particles and integrating over momenta gives the total cross section,
\begin{equation}
\sigma_{\rm T} =
\sum_{m}\frac{1}{m!}
\int\!\big[d\{p\}_{m}\big]\
\rho(\{p\}_{m},t)
\;\;.
\end{equation}
At the final time, $t_{\rm f}$, the value of the measurement function is
\begin{equation}
\sigma[F] = 
\sum_{m}\frac{1}{m!}
\int\!\big[d\{p\}_{m}\big]\
\rho(\{p\}_{m},t_{\rm f})\,
F(\{p\}_{m})
\;\;.
\end{equation}

The possible functions $\rho$ (at a given time $t$) form a vector space, so that $\rho$ at time $t$ can be considered to be a vector $\sket{\rho(t)}$. We use rounded brackets here. The notation $\ket{\psi}$ is reserved for a quantum state, while $\sket{\rho}$ denotes a state in the sense of statistical mechanics. We therefore call it a statistical state. The inner product is\footnote{Note that there is no * here.}
\begin{equation}
\sbrax{A}\sket{B} = \sum_{m}\frac{1}{m!}
\int\!\big[d\{p\}_{m}\big]\
A(\{p\}_{m})\,
B(\{p\}_{m})
\;\;.
\end{equation}
We can define basis vectors $\sket{\{p\}_{m}}$ in this space so that
\begin{equation}
\rho(\{p\}_{m},t) =
\sbrax{\{p\}_{m}}\sket{\rho(t)}
\;\;.
\end{equation}
With these definitions, there is a completeness relation
\begin{equation}
1 = \sum_{m}\frac{1}{m!}
\int\!\big[d\{p\}_{m}\big]\
\sket{\{p\}_{m}}\sbra{\{p\}_{m}}
\;\;.
\end{equation}
The measurement function $F$ can also be considered to be a vector, $\sbra{F}$. Thus
\begin{equation}
\sigma[F] = \sbrax{F}\sket{\rho(t_{\rm f})}
\;\;.
\end{equation}
There is a special vector $\sbra{1}$ with
\begin{equation}
\sbrax{1}\sket{\{p\}_{m}} = 1
\;\;.
\end{equation}
This vector represents the totally inclusive measurement function corresponding to the total cross section,
\begin{equation}
\sigma_{\rm T} = \sbrax{1}\sket{\rho(t)}
\;\;.
\end{equation}

Now we are ready to discuss the evolution of the statistical state. We take the evolution to be given by a linear operator ${\cal U}(t,t')$, with
\begin{equation}
\sket{\rho(t)} = {\cal U}(t,t') \sket{\rho(t')}
\;\;.
\end{equation}
Here ${\cal U}(t,t) = 1$. These operators have the group composition property
\begin{equation}
{\cal U}(t_3,t_2)\, {\cal U}(t_2,t_1)  = {\cal U}(t_3,t_1)\;\;.
\end{equation}

The class of evolution operators that we will use is defined by two operators. The first is an infinitesimal generator of evolution or hamiltonian, ${\cal H}_\LI(t)$. We can specify ${\cal H}_\LI(t)$ by giving its action on an arbitrary state $\sket{\rho}$,
\begin{equation}
\label{eq:simplesplit}
{\cal H}_\LI(t)
\sket{\rho}
\;\;.
\end{equation}
In a lowest order shower, the operator ${\cal H}_\LI(t)$ describes parton splitting, changing a state with $m$ particles to one with $m+1$ particles. One of the particles in $\{p\}_{m}$, say particle $l$, is removed and replaced by two, with momenta $\hat p_l$ and $\hat p_{m+1}$. The quantum amplitude after the splitting is approximately
\begin{equation}
\ME(\{\hat p\}_{m+1}) \approx \ME(\{p\}_{m}) \times 
\frac{g}{2\hat p_l\!\cdot\! \hat p_{m+1}}
\;\;.
\label{eq:phi3factorization}
\end{equation}
Here, to precisely define the right-hand side, one needs to redefine the momenta so that the mother parton has a momentum $p_{l} \approx \hat p_l + \hat p_{m+1}$ that is nevertheless on-shell, $p_{l}^2 = 0$. Thus the momenta $\{p\}_m$ are functions of the momenta $\{\hat p\}_{m+1}$. We omit a discussion here of the various ways to define this momentum mapping. More important for now is the idea that the quantum matrix element factorizes in the form (\ref{eq:phi3factorization}) when $2 \hat p_l \cdot \hat p_{m+1}$ is much smaller than any of the dot products among the momenta in $\{p\}_m$. This factorization is at the heart of the reason why parton shower Monte Carlo programs give useful approximations. For the statistical splitting function in Eq.~(\ref{eq:simplesplit}), we need the square of the quantum amplitude. Thus we want
\begin{equation}
\label{eq:simplesplit2}
\sbra{\{\hat p\}_{m+1}} 
{\cal H}_\LI(t)
\sket{\rho}
= 
\sum_l
\delta\!\left(t - \log\left(\frac{Q_0^2}{2\hat p_l \!\cdot\! \hat p_{m+1}}
\right)\right)
\left[\frac{g}{2\hat p_l \!\cdot\! \hat p_{m+1}}\right]^2
\sbrax{\{p\}_{m}} 
\sket{\rho}
\;\;.
\end{equation}
We have inserted the definition of the Monte Carlo time, $t$, that we here imagine using. For our present pedagogical purposes, the details of the definition of ${\cal H}_\LI(t)$ are not so important. What is important is that it reflects the factorization (\ref{eq:phi3factorization}).

The second operator used in the construction of ${\cal U}(t',t)$ is a no-change operator ${\cal N}(t,t')$ with ${\cal N}(t,t) = 1$ and
\begin{equation}
{\cal N}(t_3,t_2)\, {\cal N}(t_2,t_1)  = {\cal N}(t_3,t_1)\;\;.
\label{eq:Ngroupcomposition0}
\end{equation}
The no-change operator leaves the basis states unchanged except for multiplying each of them by an eigenvalue $\Delta$:
\begin{equation}
{\cal N}(t',t)
\sket{\{p\}_{m}} =
\Delta(t',t;\{p\}_{m})
\sket{\{p\}_{m}}
\;\;.
\label{eq:Neigenvalue0}
\end{equation}
\FIGURE{
\includegraphics[width = 10 cm]{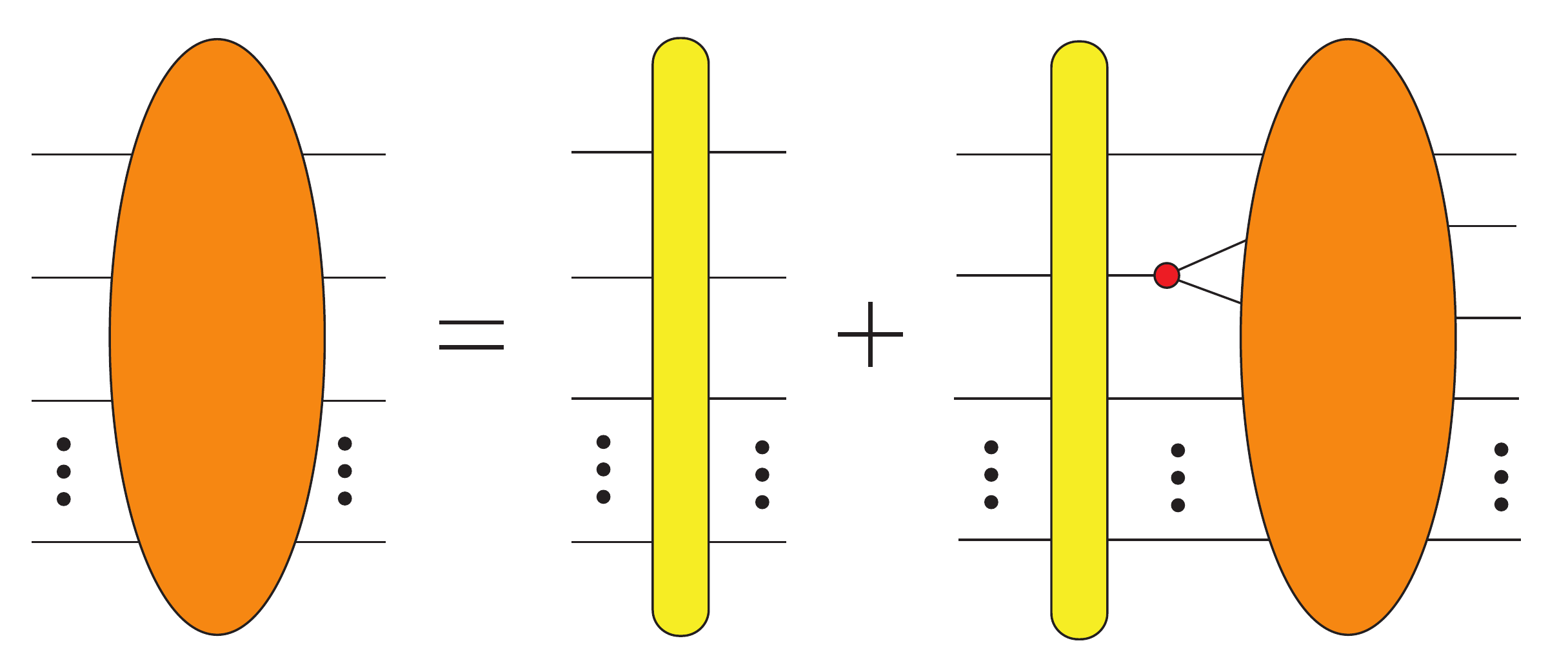}
\caption{Illustration of Eq.~(\ref{eq:evolution0}). The wide ovals represent the evolution operator ${\cal U}$ while the narrow ovals represent the no change operator ${\cal N}$, which provides the Sudakov exponentials, and the small circle is a parton splitting. Shower time $t$ runs from left to right.}
\label{fig:MainEqn}
}

The evolution operator ${\cal U}$ is expressed in terms of the hamiltonian and the no-change operators by (see Fig.~\ref{fig:MainEqn})
\begin{equation}
\label{eq:evolution0}
{\cal U}(t_3,t_1) = {\cal N}(t_3,t_1)
+ \int_{t_1}^{t_3}\! dt_2\ 
{\cal U}(t_3,t_2)\,{\cal H}_\LI(t_2)\,{\cal N}(t_2,t_1)
\;\;.
\end{equation}
This equation is interpreted as saying that either the system evolves
without splitting from $t_1$ to $t_3$, or else it evolves without splitting until an intermediate time $t_2$, splits at $t_2$, and then evolves (possibly with further splittings) from $t_2$ to $t_3$.

We need one more ingredient. We wish to construct the evolution so that it leaves the total cross section invariant, $\sbra{1} {\cal U}(t',t) \sket{\rho} = \sbrax{1} \sket{\rho}$. This should hold for every state $\sket{\rho}$, so
\begin{equation}
\label{eq:unitarity0}
\sbra{1} {\cal U}(t',t)  = \sbra{1}
\;\;.
\end{equation}
This assumption defines what ${\cal N}(t_3,t_1)$ has to be. Multiply Eq.~(\ref{eq:evolution0}) on the left by $\sbra{1}$ and on the right by $\sket{\{p\}_{m}}$. Then, using Eqs.~(\ref{eq:unitarity0}) and (\ref{eq:Neigenvalue0}) we have
\begin{equation}
\label{eq:evolution01}
1 = \Delta(t_3,t_1;\{p\}_{m})
+ \int_{t_1}^{t_3}\! dt_2\ 
\sbra{1}{\cal H}_\LI(t_2)\sket{\{p\}_{m}}
\,\Delta(t_2,t_1;\{p\}_{m})
\;\;.
\end{equation}
If we now differentiate with respect to $t_3$, we get
\begin{equation}
\label{eq:evolution02}
\frac{d}{dt_3}\,\Delta(t_3,t_1;\{p\}_{m}) =
- 
\sbra{1}{\cal H}_\LI(t_3)\sket{\{p\}_{m}}
\,\Delta(t_3,t_1;\{p\}_{m})
\;\;.
\end{equation}
The solution of this with the initial condition $\Delta(t_1,t_1;\{p\}_{m}) = 1$ is
\begin{equation}
\Delta(t_3,t_1;\{p\}_{m}) =
\exp\left(-\int_{t_1}^{t_3} d\tau\ 
\sbra{1}{\cal H}_\LI(\tau)
\sket{\{p\}_{m}}\right)
\;\;.
\label{eq:sudakov0}
\end{equation}
This result has a simple interpretation. The quantity $\sbra{1}{\cal H}_\LI(\tau)\sket{\{p\}_{m}}$ is the total probability for one of the partons in the state $\{p\}_{m}$ to split at time $\tau$. The exponential, known as the Sudakov factor, is the probability that none of these partons splits between $t_1$ and $t_3$.

This operator notation provides a convenient way to express the essence of standard shower Monte Carlo algorithms. Different algorithms differ in how the evolution variable $t$ is defined and in exactly what the splitting operator ${\cal H}_\LI(t)$ is.

At its heart, the shower Monte Carlo idea is that soft interactions factor from hard interactions in QCD. After being produced in a hard interaction, partons in QCD travel a long way before undergoing much softer interactions such as splitting. What subsequent splitting does occur does not much change the basic hard matrix element for producing (almost) on-shell partons. There are some complications, however. On shell partons carry both spin and color. Furthermore, soft gluons can transmit color changes over long distances. In the subsequent sections, we will extend the meaning of the symbols used here so as to accommodate spin and color.

\section{Structure of the calculation}
\label{sec:structure}

In this section we introduce notational conventions and a general structure for the calculation that we will use later in the paper.

\subsection{The space of quantum parton states}
\label{sec:quantumstates}

In order to describe showers, we need a notation for the description of quantum states consisting of two initial state partons and $m$ final state partons. The partons are labeled by an index such as $i$ that takes values ``${\mathrm a}$'' or ``${\mathrm b}$'' for the initial state partons and $1, 2, \dots, m$ for the final state partons. Each parton is described by a momentum $p$, a flavor $f \in \{{\rm g},{\rm u},\bar {\rm u}, {\rm d}, \bar {\rm d}, ...\}$, a spin index $s$ and a color index $c$. We denote the quantum numbers of such a state by
\begin{equation}
\{p,f,s,c\}_{m}\equiv
\{\eta_{\La}, a, s_\La, c_\La;
  \eta_{\Lb}, b, s_\Lb, c_\Lb; 
p_{1}, f_{1},s_{1}, c_{1}; ...;p_{m}, f_{m}, s_{m}, c_{m}\}
\;\;.
\end{equation}
Here there is a special notation with respect to the incoming partons. The momentum fractions of the incoming partons are denoted by $\eta_\La$ and $\eta_\Lb$, defined below. We call the flavor of parton ``a'' simply $a$ and we call the flavor of parton ``b'' simply $b$. This notation is useful for designating the parton distribution functions, $f_{a/A}(\eta_\La,\mu_\LF^2)$ and $f_{b/B}(\eta_\Lb,\mu_\LF^2$). For the purpose of describing backward evolution of the initial state partons, we will often need the antiflavors of the incoming partons. We use $f_\La$ and $f_\Lb$ for these,
\begin{equation}
\begin{split}
\label{eq:initialantiflavors}
f_\La ={}& -a
\;\;,
\\
f_\Lb ={}& -b
\;\;.
\end{split}
\end{equation}
Here our notation is $-{\rm u} = \bar {\rm u}$, $-\bar{\rm u} = {\rm u}$, $-{\rm g} = {\rm g}$, {\it etc}.

The final state partons are always on-shell. Our kinematics allows parton masses, with
\begin{equation}
\label{eq:partonmass}
p_j^2 = m^2(f_j)
\;\;.
\end{equation}
To describe the momenta of the initial state partons, we start by defining $p_\LA$ and $p_\LB$ to be massless approximations to the momenta of the two incoming hadrons
\begin{equation}
\begin{split}
p_\LA^2 ={}& 0
\;\;,
\\
p_\LB^2 ={}& 0
\;\;,
\\
2p_\LA \!\cdot\! p_\LB ={}& s
\;\;.
\end{split}
\end{equation}
The initial state partons are on-shell. In general, they can have masses but, with a small modification of the notation, their masses could be set to zero.\footnote{There are several possibilities for the treatment of masses of initial state partons and there are some subtle issues associated with the choice. We mention some of these issues in Sec.~\ref{sec:conclusions}.} In any case, we take the initial state partons to have zero transverse momentum. Thus
\begin{equation}
\begin{split}
\label{eq:etadef}
p_\La ={}& \eta_\La p_\LA + \frac{m^2(f_\La)}{\eta_\La s}\,p_\LB
\;\;,
\\
p_\Lb ={}& \eta_\Lb p_\LB + \frac{m^2(f_\Lb)}{\eta_\Lb s}\,p_\LA
\;\;.
\end{split}
\end{equation}
This defines the momentum fractions $\eta_\La$ and $\eta_\Lb$.

In the event that we include parton masses, the hardness scale $Q_0^2$ at which the parton shower is initiated should be much bigger than the mass of any parton that is included as a possible constituent of the incoming hadrons. For example, if the hard process were $t + \bar t$ production near threshold, then top quarks should not be used as possible initial state partons. Thus we demand that
\begin{equation}
\label{eq:Q0condition}
Q_0^2 > 4 m_{\rm H}^2
\;\;,
\end{equation}
where $m_{\rm H}$ is the mass of the heaviest quark that is included as an initial state parton, typically the $b$ quark. In any reasonable application of the formalism of this paper, the ``$>$'' here will be ``$\gg$.'' Equation (\ref{eq:Q0condition}) suffices to make certain kinematic formulas in the paper work. At any stage in the shower, we define $p_\La + p_\Lb = Q$. At the first step, the starting hard scattering, with a sensible definition of the starting hardness scale we must have $Q^2 \ge Q_0^2$. The values of $Q^2$ increase as the shower develops. Thus at any stage we will have
\begin{equation}
\label{eq:Qcondition}
(p_\La + p_\Lb)^2 > 4 m_{\rm H}^2
\;\;,
\end{equation}

We will impose a kinematic restriction on the momentum fractions,
\begin{equation}
\begin{split}
\label{eq:etalimits0}
\frac{m^2(f_\Lb)}{\eta_\Lb s} <{}  \eta_\La 
\;\;,
\hskip 2 cm
\frac{m^2(f_\La)}{\eta_\La s} <{}  \eta_\Lb 
\;\;.
\end{split}
\end{equation}
These limits require that the momentum in the system in the direction of $p_\LA$ comes mainly from parton ``a,'' and the same for $p_\LA \leftrightarrow p_\LB$,  $\La \leftrightarrow \Lb$. We can be sure that both conditions hold by requiring
\begin{equation}
\label{eq:etalimits}
\eta_\La\eta_\Lb s > m_{\rm H}^2
\;\;.
\end{equation}
The reason for imposing this condition is as follows. Given Eq.~(\ref{eq:etadef}), there are two choices for $\eta_\La\eta_\Lb s$ that yield the same value of $(p_\La + p_\Lb)^2$. As long as condition (\ref{eq:Qcondition}) holds, the larger of the two choices for $\eta_\La\eta_\Lb s$ satisfies Eq.~(\ref{eq:etalimits}). Imposing Eq.~(\ref{eq:etalimits}) eliminates the other solution, in which  parton ``a'' moves in approximately the $p_\LB$ direction and parton ``b'' moves in approximately the $p_\LA$ direction, creating a large value for $(p_\La + p_\Lb)^2$.

The upper limit on $\eta_\La$ and $\eta_\Lb$ is 1. We note here that this is an approximation. To discuss this, let $n_\LA$ and $n_\LB$ be dimensionless lightlike vectors in the directions of $p_\LA$ and $p_\LB$ respectively, normalized to $n_\LA \cdot n_\LB = 1$. Then consider, for example, the limit on $\eta_\La$. The total momentum in the direction of $n_\LA$ of the final state particles is 
\begin{equation}
\label{eq:toomuch}
{(p_\La + p_\Lb)\cdot n_\LB}
=
\left[\eta_\La + \frac{m^2(f_\Lb)}{\eta_\Lb s}\right]\,
{p_\LA\cdot n_\LB}
\;\;.
\end{equation}
This can be bigger than the available momentum $p_\LA\cdot n_\LB$ if $\eta_\La$ is very close to 1. A remedy for this would be to redefine $p_\LA$ and $p_\LB$ in our formulas. Suppose that the exact hadron momenta are $P_\LA$ and $P_\LB$. Then if we put
\begin{equation}
\begin{split}
p_\LA ={}& \lambda\ {P_\LA \cdot n_\LB}\ n_\LA
\;\;,
\\
p_\LB ={}& \lambda\ {P_\LB \cdot n_\LA}\ n_\LB
\;\;,
\end{split}
\end{equation}
then there is a value of $\lambda$ that makes $(p_\LA + p_\LB)^2 = (P_\LA + P_\LB)^2$. By taking a value of $\lambda$ that is a little smaller than this, one can ensure that the momentum in the final state in the directions of $n_\LA$ and $n_\LB$ is not more than was present in the initial state. It is this momentum that is available for the ``underlying event.'' With this adjustment, the value of $s$ in our formulas is a little less than the true c.m. squared energy, $(P_\LA + P_\LB)^2$.

Our notation with respect to spin and color is meant to be flexible. A standard helicity basis will work for spin. For color, we begin with a straightforward basis in which each parton $i$ has a color index $c_i$ that can take values 1,2,3 for quarks and antiquarks and 1,\dots,8 for gluons. Later, we will want to consider the subspace of the whole color space in which the parton state is a singlet under the $SU(3)$ color group. We will choose a basis for this subspace. Using this basis, we will still have labels that we can call $\{c\}_m$, but the new labels will describe the color links among the partons rather than individual color indices $c_i$ for the individual partons. The notation $\{p,f,c,s\}_{m}$ is thus supposed to include the possibility of any representation of the colors of the $m+2$ partons.

\subsection{The density matrix}
\label{sec:densitymatrix}

A matrix element used in the computation of a cross section can be thought of as having the form of a function of the momenta and flavors that carries indices for spin and color,
\begin{equation}
\ME(\{p,f\}_{m})
^{c_\La,c_\Lb, c_1,\dots,c_m}_{s_\La, s_\Lb, s_1,\dots,s_m}
\;\;.
\end{equation}
Here we denote the functions $\ME$ for different numbers $m$ of final state partons by the same name, simply $\ME(\{p,f\}_{m})$ rather than $\ME_m(\{p,f\}_{m})$. The array $\ME$ can be thought of as a vector in spin and color space,
\begin{equation}
\ket{\ME(\{p,f\}_{m})}
\;\;.
\end{equation}
The inner product $\brax{\ME'}\ket{\ME}$ denotes multiplying ${\ME'}^*$ by $\ME$ and summing over the spins and colors.

An observable $F$ can be specified by giving a set of functions $F(\{p,f\}_{m})$ that are linear operators on the color-spin space. (In many important cases, $F(\{p,f\}_{m})$ is a function times the unit operator on color-spin space. If $F$ is be simply made from theta functions defining final state cuts, then $\sigma[F]$ is the cross section to find the final state partons within the cuts.) With this notation the cross section for an observable $F$ takes the form\footnote{This formula contains a parton flux factor $2\eta_{\La}\eta_{\Lb}p_\LA\cdot p_\LB$ that corresponds to massless partons. The flux factor for scattering of free massive particles is more complicated. However, $(p_{\La}+p_{\Lb})^2$ is always bigger than the initial hard scale $Q_0^2$ and the formalism of this paper is valid only when $Q_0^2$ is much larger than the masses of any initial state partons. For this reason, we use the flux factor for massless parton scattering.}
\begin{equation}
\label{eq:opF}
\begin{split}
\sigma[F] = 
\sum_{m}&\frac{1}{m!}\int\big[d\{p,f\}_{m}\big]\,
\frac{f_{a/A}(\eta_{\La},\mu^2_\LF)\,
f_{b/B}(\eta_{\Lb},\mu^2_\LF)}
{4n_\Lc(a) n_\Lc(b)\,2\eta_{\La}\eta_{\Lb}p_\LA\!\cdot\!p_\LB}\,
\\
&\!\times
\bra{\ME(\{p,f\}_{m})}
F(\{p,f\}_{m})
\ket{\ME(\{p,f\}_{m})}
\;\;.
\end{split}
\end{equation}
Here the functions $f$ are parton distribution functions while $n_\Lc(a)$ is the number of colors that a parton of flavor $a$ can have, $N_\Lc = 3$ for a quark or antiquark, $N_\Lc^2 -1 = 8$ for a gluon. The factor $4n_\Lc(a) n_\Lc(b)$ turns the sum over spins and colors for the initial state partons into an average over spins and colors. We have indicated the appropriate integrations over momenta by
\begin{equation}
\label{eq:pfmeasure}
\begin{split}
\int \big[d\{p,f\}_{m}\big] g(\{p,f\}_{m})\equiv{}& 
\prod_{i=1}^m
\left\{\sum_{f_i}\int  \frac{d^4p_i}{(2\pi)^{4}}\,
2\pi\delta_{+}(p_{i}^{2} - m^2(f_i))\right\}
\sum_{a}\int_{0}^{1}\!d\eta_{\La}
\sum_{b}\int_{0}^{1}\!d\eta_{\Lb}
\\& \times
(2\pi)^{4}
\delta\bigg(p_\La + p_\Lb-\sum_{i=1}^{m}p_{i}\bigg)\
\theta\big(
m_{\rm H}^2 < \eta_\La\eta_\Lb s
\big)
\\&\times
g(\{p,f\}_{m})
\;\;.
\end{split}
\end{equation}
Here $g(\{p,f\}_{m})$ is an arbitrary function. 

The final state particles carry labels $i \in \{1,\dots,m\}$. Then particle $i$ has momentum, flavor, spin, and color given by $\{p_i,f_i,s_i,c_i\}$. One can arrange the definitions such that the amplitude $M$ is symmetric under interchange of the labels. However, we do not necessarily do so. Instead, the notation allows for a general labeling scheme.\footnote{Just to take a trivial example, in a ${\rm u}\bar {\rm u} {\rm g}$ state, the label 1 might be assigned to the up quark, 2 to the anti-up quark and 3 to the gluon. Of course, it will not work in general to use the flavors as labels because one can have two final state partons with the same label.} A measurement function $F$ must be symmetric under interchange of labels, since the labels are not physical. Two amplitudes $M$ that become the same if the labels are symmetrized are equivalent.

We will find it useful to rewrite $\sigma[F]$ in the form of a trace over the spin and color space,
\begin{equation}
\label{eq:sigmaFtrace}
\begin{split}
\sigma[F] = 
\sum_{m}&\frac{1}{m!}\int\big[d\{p,f\}_{m}\big]\,
{\rm Tr}\{ \rho(\{p,f\}_{m})
F(\{p,f\}_{m})
\}
\;\;,
\end{split}
\end{equation}
where 
\begin{equation}
\label{eq:rhodef1}
\rho(\{p,f\}_{m}) = 
\ket{\ME(\{p,f\}_{m})}
\frac{f_{a/A}(\eta_{\La},\mu^{2}_{F})
f_{b/B}(\eta_{\Lb},\mu^{2}_{F})}
{4n_\Lc(a) n_\Lc(b)\,2\eta_{\La}\eta_{\Lb}p_\LA\!\cdot\!p_\LB}\,
\bra{\ME(\{p,f\}_{m})}
\;\;.
\end{equation}
Thus $\rho$ is the density operator in color\,$\otimes$\,spin space. It is illustrated in Fig.~\ref{fig:rho1}. The density operator, for momentum as well as spin, is widely used as the basis of quantum statistical mechanics. It was introduced for the spin space in parton showers by Collins \cite{JCCspin} and is used in \textsc{Herwig} for the heavy partner particles in supersymmetry \cite{HerwigSpin}.

\FIGURE{
\includegraphics[width = 10 cm]{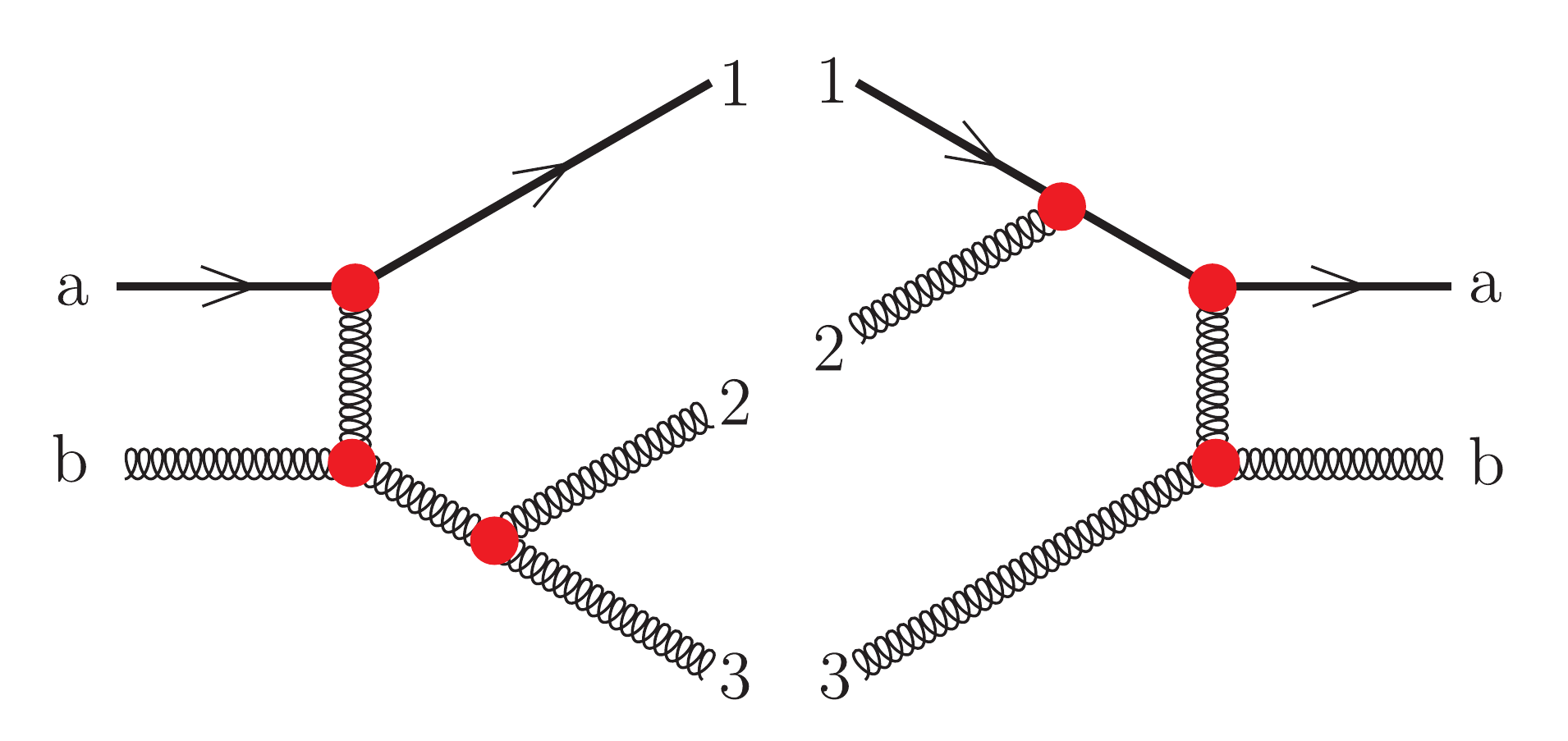}
\caption{Illustration of a  contribution to $\rho$, Eq.~(\ref{eq:rhodef1}). The Feynman graph on the left is a contribution to $\ket{\ME(\{p,f\}_{m})}$ and the Feynman graph on the right is a contribution to $\bra{\ME(\{p,f\}_{m})}$. The factor with parton distributions is not illustrated. The momenta and flavors of the labeled partons match between  $\ket{\ME(\{p,f\}_{m})}$ and $\bra{\ME(\{p,f\}_{m})}$, but the partons on the left have colors and spins $\{s,c\}_{m}$ while those on the right have possibly different colors and spins $\{s',c'\}_{m}$. Additionally, $\rho$ can contain quantum interference between different graphs, as illustrated here. We imagine that eventually the partons will evolve to form a final state in the middle and an initial state on the left and right.
}
\label{fig:rho1}
}

We can expand $\rho(\{p,f\}_{m})$ in basis states $\ket{\{s,c\}_{m}}$ for the color\,$\otimes$\,spin space,
\begin{equation}
\label{eq:rhodef2}
\rho(\{p,f\}_{m})
= \sum_{s,c}\sum_{s',c'}
\ket{\{s,c\}_{m}}\,
\rho(\{p,f,s',c',s,c\}_{m})\,
\bra{\{s',c'\}_{m}}
\;\;.
\end{equation}
Here $\rho(\{p,f,s',c',s,c\}_{m})$ is a function that depends on the momenta and flavors $\{p,f\}_{m}$, the labels $\{s,c\}_{m}$ for the quantum ``ket'' state and the labels $\{s',c'\}_{m}$ for the conjugate quantum ``bra'' state. We denote the state labels collectively by $\{p,f,s',c',s,c\}_{m}$. Thus $\rho(\{p,f,s',c',s,c\}_{m})$ a function giving the matrix elements of the density matrix. We find it convenient to base our treatment on this function.

Later, we will choose an orthonormal basis for the spin space, so that 
\begin{equation}
\brax{\{s'\}_{m}}\ket{\{s\}_{m}} =
\delta^{\{s'\}_{m}}_{\{s\}_m}
\;\;.
\end{equation}
For the color space, we will find it convenient to use a basis in which $\brax{\{c\}_{m}} \ket{\{c\}_{m}}$ is approximately but not exactly 1 and where $\brax{\{c'\}_{m}} \ket{\{c\}_{m}}$ is not generally zero for $\{c\}_{m} \ne \{c'\}_{m}$. With a non-orthogonal basis, we need to be a little careful about the notation. Suppose that we expand a vector in such a basis,
\begin{equation}
\ket{\psi} = \sum_{\{c\}_m} \ket{\{c\}_m}\, a({\{c\}_m})
\;\;.
\label{expansionexample}
\end{equation}
A convenient representation for the expansion coefficients $a({\{c\}_m})$ is obtained by taking matrix elements with elements of the dual basis $\ket{{\{c\}_m}}\dualR$ defined\footnote{Readers familiar with general relativity can think of $\brax{\{c\}_m}\ket{a}$ as the covariant components $a_c$ of $a$ and $\dualL\brax{\{c\}_m}\ket{a}$ as the contravariant components $a^c$. One can lower indices using the metric tensor $g_{cc'}$, analogous to $\brax{\{c\}_m}\ket{\{c'\}_m}$. The inverse matrix $\dualL\brax{\{c\}_m}\ket{\{c'\}_m}\dualR$ is analogous to $g^{cc'}$.} by 
\begin{equation}
\dualL\brax{{\{c'\}_m}}\ket{{\{c\}_m}} = 
\delta^{\{c'\}_{m}}_{\{c\}_m}
\;\;.
\end{equation}
Using the dual basis we can write
\begin{equation}
\dualL\brax{{\{c\}_m}}\ket{\psi}
= a({\{c\}_m})
\;\;.
\end{equation}
The expansion of any vector in the basis $\ket{\{c\}_m}$ can conveniently be obtained directly by using the completeness relation
\begin{equation}
\label{eq:dualcompleteness1}
1 = \sum_{\{c\}_m} \ket{\{c\}_m}\
\dualL\bra{{\{c\}_m}}
\;\;.
\end{equation}
If we want to expand a bra vector in the basis $\bra{\{c\}_m}$ we can use the completeness relation
\begin{equation}
\label{eq:dualcompleteness2}
1 = \sum_{\{c\}_m} \ket{{\{c\}_m}}\dualR\
\bra{{\{c\}_m}}
\;\;.
\end{equation}
This is particularly useful with respect to operators. Let $O$ be a linear operator on the color space. If $\ket{\psi}$ has expansion coefficients $a({\{c\}_m})$ defined by Eq.~(\ref{expansionexample}) and $\ket{\psi'} = O\ket{\psi}$ has expansion coefficients $a'({\{c'\}_m})$ then $O$ is conventionally described by the matrix defined by
\begin{equation}
a'({\{c'\}_m}) = \sum_{\{c\}_m} o(\{c'\}_m,\{c\}_m)
\,a({\{c\}_m})
\;\;.
\end{equation}
We can write this as
\begin{equation}
o(\{c'\}_m,\{c\}_m) = 
\dualL\bra{{\{c'\}_m}}
O \ket{{\{c\}_m}}
\;\;.
\end{equation}

\subsection{Statistical states}

The set of functions $\rho(\{p,f,s',c',s,c\}_{m})$ forms a vector space, which we can call the statistical state space (as distinct from the quantum state space). We can call the vector corresponding to this function simply $\sket{\rho}$. Note the rounded brackets instead of angle brackets that we use for quantum states, as in ${\ket\psi}$. We can define basis vectors $\sbra{\{p,f,s',c',s,c\}_{m}}$ for the statistical state space\footnote{More precisely, the bra vectors $\sbra{F}$ are vectors in the dual space to the ket vectors $\sket\rho$, that is the space of linear functions on the vectors $\sket\rho$.} so that
\begin{equation}
\label{eq:rhoket}
\rho(\{p,f,s',c',s,c\}_{m}) = \sbrax{\{p,f,s',c',s,c\}_{m}}\sket{\rho}
\;\;.
\end{equation}
There are also ket basis vectors such that the completeness relation for the basis states is
\begin{equation}
\label{eq:completeness}
  1 = \sum_m \frac{1}{m!}\int \big[d\{p,f,s',c',s,c\}_{m}\big]\,
  \sket{\{p,f,s',c',s,c\}_{m}}
  \sbra{\{p,f,s',c',s,c\}_{m}}
  \;\;,
\end{equation}
where $\big[d\{p,f,s,s',c,c'\}_{m}\big]$ is an extension of the integration measure Eq.~(\ref{eq:pfmeasure})
\begin{equation}
\label{eq:pfssccmeasure}
\begin{split}
\int \big[d\{p,f,s',c',s,c\}_{m}\big] \equiv{}& 
\int \big[d\{p,f\}_{m}\big]
\sum_{s_{\rm a},s'_{\rm a},c_{\rm a},c'_{\rm a}}
\sum_{s_{\rm b},s'_{\rm b},c_{\rm b},c'_{\rm b}}\
\prod_{i=1}^m
\left\{
\sum_{s_i,s'_i,c_i,c'_i}
\right\}
\;\;.
\end{split}
\end{equation}
The corresponding inner product of basis states is 
\begin{equation}
\sbrax{\{p,f,s',c',s,c\}_{m}}
\sket{\{\tilde p,\tilde f,\tilde s',\tilde c',\tilde s,\tilde c\}_{\tilde m}}
= \delta_{m,\tilde m}\
\delta(\{p,f,s',c',s,c\}_{m};
   \{\tilde p,\tilde f,\tilde s',\tilde c',\tilde s,\tilde c\}_{m})
   \;\;,
\label{eq:innerproduct}
\end{equation}
where the function $\delta$ is a generalization of the ordinary delta-function that is defined by
\begin{equation}
\begin{split}
\frac{1}{m!}\int \big[d\{p,f,s',c',s,c\}_{m}&\big]\
\delta(\{p,f,s',c',s,c\}_{m};
   \{\tilde p,\tilde f,\tilde s',\tilde c',\tilde s,\tilde c\}_{m})\
   h(\{p,f,s',c',s,c\}_{m})
\\
={}&  h(\{\tilde p,\tilde f,\tilde s',\tilde c',\tilde s,\tilde c\}_{m})
\;\;.
\end{split}
\end{equation}
Here $h$ is any well behaved function of the variables indicated, defined on the integration surface.

Let us define a vector corresponding to a measurement function $F$ using\footnote{Note that this equation for the measurement function has a different structure from the equation used to define the statistical state vector $\sket{\rho}$, $\rho(\{p,f\}_{m}) = \ket{\{s,c\}_{m}}\sbrax{\{p,f,s',c',s,c\}_{m}}\sket{\rho}\bra{\{s',c'\}_{m}}$.}
\begin{equation}
\sbrax{F}\sket{\{p,f,s',c',s,c\}_{m}}
= \bra{\{s',c'\}_{m}}F(\{p,f\}_{m})\ket{\{s,c\}_{m}}
\;\;.
\label{eq:braFdef}
\end{equation}
Then, using the completeness relation (\ref{eq:completeness}), the cross section (\ref{eq:sigmaFtrace}) corresponding to a measurement function $F$ can be expressed using Eqs.~(\ref{eq:rhodef2}) and (\ref{eq:rhoket}) as
\begin{equation}
\sigma[F] = 
\sbrax{F}\sket{\rho}
\;\;.
\label{eq:sigmaF}
\end{equation}

In the case that $F$ consists of a function $F(\{p,f\}_{m})$ times a unit operator in spin-color space, the inner product $\sbrax{F}\sket{\rho}$ is
\begin{equation}
\label{eq:sigmaFsimple}
\begin{split}
\sbrax{F}\sket{\rho} ={}&
\sum_m\frac{1}{m!}
\int \big[d\{p,f,s',c',s,c\}_{m}\big]
F(\{p,f\}_{m})\
\brax{\{s'\}_{m}}\ket{\{s\}_{m}}
\brax{\{c'\}_{m}}\ket{\{c\}_{m}}
\\&\times
\rho(\{p,f,s',c',s,c\}_{m})
\;\;.
\end{split}
\end{equation}

\subsection{The resolution scale}
\label{sec:resolution}

We now need to introduce a resolution scale into our equations. We first discuss the resolution scale of the observable. In Eq.~(\ref{eq:sigmaF}), let us suppose that the observable represented by the functions $F(\{p,f\}_{m})$ is infrared safe. To be precise about what this means, we first demand that the functions $F(\{p,f\}_{m})$ be smooth functions of the momentum variables and that they be invariant under label interchanges. Then we consider a list of parton variables $\{\hat p,\hat f\}_{m+1}$ for $m+1$ partons and suppose that $\hat p_{m+1}$ becomes collinear with the momentum $\hat p_l$ of parton $l$. Then we can consider the list of parton momenta $\{p,f\}_{m}$ where $p_j = \hat p_j$ and $f_j = \hat f_j$ for $j \ne l$ while $p_l = \hat p_m + \hat p_l$ and $f_l = \hat f_m + \hat f_l$ (with the obvious definition of adding flavors). That is, the partons with variables $\{\hat p,\hat f\}_{m+1}$ could have arisen from partons with variables $\{p,f\}_{m}$ by the collinear splitting of parton $l$ into new partons with labels $l$ and $m+1$. Then infrared safety requires that
\begin{equation}
F(\{\hat p,\hat f\}_{m+1}) \to F(\{p,f\}_{m})
\end{equation}
in the limit. This should also hold if parton $m+1$ becomes soft, $\hat p_{m+1} \to 0$. It should also hold with a suitable adjustment of the notation for a splitting of one of the initial state partons. This is, so far, just the standard definition of infrared safety. It allows us to have perturbatively calculable cross sections. Now let us extend the definition to include a scale. We can say that the observable is infrared safe at scale $\mu^2$ if 
\begin{equation}
F(\{\hat p,\hat f\}_{m+1}) \approx F(\{p,f\}_{m})
\end{equation}
when $|2 \hat p_{m+1}\cdot \hat p_l| < \mu^2$. To be really precise, we should specify how good this approximation has to be, but this will not matter for our purposes. Strictly speaking an ``infrared safe observable'' without further qualification is one that is infrared safe at any scale, no matter how small. However, what is usually meant is that it is infrared safe at a scale not much smaller than the scale $Q_0^2$ of the hardest interaction in the problem. What we want to do here is to specify the splitting scale at which the observable is sensitive to the splitting. That scale could be much smaller than $Q_0^2$.

Now we introduce the concept of the density operator $\rho$ evaluated at resolution scale $\mu^2$. The idea is that interactions with scales greater than $\mu^2$ are included in $\ket{M}\bra{M}$ while interactions with scales smaller than $\mu^2$ are integrated out (for final state interactions) or included in the parton distributions (for initial state interactions).\footnote{This is the idea of the standard factorization theorem \cite{factorization}. However, we here go beyond anything that has been proved.} We can describe this in a rough way as follows. In each cut Feynman diagram, each integration region for final state partons that produces a collinear or soft divergence can be described as a region in which some group of partons with labels $i$ become collinear to a given direction of a mother parton or some become soft, so that $(\sum_i p_i)^2 \to 0$. Divide this region into subregions with $(\sum_i p_i)^2 < \mu^2$ and $(\sum_i p_i)^2 > \mu^2$. In the $(\sum_i p_i)^2 < \mu^2$ region we can combine the partons $i$ into a single effective parton for purposes of calculating the observable. That is, the partons $i$ are ``unresolved.'' This leads to a free integration over this region using a constant $F$ for fewer partons. Adding these real emission integrals to the corresponding virtual diagrams and counterterms from the parton distributions gives a finite result containing logarithms of $\mu^2/Q_0^2$.

A more intuitive way of thinking about this is to imagine writing the Feynman diagrams in a coordinate space representation, in which we integrate over the positions $x_i^\mu$ of the interaction vertices relative to the position of the hard interaction, which is determined to within $1/Q_0^2$. Then we can restrict these integrations to $|x_i^2| < 1/\mu^2$ before integrating over the final state momenta.

In order that the parton distributions in Eq.~(\ref{eq:rhodef1}) include initial state interactions at all scales smaller than $\mu^2$, the factorization scale $\mu_\LF^2$ at which the parton distributions are evaluated should be $\mu^2$. 

With this meaning of $\rho$ evaluated at resolution scale $\mu^2$, $\sbrax{F}\sket{\rho}$ is invariant under $\mu^2 \to \mu^2 + \delta \mu^2$ as long as $F$ is infrared safe at a scale equal to $\mu^2$ or larger. However, if one looks at $\sket{\rho}$ with a resolution scale smaller than $\mu^2$ then $\sbrax{F}\sket{\rho}$ does see the effect of changing $\mu^2$.

Our object in this paper is to construct an approximate version of $\sket{\rho}$ as a function of the resolution scale. The idea is to construct $\sket{\rho}$ using a parton shower, starting from $\mu^2 = Q_0^2$ and evolving to smaller values of $\mu^2$, down to a final infrared cutoff. This intuitively appealing idea has been inherent in the idea of a parton shower since the earliest days. It would be very useful to have a precise field theoretic definition of $\sket{\rho}$ as a function of resolution scale. However, this is beyond our scope in this paper.

One finds that the logarithm of the resolution scale $\mu^2$ is more useful as a variable than $\mu^2$ itself. Therefore we define Monte Carlo time $t$ by
\begin{equation}
\label{eq:mudef}
\mu^2 = Q_0^2\, e^{-t}
\end{equation}
and write the density operator with this resolution scale as $\sket{\rho(t)}$. Then also the parton distributions are evaluated at factorization scale
\begin{equation}
\label{eq:muFdef}
\mu_{\LF}^2 = Q_0^2\, e^{-t}
\;\;.
\end{equation}

\subsection{Parton shower evolution}
\label{sec:evolution}

We are now prepared to set up a quite general framework for describing a
parton shower. We take the framework as a set of axioms that we hope are reasonably intuitive. Later, we relate the operators that occur to the structure of Green functions at tree level in QCD.

We use the evolution variable $t$ that specifies the resolution scale of $\rho$ according to Eq.~(\ref{eq:mudef}). Thus $t$ starts at zero and increasing $t$ corresponds to decreasing virtuality. One stops evolution at an infrared cutoff $t_{\rm f}$ at which the use of an evolution based on perturbation theory is no longer appropriate. For instance, $t_{\rm f}$ might correspond to a $1\ {\rm GeV}^2$ virtuality. The evolving shower is represented by a state $\sket{\rho(t)}$ that begins with an initial state $\sket{\rho(0)}$. The evolution is given by a linear operator ${\cal U}(t,t')$, with
\begin{equation}
\sket{\rho(t)} = {\cal U}(t,t') \sket{\rho(t')}
\;\;.
\end{equation}
Here ${\cal U}(t,t) = 1$. These operators have the group composition property
\begin{equation}
{\cal U}(t_3,t_2)\, {\cal U}(t_2,t_1)  = {\cal U}(t_3,t_1)\;\;.
\end{equation}

The class of evolution operators that will use is defined by two operators, ${\cal H}_{\LI}(t)$ and ${\cal V}(t)$, according to the differential equation
\begin{equation}
\label{eq:evolutionplain}
\frac{d}{dt}\,{\cal U}(t,t') = 
[{\cal H}_\LI (t) - {\cal V}(t)]\,{\cal U}(t,t')
\;\;,
\end{equation}
with initial condition ${\cal U}(t,t) = 1$. 

The first operator, ${\cal H}_{\LI}(t)$, represents parton interactions and, in general, changes the number of partons and their momenta. We specify ${\cal H}_{\LI}(t)$ by giving its matrix elements
\begin{equation}
\sbra{\{\hat p,\hat f,\hat s',\hat c',\hat s,\hat c\}_{m'}} 
{\cal H}_\LI(t)
\sket{\{p,f,s',c',s,c\}_{m}}
\;\;.
\end{equation}
In a lowest order shower, which we consider in this paper, the operator ${\cal H}_{\LI}(t)$ describes $1 \to 2$ parton splitting, changing a state with $m$ final state partons to one with $m+1$ final state partons.

The second operator that controls evolution, ${\cal V}(t)$, describes the effect of virtual graphs and the ``unresolved'' part of real emission graphs. In a lowest order shower, we do not account for the virtual graphs exactly, but rather account for only the infrared singular part of the virtual graphs, which can be deduced from the real emission graphs. That is, ${\cal V}(t)$ is determined from ${\cal H}_\LI (t)$. The operator ${\cal V}(t)$ does not change the number of partons or their flavors or spins, but can change their color states.

We construct the shower algorithm in such a way that it conserves probability in a certain sense. If we were dealing with $e^+ + e^- \to {\it hadrons}$, we would demand that the development of the shower does not change the total cross section. For hadron-hadron collisions, the total cross section does not have a well defined perturbative expansion. However, if we start with a state $\sket{\rho(0)}$ that is ``hard'' in the sense that $\sbrax{\{p,f,s',c',s,c\}_{m}} \sket{\rho(0)}$ is non-zero only for parton configurations with a large transverse energy, then we can demand that as this state evolves into the shower, the contribution from $\sket{\rho(t)}$ to the total cross section does not change. The observable that measures the total cross section is  
\begin{equation}
F_1(\{p,f\}_{m}) =  1
\;\;.
\end{equation}
We will call the vector corresponding to $F_1$ simply $\sbra{1}$. Using Eq.~(\ref{eq:braFdef}), the inner product of $\sbra{1}$ with a basis state is
\begin{equation}
\sbrax{1}\sket{\{p,f,s',c',s,c\}_{m}}
= 
\brax{\{s'\}_{m}}\ket{\{s\}_{m}}\,
\brax{\{c'\}_{m}}\ket{\{c\}_{m}}\,
\;\;.
\label{eq:braF1def}
\end{equation}

The statement that shower evolution leaves contributions to the total cross section invariant is
\begin{equation}
\label{eq:unitarity}
\sbra{1} {\cal U}(t',t) \sket{\rho} = \sbrax{1} \sket{\rho}
\end{equation}
for any (suitably hard) state $\sket{\rho}$. This requirement leads to  a relation between the matrix elements of ${\cal V}$, and ${\cal H}_{\LI}$. To derive this relation we multiply Eq.~(\ref{eq:evolutionplain}) on the left by $\sbra{1}$ and on the right by $\sket{\rho}$. After using Eq.~(\ref{eq:unitarity}), we get
\begin{equation}
\label{eq:evolution2}
\begin{split}
0 ={}&
\sbra{1}[{\cal H}_{\LI}(t) - {\cal V}(t)]\,{\cal U}(t,t')\sket{\rho}
\;\;.
\end{split}
\end{equation}
Since this holds for any suitably hard state $\sket{\rho}$ we have
\begin{equation}
\label{eq:evolution3}
\begin{split}
0 ={}&
\sbra{1}[{\cal H}_{\LI}(t) - {\cal V}(t)]
\;\;.
\end{split}
\end{equation}
We multiply on the right by $\sket{\{p,f,s',c',s,c\}_{m}}$ to obtain  
\begin{equation}
\label{eq:evolution4}
\sbra{1} {\cal V}(t)\sket{\{p,f,s',c',s,c\}_{m}} =
\sbra{1}{\cal H}_{\LI}(t)\sket{\{p,f,s',c',s,c\}_{m}}
\;\;.
\end{equation}

At this point, we need to discuss the structure of the parton splitting operator ${\cal H}_{\LI}$. In subsequent sections, we derive the form of ${\cal H}_\LI(t)$ based on the structure of QCD tree level matrix elements in the limit that two of the $m+1$ partons become massless and collinear, one becomes massless and collinear with one of the beam directions, or one (a gluon) becomes soft. In this limit, the matrix elements take a factored form, ${\it hard}\otimes({\it soft\,\&\,collinear})$. This factorization leads to the definition\footnote{The definition is not unique because there is freedom to choose what to do away from the soft and collinear limits} of ${\cal H}_\LI(t)$ in Sec.~\ref{sec:HIoft}. At the moment, what we need is the structure of $\sbra{1}{\cal H}_\LI(t)\sket{\{p,f,s',c',s,c\}_{m}}$, which represents the inclusive splitting probability at splitting scale $t$. We will find
\begin{equation}
\label{eq:1H}
\sbra{1}{\cal H}_\LI(t)\sket{\{p,f,s',c',s,c\}_{m}}
= 2\,\brax{\{s'\}_{m}}\ket{\{s\}_{m}}\,
\bra{\{c'\}_{m}}h(t,\{p,f\}_{m})\ket{\{c\}_{m}}
\;\;,
\end{equation}
where the function $h(t,\{p,f\}_{m})$ is given in Sec.~\ref{sec:inclusiveH}, Eq.~(\ref{eq:hdef}). The important point is that there is a trivial spin structure and a non-trivial color structure.

We take the operator ${\cal V}(t)$ to operate only on the color space and define its action in terms of its matrix elements, for which we use the notation 
$\sbra{\{\hat c', \hat c\}_{m}}{\cal V}(t; \{p,f\}_{m}) \sket{\{c',c\}_{m}}$. The definition of ${\cal V}(t)$ in terms of these matrix elements is
\begin{equation}
\begin{split}
{\cal V}(t)\sket{\{p,f,s',c',s,c\}_{m}} = {}&  
\sum_{\{\hat c',\hat c\}_{m}} 
\sket{\{p,f,s',\hat c',s,\hat c\}_{m}}
\sbra{\{\hat c', \hat c\}_{m}}
{\cal V}(t; \{p,f\}_{m})\sket{\{c',c\}_{m}}
\;\;.
\end{split}
\end{equation}
Thus, using Eq.~(\ref{eq:braF1def}),
\begin{equation}
\begin{split}
\label{eq:1V}
\sbra{1}{\cal V}(t)\sket{\{p,f,s',c',s,c\}_{m}}
={}&
\sum_{\{\hat c',\hat c\}_{m}} 
\brax{\{s'\}_m}\ket{\{s\}_m}
\brax{\{\hat c'\}_m}\ket{\{\hat c\}_m}
\\&\quad\times
\sbra{\{\hat c', \hat c\}_{m}}
{\cal V}(t; \{p,f\}_{m})\sket{\{c',c\}_{m}}
\;\;.
\end{split}
\end{equation}
If we insert Eq.~(\ref{eq:1H}) and Eq.~(\ref{eq:1V}) into Eq.~(\ref{eq:evolution4}) and cancel the spin factors, we get
\begin{equation}
\begin{split}
\label{eq:calVrequirement}
\sum_{\{\hat c',\hat c\}_{m}} 
\brax{\{\hat c'\}_m}\ket{\{\hat c\}_m}
&
\sbra{\{\hat c', \hat c\}_{m}}
{\cal V}(t; \{p,f\}_{m})\sket{\{c',c\}_{m}} 
\\
={}& 
2\,
\bra{\{c'\}_{m}}h(t,\{p,f\}_{m})\ket{\{c\}_{m}}
\;\;.
\end{split}
\end{equation}
There is a simple way to satisfy this equation. We define
\begin{equation}
\label{eq:calVdef}
\begin{split}
\sbra{\{\hat c',\hat c\}_{m}}
{\cal V}(t,\{p,f\}_{m})\sket{\{c',c\}_{m}} 
={}& 
\dualL\bra{\{\hat c\}_m}
h(t,\{p,f\}_{m})
\ket{\{c\}_m}\
\delta^{\{\hat c'\}_m}_{\{c'\}_m}
\\ & +
\delta^{\{\hat c\}_m}_{\{c\}_m}\
\bra{\{c'\}_m}
h(t,\{p,f\}_{m})
\ket{\{\hat c'\}_m}\dualR
\;\;.
\end{split}
\end{equation}
What we have done here is to decompose $\sbra{\{\hat c',\hat c\}_{m}}
{\cal V}(t,\{p,f\}_{m})\sket{\{c',c\}_{m}}$ into two terms. In the first term, nothing happens on the bra side of the density matrix but there is a virtual correction on the ket side, while in the second term nothing happens on the ket side of the density matrix but there is a virtual correction on the bra side.\footnote{What ${\cal V}$ contains is the singular parts of the virtual corrections, which are related to the collinear and soft singularities of the real emission diagrams, plus an ``unresolved'' contribution from the real emission diagrams. Thus we obtain ${\cal V}$ from ${\cal H}$.}  
With this definition, when we perform the sums over $\{\hat c\}_m$ and $\{\hat c'\}_m$ in Eq.~(\ref{eq:calVrequirement}) using Eq.~(\ref{eq:dualcompleteness1}) we see that Eq.~(\ref{eq:calVrequirement}) is satisfied.

We note that one could imagine solving Eq.~(\ref{eq:evolutionplain}) numerically in the form
\begin{equation}
\label{eq:evoltionnum}
{\cal U}(t+\Delta t,0)\sket{\rho(0)} = 
[1 - {\cal V}(t)\Delta t]\,{\cal U}(t,0)\sket{\rho(0)}
+ {\cal H}_{\rm I}(t)\Delta t\,{\cal U}(t,0)\sket{\rho(0)}
\;\;.
\end{equation}
That is, one could use small time steps in which either one of the partons splits or else no parton splits and the weights for different color states are readjusted. However, this is not the way that shower evolution is typically constructed. 

To proceed down a more traditional path, we define two other operators, ${\cal V}_{\LE}(t)$ and ${\cal V}_{\LS}(t)$ with sum
\begin{equation}
\label{eq:calVdecomposed}
{\cal V}_{\LE}(t) + {\cal V}_{\LS}(t) = {\cal V}(t)
\;\;.
\end{equation}
The distinction between ${\cal V}_\LE(t)$ and ${\cal V}_\LS(t)$ lies in how we treat them within shower generation: ${\cal V}_\LE(t)$ is exponentiated and ${\cal V}_\LS(t)$ is subtracted. We express the solution of Eq.~(\ref{eq:evolutionplain}) in the form
\begin{equation}
\label{eq:evolution}
{\cal U}(t,t') = {\cal N}(t,t')
+ \int_{t'}^{t}\! d\tau\ 
{\cal U}(t,\tau)\,
[{\cal H}_{\LI}(\tau)
- {\cal V}_{\LS}(\tau)]
\,{\cal N}(\tau,t')
\;\;,
\end{equation}
where the operator ${\cal N}(t,t')$ is the time ordered exponential of the operator ${\cal V}_{\LE}(t)$,
\begin{equation}
\label{eq:calN}
{\cal N}(t, t') = \mathbb{T}
\exp\left\{-\int_{t'}^{t}\!d\tau\,{\cal V}_{\LE}(\tau)\right\}
\;\;. 
\end{equation}
Here $\mathbb T$ represents the ordering in evolution time $t$. The operator ${\cal N}(t, t')$ is a generalization of the standard Sudakov exponential in parton shower Monte Carlo programs. It has the group multiplication property
\begin{equation}
{\cal N}(t_3,t_2)\, {\cal N}(t_2,t_1)  = {\cal N}(t_3,t_1)
\label{eq:Ngroupcomposition}
\end{equation}
and satisfies the differential equation
\begin{equation}
\label{eq:Nevolution}
\frac{d}{dt}\,{\cal N}(t,t')
= -{\cal V}_{\LE}(t)\,{\cal N}(t,t')\;\;.
\end{equation}
Eq.~(\ref{eq:evolution}) is interpreted as saying that either the system evolves
without splitting from $t'$ to $t$, or else it evolves without splitting until an intermediate time $\tau$, splits or undergoes a color change at $\tau$, and then evolves (possibly with further splittings or color changes) from $\tau$ to $t$. The first term contains a summation of effects from virtual splittings. In the second term we have a parton splitting contribution along with a subtraction that arises from the part of the virtual splitting contribution that was not summed to form part of ${\cal N}(t,t')$.

Now the operator ${\cal V}(t)$ is completely defined but we still need to define ${\cal V}_{\LE}(t)$ and ${\cal V}_{\LS}(t)$. Here we have some freedom. There are at least three obvious choices:
\begin{enumerate}
\item We could define ${\cal V}_{\LE}(t)=0$. Then ${\cal V}_{\LS}(t) = {\cal V}(t)$. This choice leads to a trivial Sudakov exponential, ${\cal N}(t,t') = 1$. This is similar to what one does in fixed order calculations when the singularities of the real emission graphs with $m+1$ final state partons are removed by the subtraction terms with $m$ partons. This is not useful in the context of a parton shower.
\item We could define ${\cal V}_{\LS}(t)=0$. Then ${\cal V}_{\LE}(t) = {\cal V}(t)$. This means we exponentiate the whole virtual splitting operator. It is the most ``{shower way}'' to organize the parton evolution. The integral of ${\cal V}(t)$ over a range of $t$ produces large logarithms and all of these logarithms appear in the Sudakov exponent. Since ${\cal V}(t)$ is a non-diagonal matrix in color space, the implementation of this choice may present difficulties. 
\item Alternatively we can define ${\cal V}_{\LE}(t)$ to be the diagonal in color as follows:
\begin{equation}
\begin{split}
\label{eq:VEdef}
\sbra{\{\hat c',\hat c\}_{m}}
{\cal V}_{\LE}(t,\{p,f\}_{m})
\sket{\{c',c\}_{m}} 
={}&
\delta^{\{\hat c\}_m}_{\{c\}_m}\,\delta^{\{\hat c'\}_m}_{\{c'\}_m}
\\ & \times
\bigl[\bra{\{c\}_m}
h(t,\{p,f\}_{m})
\ket{\{c\}_m}
\\ & \quad
+\bra{\{c'\}_m}
h(t,\{p,f\}_{m})
\ket{\{c'\}_m}
\bigr] 
\;\;.
\end{split}
\end{equation}
Then ${\cal V}_{\LS}(t)$ is
\begin{equation}
\begin{split}
\label{eq:VS}
{\cal V}_{\LS}(t,\{p,f\}_{m})
= {\cal V}(t,\{p,f\}_{m}) - {\cal V}_{\LE}(t,\{p,f\}_{m})
\;\;.
\end{split}
\end{equation}
With this choice, the computation of the operator ${\cal N}(t,t')$ is simple because the basis vectors in the statistical space are eigenvectors of the operator ${\cal V}_{\LE}(t)$. 

With this alternative, we do not exponentiate everything. However, when we study color in the following sections, we will see with this choice that (with the color basis that we will choose) ${\cal V}_{\LS}(t)$ is small compared to ${\cal V}_{\LE}(t)$. First, we have
\begin{equation}
\begin{split}
\bra{\{c\}_m}h(t,\{p,f\}_{m})\ket{\{c\}_m}
={}&
\sum_{\{\tilde c\}_m}
\brax{\{c\}_m}\ket{\{\tilde c\}_m}\
\dualL\bra{\{\tilde c\}_m}h(t,\{p,f\}_{m})\ket{\{c\}_m}
\;\;,
\\
\bra{\{c'\}_m}h(t,\{p,f\}_{m})\ket{\{c'\}_m}
={}&
\sum_{\{\tilde c\}_m}
\bra{\{c'\}_m}h(t,\{p,f\}_{m})\ket{\{\tilde c\}_m}\dualR\
\brax{\{\tilde c\}_m}\ket{\{c'\}_m}
\;\;.
\end{split}
\end{equation}
We will see in Eq.~(\ref{eq:offdiagonalcolors}) that the matrix $\brax{\{c\}_m}\ket{\{\tilde c\}_m}$ is the unit matrix except for $1/N_{\Lc}^{2}$ corrections. Thus the difference between using dual basis vectors $\ket{\{\tilde c\}_m}\dualR$ in Eq.~(\ref{eq:calVdef}) and ordinary basis vectors $\ket{\{c'\}_m}$ is not important in the large $N_\Lc$ limit. Second, we will see in Sec.~\ref{sec:inclusiveH} that the matrix $\bra{\{\hat c\}_m} h(t,\{p,f\}_{m}) \ket{\{c\}_m}$ is almost diagonal in the sense that its off-diagonal matrix elements are suppressed compared to its diagonal matrix elements by factors of $1/N_{\Lc}^{2}$. Thus the part of the virtual contribution that is not exponentiated is small. This small part is not neglected, but we can leave it out of the Sudakov exponent and treat it as a subtraction instead. Effectively, this means that we treat $1/N_{\Lc}^{2}$ as a small parameter in addition to $\alpha_\Ls$.

\end{enumerate}

\section{Momentum and flavor mapping}
\label{sec:mapping}

In this and the following sections, we explore how to define the splitting operator ${\cal H}_I$. The first issue to examine is the momentum mapping. We begin with an $m$ parton state with momenta $\{p\}_m$. One of the partons, with label $l \in \{\La,\Lb,1,\dots,m\}$, splits. After the splitting, we have an $m+1$ parton state with momenta $\{\hat p\}_{m+1}$. Our notation is that parton $l$ splits into partons with labels $l$ and $m+1$, while the other partons keep their labels. The momenta $\{\hat p\}_{m+1}$ after splitting are determined by the momenta $\{p\}_m$ and a momentum splitting variable that we call $\zeta_{\rm p}$, which defines the momenta of the daughter partons. There is also a flavor splitting variable, $\zeta_{\rm f}$, which tells the daughter flavors. 

In this section, we first describe the splitting of a final state parton, then move on to the somewhat more complicated splitting of an initial state parton. For the final state splitting we first describe how $\{\hat p,\hat f\}_{m+1}$ is determined from $\{p,f\}_m$ and $\{\zeta_{\rm p},\zeta_{\rm f}\}$. Then we state the inverse transformation, from $\{\hat p,\hat f\}_{m+1}$ to $\{p,f\}_m$ and $\{\zeta_{\rm p},\zeta_{\rm f}\}$. Finally, we deduce the jacobian for this change of variables. We will then be ready to do the same thing for the splitting of an initial state parton.

There are many ways to define the momentum mapping $\{p\}_m \leftrightarrow \{\hat p\}_{m+1}$. One of the most successful is that of Catani and Seymour \cite{CataniSeymour}. This may be called a local mapping: the momenta of most of the partons are left unchanged, while the momenta of two partons are mapped into the momenta of three partons, $(p_l,p_k) \leftrightarrow (\hat p_l, \hat p_{m+1}, \hat p_k)$. Here parton $k$ is a spectator parton, chosen as one of the partons that is color connected to parton $l$. (There is an exception to this rule in the case of initial state splittings.) The antenna factorization of Ref.~\cite{antenna} also uses a local mapping. We use a global mapping, in which all of the partons participate, as in Ref.~\cite{phasespacemap}. This way, each parton has to contribute only a little momentum. We also include quark masses in the kinematics.

\subsection{Splitting a final state parton}
\label{sec:FSsplittingkinematics}

We begin by defining what happens to the parton flavors when a final state parton $l$ splits. The partons with indices other than $l$ and $m+1$ keep their flavors,
\begin{equation}
\hat f_j = f_j 
\;\;,\hskip 1 cm j \notin\{l,m+1\}
\;\;.
\end{equation}
What happens to partons $l$ and $m+1$ is given by the value of the variable $\zeta_{\rm f} = (\hat f_l,\hat f_{m+1})$. The flavor splitting variable takes values in a set $\Phi_{l}(f_l)$ that depends on the flavor of the mother parton. If parton $l$ is a quark or antiquark, then the set $\Phi_{l}(f_l)$ has only one element, 
\begin{equation}
\label{eq:setSfnotgl}
\Phi_{l}(f_l) = \{(f_l,{\rm g})\}
\;\;, \hskip 1 cm  f_l \ne {\rm g} 
\;\;.
\end{equation}
Here, we have used the freedom to assign labels in order to assign the label $l$ to the daughter quark or antiquark and the label $m+1$ to the gluon. If parton $l$ is a gluon, then $\zeta_{\rm f}$ can be a pair of gluons or any choice of $(q,\bar q)$ flavors,
\begin{equation}
\label{eq:setSfgl}
\Phi_{l}({\rm g}) = 
\{({\rm g},{\rm g}),({\rm u},\bar {\rm u}), ({\rm d},\bar {\rm d}),\dots\} 
\;\;.
\end{equation}
In the case of a ${\rm g} \to q + \bar q$ splitting, we again use the freedom to assign labels in order to assign the label $l$ to the daughter quark the label $m+1$ to the daughter antiquark.

We now turn to the momenta. Parton $l$ has momentum $p_l$ with $p_l^2 = m^2(f_l)$ and splits into two partons $l$ and $m+1$ with momenta $\hat p_l$ and $\hat p_{m+1}$ respectively. The daughter partons are on-shell: $\hat p_l^2 = m^2(\hat f_l)$ and $\hat p_{m+1}^2 = m^2(\hat f_{m+1})$. We always have $(\hat p_l + \hat p_{m+1})^2 \ge m^2(f_l)$.

We need a bit of notation. Let $Q$ be the total momentum of the final state partons,
\begin{equation}
Q \equiv \sum_{j = 1}^m p_j = p_\La + p_\Lb
\;\;.
\end{equation}
Define
\begin{equation}
\begin{split}
\label{eq:albldef}
a_l ={}& \frac{Q^{2}}{2p_{l}\!\cdot\! Q}
\;\;,
\\
b_l ={}& \frac{m^2(f_l)}{2p_{l}\!\cdot\! Q}
\;\;.
\end{split}
\end{equation}
Note that $a_l + b_l \ge 1$. To see this, let $K = Q - p_l$ be the total momentum of the final state spectator partons, with $K^2 \ge 0$. Then 
\begin{equation}
\begin{split}
0 \le {}& \frac{K^{2}}{2p_{l}\!\cdot\! Q}
\\
={}& \frac{Q^2 - 2 p_l\!\cdot\! Q + m^2(f_l)}{2p_{l}\!\cdot\! Q}
\\
={}& a_l + b_l - 1
\;\;.
\end{split}
\end{equation}

In order to define the momentum mapping, we first determine the total momentum 
\begin{equation}
\label{eq:Pldef}
P_l = \hat p_l + \hat p_{m+1}
\end{equation}
of the daughters of parton $l$. We take $P_l$ to be a linear combination of $p_l$ and $Q$,
\begin{equation}
\label{eq:FSsplitting}
P_l = \lambda p_l + \frac{1 - \lambda + y}{2 a_l}\ Q
\;\;.
\end{equation}
There are two parameters in this definition. The first, $y$, is a measure of the virtuality, $P_l^2 - m^2(f_l)$, of the splitting. The second, $\lambda$, is a function of $y$ that we will determine presently.

For an exactly collinear splitting or the emission of gluon with momentum $\hat p_{m+1} = 0$, we have $P_l = p_l$. Away from these limits, the spectator partons will have to donate some momentum in order to allow $P_l \ne p_l$. We elect to leave the momenta of the initial state partons unchanged, $\hat p_\La = p_\La$ and $\hat p_\Lb = p_\Lb$.  Instead, we choose to obtain the needed momentum from the final state spectator partons by letting the momenta after the splitting be related to the momenta before the splitting by a Lorentz transformation,
\begin{equation}
\label{eq:FSboost}
\hat p_j^\mu = \Lambda^\mu_{\ \nu}\ p_j^\nu
\;\;,\hskip 1 cm j\notin\{l,m+1\}
\;\;.
\end{equation}
With this method of transferring momentum, each parton donates a share of the needed momentum, with low momentum partons donating only a little momentum. 

The total momentum of the final state spectator partons before the splitting is
\begin{equation}
K = Q - p_l
\;\;.
\end{equation}
Since the momenta of the initial state partons remains the same, $\hat Q = \hat p_\La + \hat p_\Lb$ is the same as $Q$. The total momentum of the final state spectator partons after the splitting is then
\begin{equation}
\label{eq:hatKdef}
\hat K = Q - P_l
\;\;.
\end{equation}
Since each final state spectator is changed by a Lorentz transformation, we have
\begin{equation}
\label{eq:FSboostonK}
\hat K^\mu = \Lambda^\mu_{\ \nu}\ K^\nu
\;\;.
\end{equation}
In fact, there is a Lorentz transformation that does this, namely
\begin{equation}
\label{eq:Lambdadef}
\Lambda(\hat K,K)^\mu_{\ \nu}
= g^\mu_\nu
- \frac{2 (\hat K + K)^\mu (\hat K + K)_\nu}{(\hat K + K)^2}
+ \frac{2\hat K^\mu K_\nu}{K^2}
\;\;,
\end{equation}
provided that $\hat K^2 = K^2$. Thus $P_l$ must lie on the hyperbola $(Q-P_l)^2 = (Q-p_l)^2$ in the $Q$-$p_l$ plane, as illustrated in Fig.~\ref{fig:mapping}.

In the case in which the momenta $K$ and $\hat K$ are carried by a single massless spectator, $\hat K$ is parallel to $K$. In this case, one can use an alternative representation of the boost in Eq.~(\ref{eq:Lambdadef}) that remains well defined when $K^2 = (\hat K + K)^2 = 0$,
\begin{equation}
\label{eq:Lambdadefalt}
\Lambda(\hat K,K)^\mu_{\ \nu}
= g^\mu_\nu
+ \left(\frac{K\!\cdot\! n}{\hat K\!\cdot\! n} - 1\right)
n^\mu \bar n_{\nu}
+ \left(\frac{\hat K\!\cdot\! n}{K\!\cdot\! n} - 1\right)
\bar n^\mu n_{\nu}
\;\;,
\end{equation}
where $n$ and $\bar n$ are lightlike vectors in the $Q$-$p_l$ plane with $n \cdot \bar n = 1$ and $(p_l\cdot n/p_l\cdot \bar n) <  (Q\cdot n/Q\cdot \bar n)$. That is, these are the vectors along the two coordinate axes in Fig.~\ref{fig:mapping}, with $n$ directed toward the upper right and $\bar n$ directed toward the upper left in the diagram.

\FIGURE{
\hskip 2 cm\includegraphics[width = 6 cm]{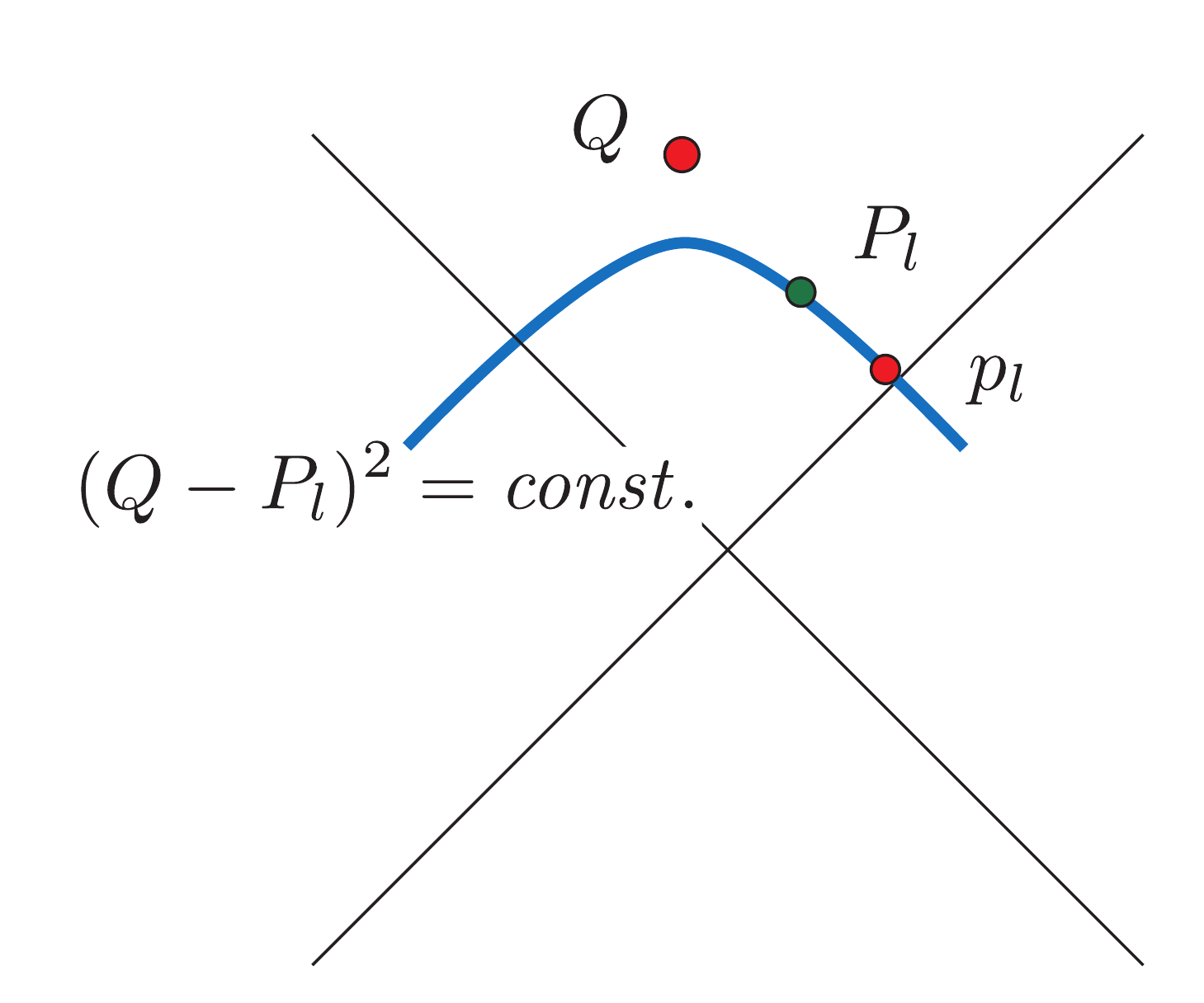}\hskip 2 cm
\caption{Momentum space mapping for a final state splitting.  Coordinate axes in the two lightlike directions in the plane of $p_l$ and $Q$ are shown. Points $p_l$ and $Q$ are shown in the case that $p_l$ is nearly lightlike. Then $P_l$ lies on the hyperbola $(Q - P_l)^2 = {\it const}.$ that passes through $p_l$.
}
\label{fig:mapping}
}

We can arrange that $\hat K^2 = K^2$ by making a proper choice of $\lambda$ in Eq.~(\ref{eq:FSsplitting}). We have
\begin{equation}
K^2 = (a_l + b_l - 1)\,2 Q\!\cdot\! p_l
\;\;,
\end{equation}
and
\begin{equation}
\hat K^2 = \frac{1}{4 a_l}\
\big[
(2a_l - 1 - y)^2 - (1 - 4 a_l b_l)\lambda^2
\big]\,
2 Q\!\cdot\! p_l
\;\;.
\end{equation}
We obtain $\hat K^2 = K^2$ if
\begin{equation}
\label{eq:lambdadef}
\lambda = 
\sqrt{\frac{(1 + y)^2 - 4 a_l (y + b_l)}{1 - 4 a_l b_l}}
\;\;.
\end{equation}

We note that $\lambda \to 1$ when $y \to 0$. As $y$ increases, $\lambda$ decreases. There is a maximum possible value of $y$, namely the value that makes $\lambda = 0$,
\begin{equation}
\label{eq:ymax}
y_{\rm max} = 
\left(
\sqrt{a_l} - \sqrt{a_l + b_l - 1}\,
\right)^2
- b_l
\;\;.
\end{equation}

The splitting parameter $y$ has a simple meaning. From Eq.~(\ref{eq:FSsplitting}), we find, using Eq.~(\ref{eq:lambdadef}),
\begin{equation}
\label{eq:FSymeaning}
y =
\frac{P_l^2 - m^2(f_l)}{2p_l\!\cdot\! Q}
\;\;.
\end{equation}
That is, $y$ is a dimensionless measure of the virtuality of the splitting. An alternative relation is
\begin{equation}
\label{eq:FSymeaningAlt}
y =
\frac{P_l^2 - m^2(f_l)}
{2P_l\!\cdot\! Q - (P_l^2 - m^2(f_l) )} 
\;\;.
\end{equation}
There are also alternative relations for $a_l$ and $b_l$,
\begin{equation}
\begin{split}
\label{eq:albldefAlt}
a_l ={}& \frac{Q^2}
{2 P_l\!\cdot\! Q + m^2(f_l) - P_l^2}
\;\;,
\\
b_l ={}& \frac{m^2(f_l)}
{2 P_l\!\cdot\! Q + m^2(f_l) - P_l^2}
\;\;.
\end{split}
\end{equation}
These relations can be derived with the use of the relation
\begin{equation}
2 P_l\!\cdot\! Q 
= (1 + y)\, 2 p_l\!\cdot\! Q
\;\;.
\end{equation}

There is also a minimum value of $y$. Since $P_l^2 > [m(\hat f_l) + m(\hat f_{m+1})]^2$, we have $y > y_{\rm min}$ where
\begin{equation}
\label{eq:ymin}
y_{\rm min} = 
\frac{[m(\hat f_l) + m(\hat f_{m+1})]^2 - m^2(f_l)}
{2 p_l\!\cdot\! Q}
\;\;.
\end{equation}
Note that if all of the partons are massless, then $y_{\rm min} = 0$. In addition, if $f_l = \hat f_l$ is a massive quark flavor and $\hat f_{m+1} = {\rm g}$, then also $y_{\rm min} = 0$.

We have seen how, given the virtuality variable $y$, we can define $P_l = \hat p_l + \hat p_{m+1}$ in the plane of $p_l$ and $Q$ such that $P_l^2 - m^2(f_l) = y\, 2p_l\cdot Q$ and so that the needed momentum donation from the spectator partons can be obtained by a Lorentz transformation. It remains to define $\hat p_l$ and $\hat p_{m+1}$ individually. This is simple. We let $\hat p_{l}$ and $\hat p_{m+1}$ be any momenta on the appropriate mass shells that sum to $P_l$.

It will prove convenient to formulate this rather abstractly. We denote the daughter parton momenta by 
\begin{equation}
\label{eq:zetapFS}
\zeta_{\rm p} \equiv (\hat p_l, \hat p_{m+1})
\;\;.
\end{equation}
We must have
\begin{equation}
\zeta_{\rm p} \in \varGamma_{l}(\{p\}_m,\zeta_{\rm f})
\;\;,
\end{equation}
where the set $\varGamma_{l}(\{p\}_m,\zeta_{\rm f})$ is described as follows,
\begin{equation}
\begin{split}
\label{eq:splitsetFS}
\varGamma_{l}(\{p\}_m,\zeta_{\rm f}) = 
\Big\{(\hat p_{l}, \hat p_{m+1})\;\;\Big|&\quad
\hat p_l^2 = m^2(\hat f_l),\quad \hat p_l\!\cdot\! Q > 0 ,
\\
&\quad \hat p_{m+1}^2 = m^2(\hat f_{m+1}),\quad \hat p_{m+1}\!\cdot\! Q > 0 ,
\\
&\quad \varepsilon_{\mu\nu\alpha\beta}
(\hat p_l^\nu + \hat p_{m+1}^\nu)\, p_l^\alpha\,  Q^\beta
 = 0 ,
\\
&\quad (Q - \hat p_l - \hat p_{m+1})^2 = (Q - p_l)^2
\Big\}\;\;.
\end{split}
\end{equation}
That is, each of $\hat p_l$ and $\hat p_{m+1}$ lies on the appropriate forward mass shell, their sum lies in the plane of $p_l$ and $Q$, and $\hat K^2 = K^2$. These conditions entail that 
\begin{equation}
[m(\hat f_l) + m(\hat f_{m+1})]^2
\le (\hat p_l + \hat p_{m+1})^2
\le
\left(\sqrt{Q^2} - \sqrt{Q^2 + m^2(f_l) - 2 p_l\!\cdot\! Q}\,\right)^2
\;\;,
\end{equation}
which corresponds to $y_{\rm min} \le y \le y_{\rm max}$.

We note that the set $\varGamma_{l}$ is a three dimensional surface in the space of momenta $(\hat p_l,\hat p_{m+1})$. One can choose three coordinates to describe this surface, for instance a virtuality variable, a momentum fraction variable, and an azimuthal angle. Different choices of coordinates may be best for different purposes, so we leave this choice open.

We give a name to this transformation of momenta and flavors:
\begin{equation}
\label{eq:Rldef}
\{\hat p, \hat f\}_{m+1} = R_l(\{p,f\}_m, \{\zeta_{\rm p},\zeta_{\rm f}\})
\;\;.
\end{equation}

In parton splittings close to the collinear or soft limits, partons lose energy when they split. It is of interest to see how this property carries over to splittings that are not close to the limit. Here, the choice of the mapping $R_l$ may be considered to be part of a model for shower evolution that could be sensible or perhaps not so sensible. An investigation of properties of the model is thus of some significance. 

Using $K^2$ = $\hat K^2$, we derive
\begin{equation}
Q\!\cdot\! p_l - Q\!\cdot\! \hat p_l = \hat p_{m+1}\!\cdot\! \hat K
+ \frac{1}{2}\,\big[m^2(f_l) - m^2(\hat f_l) + m^2(\hat f_{m+1})\big]
\;\;.
\end{equation}
This is especially interesting in the case that parton $l$ is a quark and parton $m+1$ is a gluon. Then the term involving masses vanishes. We note that $\hat p_{m+1}\cdot \hat K \ge 0$ because both $\hat p_{m+1}$ and $\hat K$ lie inside or on the forward light cone. Thus the energy of the quark, as measured in the frame in which $Q$ is at rest, is bigger before the splitting than after the splitting. By emitting bremsstrahlung, the quark slows down. If evolution of the final state were to continue long enough, the quark would slow to a stop in the $\vec Q = 0$ frame. Then radiation from that quark would cease. At any point in the shower evolution, there can be an initial state splitting, discussed below in Sec.~\ref{sec:ISsplittingkinematics}. This changes $p_\La$ or $p_\Lb$ and thus $p_\La + p_\Lb = Q$. Now final state quarks tend to come to rest in the new $\vec Q = 0$ frame. We judge that a tendency for quarks to slow down (and, similarly, for gluons to lose energy) is reasonably sensible. Of course, the partons should not be allowed to shower indefinitely. At some resolution scale, a perturbative model for showering is simply wrong and a process by which partons combine to form hadrons is needed.

\subsection{Combining two final state partons}
\label{sec:FScombinationkinematics}

It is significant (and useful) that this transformation has an inverse. Let start with $\{\hat p\}_{m+1}$ and determine $\{p\}_{m}$ and $\{\zeta_{\rm p},\zeta_{\rm f}\}$. 

The splitting variable for the momenta is given by the momenta of the daughter partons, $\zeta_{\rm p} = (\hat p_l, \hat p_{m+1})$. From $\{\hat p\}_{m+1}$ we determine
\begin{equation}
Q = \sum_{j=1}^{m+1} \hat p_j = \hat p_\La + \hat p_\Lb
\;\;.
\end{equation}
Then Eq.~(\ref{eq:FSymeaningAlt}) gives $y$, Eq.~(\ref{eq:albldefAlt}) gives $a_l$ and $b_l$, and Eq.~(\ref{eq:lambdadef}) gives $\lambda$. Since the calculation of $\lambda$ involves taking the square root of $\lambda^2$, we should check that $\lambda^2 >0$. For this purpose, we can express $\lambda^2$ in terms of dot products of vectors as
\begin{equation}
\lambda^2 = \frac{4  [(Q\!\cdot\! P_l)^2 - Q^2 P_l^2 ]
+  m^2(f_l)\,[4P_l\!\cdot\! \hat K  + m^2]}
{4 Q\!\cdot\!\hat K\ [P_l\!\cdot\!\hat K + P_l^2 - m^2(f_l)]
+ (P_l^2 - m^2(f_l))^2}
\;\;.
\end{equation}
Since $Q$, $P_l$, and $\hat K = Q-P_l$ lie inside or on the positive lightcone and $P_l^2 - m^2(f_l) > 0$, both the numerator and the denominator are non-negative. With $y$, $\lambda$, $a_l$ and $b_l$ at hand, one can calculate the lightlike momentum $p_l$ by rearranging Eq.~(\ref{eq:FSsplitting}),
\begin{equation}
\label{eq:FSsplittingbis}
p_l =
\frac{1}{\lambda}
(\hat p_l + \hat p_{m+1})
- \frac{1 - \lambda + y}{2 \lambda a_l}\ Q
\;\;.
\end{equation}

We now have $p_l$. We define $p_\La = \hat p_\La$ and $p_\Lb = \hat p_\Lb$. This leaves the $p_j$ for $j \notin \{l,\La,\Lb\}$. For this, we need the inverse Lorentz transformation to Eq.~(\ref{eq:FSboost}). From $K = Q - p_l$ and $\hat K = Q - \hat p_l - \hat p_{m+1}$, we construct $\Lambda(K,\hat K)^\mu_{\ \nu}$ using Eq.~(\ref{eq:Lambdadef}) or Eq.~(\ref{eq:Lambdadefalt}) with the roles of $\hat K$ and $K$ interchanged. Then
\begin{equation}
\label{eq:FSboostInverse}
p_j^\mu = \Lambda(K,\hat K)^\mu_{\ \nu}\ \hat p_j^\nu
\;\;,\hskip 1 cm  j\notin \{l,\La,\Lb\} 
\;\;.
\end{equation}

The transformation of the flavors is simple. The splitting variable $\zeta_{\rm f}$ is given by the flavors of the daughter particles, $(\hat f_l,\hat f_{m+1})$. The flavor of the mother parton is 
\begin{equation}
f_l = \hat f_l + \hat f_{m+1}
\;\;,
\end{equation}
with the obvious definition of adding flavors, as in ${\rm d} + {\rm g} = {\rm d}$ and ${\rm u} + \bar {\rm u} = {\rm g}$. The flavors of the other partons are unchanged
\begin{equation}
f_j = \hat f_j 
\;\;, \hskip 1 cm  j \notin \{l,m+1\} 
\;\;.
\end{equation}

We give a name to this transformation of momenta and flavors,
\begin{equation}
\label{eq:Qldef}
\{\{p,f\}_m, \{\zeta_{\rm p},\zeta_{\rm f}\}\}
= Q_l(\{\hat p,\hat f\}_{m+1})
\;\;.
\end{equation}
This is the inverse transformation to $R_l$, Eq.~(\ref{eq:Rldef}).

\subsection{The integration measure for final state splitting}
\label{sec:FSjacobian}

With a suitable choice of the integration measure $d\zeta_{\rm p}$ for integrating over the splitting variables $\zeta_{\rm p}$, we can arrange that 
\begin{equation}
\begin{split}
\label{eq:jacobianFSdef}
\int [d\{\hat p, \hat f\}_{m+1}]\ &
g(\{\hat p, \hat f\}_{m+1}) 
\\
&= 
\int[d\{p,f\}_{m}] 
\sum_{\zeta_{\rm f}\in \Phi_{l}(f_l)}
\int d\zeta_{\rm p}\
\theta(\zeta_{\rm p} \in \varGamma_{l}(\{p\}_{m},\zeta_{\rm f}))\
g(\{\hat p, \hat f\}_{m+1})
\end{split}
\end{equation}
for an arbitrary function $g(\{\hat p, \hat f\}_{m+1})$. The definition that we need is
\begin{equation}
\begin{split}
\label{eq:dzetapFSdef}
d\zeta_{\rm p}\equiv {}&
dy\ \theta(y_{\rm min} < y < y_{\rm max})\
\lambda\,
\frac{p_{l}\!\cdot\! Q}{\pi}\
\\&\times
\frac{d^{4}\hat p_{l}}{(2\pi)^{4}}\,
2\pi\delta_{+}(\hat p_{l}^{2} - m^2(\hat f_l))\
\frac{d^{4}\hat p_{m+1}}{(2\pi)^{4}}\,
2\pi\delta_{+}(\hat p_{m+1}^{2} - m^2(\hat f_{m+1}))\,
\\&\times
(2\pi)^{4}\,\delta\!\left(
\hat p_l + \hat p_{m+1}
- \lambda p_l - \frac{1 - \lambda  + y}{2 a_l}\ Q
\right)
\;\;.
\end{split}
\end{equation}
Here the limits on $y$ are given in Eqs.~(\ref{eq:ymax}) and (\ref{eq:ymin}).

\subsection{Splitting an initial state parton}
\label{sec:ISsplittingkinematics}

Consider the splitting of an initial state parton, say parton ``a.'' The initial state parton  with momentum $p_\La \approx \eta_\La p_\LA$ splits to produce a new initial state parton with momentum $\hat p_\La \approx \hat\eta_\La p_\LA$ and a new final state parton with label $m+1$ and momentum $\hat p_{m+1}$. We are using the usual backwards evolution here, so that the evolution going forward in time is $\hat p_\La \to p_\La + \hat p_{m+1}$. In this subsection, we describe how $\{\hat p,\hat f\}_{m+1}$ is determined from $\{p,f\}_{m}$ and splitting variables $\{\zeta_{\rm p},\zeta_{\rm f}\}$. The splitting of the other initial state parton is described by the same formulas with $\La \leftrightarrow \Lb$.

We begin by defining what happens to the parton flavors, recalling our notation that for the initial state partons ``a'' and ``b'', $f_\La$, $\hat f_\La$, $f_\Lb$ and $\hat f_\Lb$ denote the opposite of the flavor of the physical incoming parton. The partons with indices other than ``$\La$'' and $m+1$ keep their flavors,
\begin{equation}
\hat f_j = f_j 
\;\;,\hskip 1 cm j \notin\{\La,m+1\}
\;\;.
\end{equation}
What happens to partons ``a'' and $m+1$ is given by the value of the variable $\zeta_{\rm f} = (\hat f_\La,\hat f_{m+1})$. The flavor splitting variable takes values in a set $\Phi_{\La}(f_\La)$ that depends on the flavor of the mother parton. This set is determined by the requirement that $f_\La = \hat f_\La + \hat f_{m+1}$. If parton ``a'' is a quark or antiquark, then the set $\Phi_{\La}(f_\La)$ has only two elements,
\begin{equation}
\label{eq:setSfnotgA}
\Phi_{\La}(f_\La) = 
\{(f_\La,{\rm g}), ({\rm g}, f_\La)\} 
\;\;,\hskip 1 cm f_\La \ne {\rm g}
\;\;.
\end{equation}
If parton ``a'' is a gluon, then $\zeta_{\rm f}$ can be a pair of gluons or any choice of $(q,\bar q)$ or $(q,\bar q)$ flavors,
\begin{equation}
\label{eq:setSfgA}
\Phi_{\La}({\rm g}) = 
\{({\rm g},{\rm g}),({\rm u},\bar {\rm u}), 
(\bar {\rm u},{\rm u}), ({\rm d},\bar {\rm d}), 
(\bar {\rm d},{\rm d}), \dots\} 
\;\;.
\end{equation}

We now turn to the momenta. Let parton ``a'' with momentum fraction $\eta_\La$ radiate a parton $m+1$ with momentum $\hat p_{m+1}$. We then need to define how to determine $\{\hat p\}_{m+1}$ from $\{p\}_{m}$ and $\hat p_{m+1}$. 

As discussed in Sec.~\ref{sec:quantumstates}, we can include masses for the initial state partons.\footnote{Recall that this is optional. One could just replace the masses for the initial state partons by zero, $m^2(f_\La) = m^2(\hat f_\La) = m^2(f_\Lb) = 0$. One could also set all quark masses to zero.} We take the partons to be on-shell with zero transverse momenta,
\begin{equation}
\begin{split}
p_\La ={}& \eta_\La p_\LA + \frac{m^2(f_\La)}{\eta_\La s} p_\LB
\;\;,
\\
p_\Lb ={}& \eta_\Lb p_\LB + \frac{m^2(f_\Lb)}{\eta_\Lb s} p_\LA
\;\;,
\\
\hat p_\La ={}& \hat \eta_\La p_\LA 
+ \frac{m^2(\hat f_\La)}{\hat \eta_\La s} p_\LB
\;\;.
\\
\end{split}
\end{equation}
Recall that we define $p_\LA$ and $p_\LB$ to be lightlike approximations to the incoming hadron momenta, with $2 p_\LA \cdot p_\LB = s$. The radiated parton can have a mass,
\begin{equation}
\label{eq:pmp1massshell}
\hat p_{m+1}^2 = m^2(\hat f_{m+1})
\;\;.
\end{equation}

We take the momentum fraction of parton ``b'' to remain the same,
\begin{equation}
\hat \eta_\Lb = \eta_\Lb
\;\;.
\end{equation}
The momentum fraction $\hat\eta_\La$ after the splitting will be determined by $\hat p_{m+1}$. As in the case of a final state splitting, it is not generally possible to have $\hat p_\La = p_\La + \hat p_{m+1}$ given the mass shell conditions and the possibility that the radiated parton has non-zero transverse momentum. In order to allow the approximation that both $p_\La$ and $\hat p_\La$ are on-shell with zero transverse momenta, we therefore take some momenta from the final state spectator partons by letting the momenta after the splitting be related to the momenta before the splitting by a Lorentz transformation,
\begin{equation}
\label{eq:ISboost}
\hat p_j^\mu = \Lambda^\mu_{\ \nu}\ p_j^\nu
\;\;,
\hskip 1 cm j \in \{1,\dots,m\}
\;\;.
\end{equation}
Since each final state spectator is changed by a Lorentz transformation, we have
\begin{equation}
\label{eq:ISboostonK}
\hat K^\mu = \Lambda^\mu_{\ \nu}\ K^\nu
\;\;,
\end{equation}
where $K$ is the momentum of the final state partons before the splitting,
\begin{equation}
K = p_\La + p_\Lb
\;\;,
\end{equation}
and $\hat K$ is the momentum of the final state spectators after the splitting,
\begin{equation}
\hat K = \hat p_\La + p_\Lb - \hat p_{m+1}
\;\;.
\end{equation}

In order for $K$ and $\hat K$ to be related by a Lorentz transformation, we need $\hat K^2$ = $K^2$. To see what this means, define
\begin{equation}
\hat Q(\hat \eta_\La) = \hat p_\La + p_\Lb
=
\left( 
\hat \eta_\La 
+ \frac{m^2(\hat f_\Lb)}{\eta_\Lb s}
\right) p_\LA 
+ \left(
\eta_\Lb
+\frac{m^2(\hat f_\La)}{\hat \eta_\La s}
\right)p_\LB
\;\;.
\end{equation}
Then we demand that 
\begin{equation}
\label{eq:khatsqrelation}
(\hat Q(\hat \eta_\La) - \hat p_{m+1})^2 = K^2
\;\;.
\end{equation}
The vector $\hat p_{m+1}$ determines $\hat\eta_\La$. In the space of $\hat p_{m+1}$, a surface of constant $\hat\eta_\La$ is the intersection of the hyperbola $\hat p_{m+1}^2 = m^2(\hat f_{m+1})$ with the hyperbola given by Eq.~(\ref{eq:khatsqrelation}), as illustrated in Fig.~\ref{fig:mappingIS}. Using $\hat p_{m+1}^2 = m^2(\hat f_{m+1})$ in Eq.~(\ref{eq:khatsqrelation}), we can write
\begin{equation}
\label{eq:khatsqrelation2}
2\,\hat Q(\hat \eta_\La)\!\cdot\!\hat p_{m+1} = \hat Q(\hat \eta_\La)^2 
+ m^2(\hat f_{m+1}) - K^2
\;\;.
\end{equation}
Looked at this way, a surface of constant $\hat\eta_\La$ is the intersection of the hyperbola $\hat p_{m+1}^2 = m^2(\hat f_{m+1})$ with the plane defined by Eq.~(\ref{eq:khatsqrelation2}). We will require $\hat\eta_\La < 1$. Thus the allowed region in $\hat p_{m+1}^2$ is the part of the forward mass shell with
\begin{equation}
\label{eq:khatsqlimit}
2\,\hat Q(1)\!\cdot\!\hat p_{m+1} < \hat Q(1)^2 
+ m^2(\hat f_{m+1}) - K^2
\;\;.
\end{equation}

In order to solve for $\hat \eta_\La$ given $\hat p_{m+1}$, we write $K^2$ and $\hat K^2$ in the form
\begin{equation}
\begin{split}
\label{eq:KandhatK}
K^2 ={}& \alpha \eta_\La 
- \frac{\beta }{\eta_\La} 
- \gamma
\;\;,
\\
\hat K^2 ={}& \hat \alpha \hat \eta_\La 
- \frac{\hat \beta }{\hat \eta_\La} 
- \hat \gamma
\;\;.
\end{split}
\end{equation}
Here
\begin{equation}
\begin{split}
\alpha ={}& \eta_\Lb s
\;\;, 
\\
\beta ={}& - \frac{m^2(f_\La)\, m^2(f_\Lb)}{\eta_\Lb s}
\;\;, 
\\
\gamma ={}& - m^2(f_\La) - m^2(f_\Lb)
\;\;,
\end{split}
\end{equation}
while
\begin{equation}
\begin{split}
\hat\alpha ={}& \eta_\Lb s - 2 p_\LA \!\cdot\! \hat p_{m+1}
\;\;, 
\\
\hat\beta ={}& \frac{m^2(\hat f_\La)}{s}
\left\{2 p_\LB \!\cdot\! \hat p_{m+1} - \frac{m^2(f_\Lb)}{\eta_\Lb}
\right\}
\;\;, 
\\
\hat\gamma ={}& 2 p_\Lb \!\cdot\! \hat p_{m+1}
- m^2(\hat f_\La) - m^2(f_\Lb) - m^2(\hat f_{m+1})
\;\;.
\end{split}
\end{equation}
The condition  $\hat K^2 = K^2$ now determines $\hat \eta_\La$,
\begin{equation}
\label{eq:hatetaLaresult}
\hat \eta_\La = 
\frac{1}{2\hat \alpha}
\left\{
K^2 + \hat \gamma 
+ \sqrt{(K^2 + \hat \gamma)^2 + 4 \hat \alpha\hat\beta}
\right\}
\;\;.
\end{equation}
It is a consequence of Eq.~(\ref{eq:hatetaLaresult}) and the kinematic conditions (\ref{eq:Q0condition}) and (\ref{eq:etalimits}) that
\begin{equation}
\label{eq:etasincrease}
\hat \eta_\La > \eta_\La
\;\;.
\end{equation}
We prove this in Appendix \ref{sec:etalimit}.

Having fixed $\hat\eta_{\La}$ so that $\hat K^2 = K^2$, these two momenta will be related by Eq.~(\ref{eq:ISboostonK}),
\begin{equation}
\hat K^\mu = \Lambda(\hat K,K)^\mu_{\ \nu}\ K^\nu
\;\;,
\end{equation}
where $\Lambda^\mu_{\ \nu}$ the Lorentz transformation (\ref{eq:Lambdadef}). This allows us to define the spectator momenta after the splitting to be related to the spectator momenta before the splitting by this same boost, as in Eq.~(\ref{eq:ISboost}),
\begin{equation}
\hat p_j^\mu = \Lambda(\hat K,K)^\mu_{\ \nu}\ p_j^\nu
\;\;,\hskip 1 cm j\in \{1,\dots,m\}
\;\;.
\end{equation}

\FIGURE{
\hskip 2 cm\includegraphics[width = 6 cm]{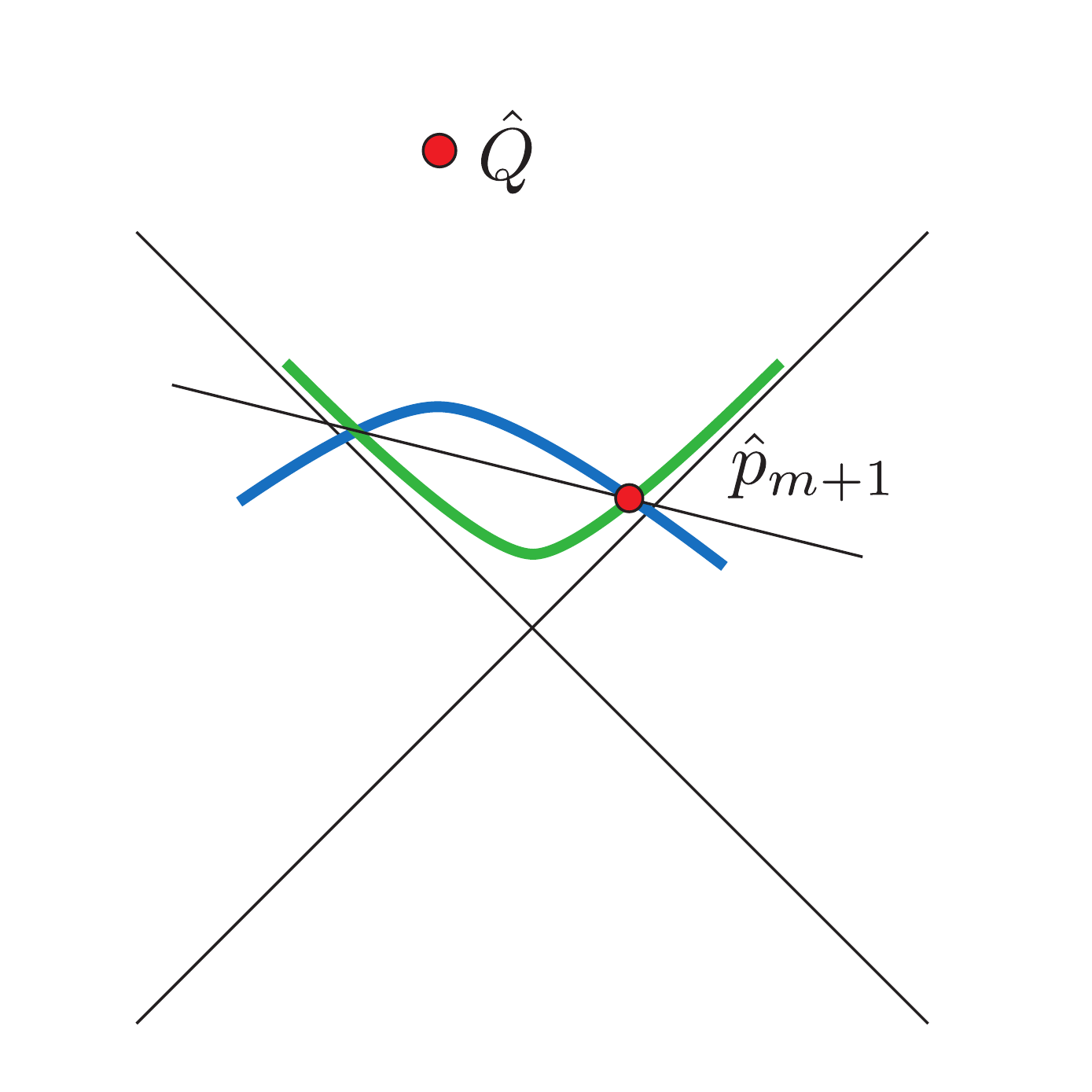}\hskip 2 cm
\caption{Momentum space for $\hat p_{m+1}$ in an initial state splitting.  Coordinate axes along $p_\LA$ and $p_\LB$ are shown, along with the point $\hat Q(\hat \eta_\La)$. We show the mass shell for $\hat p_{m+1}$ and the hyperbola $(\hat Q(\hat \eta_\La) - \hat p_{m+1})^2 = K^2$. Points $\hat p_{m+1}$ for a given value of $\hat\eta_\La$ are on the intersection of these two hyperbolas. We illustrate a point $\hat p_{m+1}$ with zero momentum transverse to $p_\LA$ and $p_\LB$, but in general $\hat p_{m+1}$ can have transverse components. The plane specified in Eq.~(\ref{eq:khatsqrelation2}) is also depicted.
}
\label{fig:mappingIS}
}

For a final state splitting, we defined a splitting variable $\zeta_{\rm p}$ in a three dimensional set $\varGamma_l$ so that $\{p\}_m$ together with $\zeta_{\rm p}$ determines $\{\hat p\}_{m+1}$. Here we use an analogous notation, with 
\begin{equation}
\label{eq:zetapIS}
\zeta_{\rm p} = (\hat p_\La, \hat p_{m+1})
\;\;.
\end{equation}
We choose
\begin{equation}
\zeta_{\rm p} \in \varGamma_\La(\{p\}_m,\zeta_{\rm f})
\;\;,
\end{equation}
where
\begin{equation}
\begin{split}
\label{eq:splitsetIS}
\varGamma_\La(\{p\}_m,\zeta_{\rm f}) = 
\Big\{(\hat p_\La, \hat p_{m+1})\;\;\Big|&\quad
\hat p_{m+1}^2 = m^2(\hat f_{m+1}),\quad \hat p_{m+1}\!\cdot\! (p_\La + p_\Lb) > 0 ,
\\&
\quad 
\hat p_\La =  \hat \eta_\La p_\LA 
+ \frac{m^2(\hat f_\La)}{\hat \eta_\La s}\, p_\LB ,
\\&
\quad
\hat K^2(\hat \eta_\La) = K^2(\eta_\La),
\quad \hat K \!\cdot\! K > 0 ,
\\&
\quad
\eta_\La < \hat\eta_\La < 1
\Big\}\;\;.
\end{split}
\end{equation}
That is, the radiated parton lies on the forward mass-shell and the new initial state parton is also on its mass shell with has zero transverse momentum. Its momentum fraction $\hat\eta_\La$ is determined by $\hat K^2 = K^2$, with $K$ inside the forward lightcone. The solution of $\hat K^2 = K^2$ is given by Eq.~(\ref{eq:hatetaLaresult}). We also need $\hat\eta_\La$ to be big enough that that the kinematic bound that was given in Eq.~(\ref{eq:etalimits}) is obeyed. This follows from $\hat \eta_\La > \eta_\La$. Finally, $\hat \eta_\La$ must be less than 1. 

As for a final state splitting, the set $\varGamma_\La$ is a three dimensional surface in the space of momenta $(\hat p_\La,\hat p_{m+1})$. One can describe this surface using three coordinates such as a virtuality variable, a momentum fraction variable, and an azimuthal angle. We leave the choice of coordinates open.

We give a name to this transformation of momenta and flavors, $R_l$ with $l = \La$:
\begin{equation}
\label{eq:Radef}
\{\hat p, \hat f\}_{m+1} = R_\La(\{p,f\}_m, \{\zeta_{\rm p},\zeta_{\rm f}\})
\;\;.
\end{equation}

\subsection{Combining an initial state parton with a final state parton}
\label{sec:IScombinationkinematics}

The transformation for splitting an initial state parton has an inverse. Let us start with $\{\hat p,\hat f\}_{m+1}$ and determine $\{p,f\}_{m}$ and $\{\zeta_{\rm p},\zeta_{\rm f}\}$.

The momentum splitting variable is simply $\zeta_{\rm p} = (\hat p_\La, \hat p_{m+1})$. For the momentum fraction of parton ``b'' before the splitting, we take
\begin{equation}
\eta_\Lb = \hat \eta_\Lb
\;\;.
\end{equation}
To determine $\eta_\La$, we simply use the representation (\ref{eq:KandhatK}) of $K^2$ and $\hat K^2$ and solve $K^2 = \hat K^2$ for $\eta_\La$ in terms of $\hat \eta_\La$. This gives
\begin{equation}
\label{eq:etaLaresult}
\eta_\La = 
\frac{1}{2\alpha}
\left\{
\hat K^2 + \gamma 
+ \sqrt{(\hat K^2 + \gamma)^2 + 4 \alpha\beta}
\right\}
\;\;.
\end{equation}

Once we have $\eta_\La$, we can construct $K = p_\La + p_\Lb$. Then from $K$ and $\hat K$ we can construct $\Lambda(K,\hat K)$, the inverse Lorentz transformation matrix to $\Lambda(\hat K, K)$ defined in Eq.~(\ref{eq:Lambdadef}) by simply using the same equation with $K \leftrightarrow \hat K$. Then we can construct the remaining momenta using
\begin{equation}
p_j^\mu = \Lambda(K,\hat K)^\mu_{\ \nu}\, \hat p_j^\nu
\;\;.
\end{equation}

The transformation of the flavors is simple. The splitting variable $\zeta_{\rm f}$ is given by the flavors of the (backwards evolution) daughter particles, $(\hat f_\La,\hat f_{m+1})$. The flavor of the mother parton is
\begin{equation}
f_\La = \hat f_\La + \hat f_{m+1}
\;\;.
\end{equation}
The flavors of the other partons are unchanged
\begin{equation}
f_j = \hat f_j 
\;\;,\hskip 1 cm j \notin \{\La,m+1\}
\;\;.
\end{equation}

We give a name to this transformation of momenta and flavors,
\begin{equation}
\label{eq:Qadef}
\{\{p,f\}_m, \{\zeta_{\rm p},\zeta_{\rm f}\}\}
= Q_\La(\{\hat p,\hat f\}_{m+1})
\;\;.
\end{equation}
This is the inverse transformation to $R_\La$, Eq.~(\ref{eq:Radef}).

\subsection{The integration measure for initial state splitting}
\label{sec:ISjacobian}

With a suitable choice of the integration measure $d\zeta_{\rm p}$ for integrating over the splitting variables $\zeta_{\rm p}$, we can arrange that 
\begin{equation}
\begin{split}
\label{eq:jacobianISdef}
\int [d\{\hat p, \hat f\}_{m+1}]\ &
g(\{\hat p, \hat f\}_{m+1}) 
\\
&= 
\int[d\{p,f\}_{m}] 
\sum_{\zeta_{\rm f}\in \Phi_{\La}(f_\La)}
\int d\zeta_{\rm p}\
\theta(\zeta_{\rm p} \in \varGamma_{\La}(\{p\}_{m},\zeta_{\rm f}))\
g(\{\hat p, \hat f\}_{m+1})
\end{split}
\end{equation}
for an arbitrary function $g(\{\hat p, \hat f\}_{m+1})$ with support in $\hat K^2 > 4 m_{\rm H}^2$. The definition that we need is
\begin{equation}
\label{eq:dzetapIS}
d\zeta_{\rm p} \equiv 
\frac{d^{4}\hat p_{m+1}}{(2\pi)^{4}}\
2\pi \delta_{+}(\hat p_{m+1}^{2} - m^2(\hat f_{m+1}))\
\frac{\alpha + \beta/\eta_\La^2}{\hat\alpha + \hat\beta/\hat\eta_\La^2}\
\;\;.
\end{equation}
The factor $(\alpha + \beta/\eta_\La^2)/(\hat\alpha + \hat\beta/\hat\eta_\La^2)$ here is just $d\hat\eta_\La/d\eta_\La$ calculated from the relation $\hat K^2 = K^2$.

\section{Spin states}
\label{sec:spinstates}

The quantum scattering amplitude $\ket{M(\{p,f\}_m)}$ is a vector in spin$\,\otimes\,$color space. Thus we can expand it in terms of spin and color basis vectors,
\begin{equation}
\ket{M(\{p,f\}_m)} =
\sum_{\{c\}_m} \ket{\{c\}_m}
\sum_{\{s\}_m} \ket{\{s\}_m}\
M(\{p,f,s,c\}_m)
\;\;.
\end{equation}
The treatment of color that appears to us to be most useful for parton showers is a bit subtle and, in particular, involves basis vectors that are not exactly conventionally normalized and are not exactly orthogonal to one another. The color basis is described in Sec.~\ref{sec:color}. In contrast, our spin basis vectors are quite standard, are orthogonal and normalized,
\begin{equation}
\brax{\{s'\}_m'}\ket{\{s\}_m} = \delta_{m',m}\,\delta_{\{s'\}_m,\{s\}_m}
\;\;.
\end{equation}
The spin labels $\{s_\La,s_\Lb,s_1,\dots,s_m\}$ represent the helicities of the corresponding particles.\footnote{Many authors follow the convention that the helicity label for an incoming particle is the negative of the particle's helicity. In contrast, our convention is that $s_\La$ and $s_\Lb$, as well as $s_1,\dots,s_m$, represent the physical helicities of the particles.}

The basis states for parton $l$ are represented for quarks or antiquarks by Dirac spinors $U(p_l,s_l)$ or $V(p_l,s_l)$, where $p_l^2 = m^2(f_l)$. For gluons, we need polarization vectors $\varepsilon(p_l,s_l)$ with $p_l^2 = 0$.
Our definition makes use an auxiliary vector $n_l$, chosen along the intersection of the positive lightcone with the plane of $p_l$ and $Q$, where $Q = \sum_{j=1}^m p_j = p_\La + p_\Lb$ is the total momentum of the final state particles. The normalization of $n_l$ is not important. We take the solution that is not close to $p_l$ in the case that $m^2(f_l)$ is small. A convenient set of choices is
\begin{equation}
\label{eq:nldef}
n_l = 
\begin{cases}
p_\LB \;\;,&  l = \La\;\;, \\
p_\LA \;\;,&  l = \Lb\;\;, \\
\displaystyle{
Q
-\frac{Q^2}
{Q\!\cdot\! p_l
+ \sqrt{(Q\!\cdot\! p_l)^2 - Q^2\, m^2(f_l)}}\
p_l
}
\;\;,&  l \in \{1,\dots,m\}\;\;.
\end{cases}
\end{equation}

For quarks and antiquarks, we use Dirac spinors $U(p,s)$ or $V(p,s)$ with $p^2 = m^2$ and $s = \pm 1/2$. We can take $V(p,s) = [\overline U(p,s)C]^T$, where $C$ is the charge conjugation matrix, $i \gamma^2 \gamma^0$ in the chiral representation of the gamma matrices, with $C^{-1} \gamma^\mu C = -(\gamma^\mu)^T$ and $C^{-1} = C^\dagger = C^T = -C$. The spinors obey $(\s{p} - m)U(p,s) = 0$ and $(\s{p} + m)V(p,s) = 0$. They are normalized to
\begin{equation}
\begin{split}
\overline U(p,s)\gamma^\mu U(p,s) ={}& 2 p^\mu
\;\;,
\\
\overline V(p,s)\gamma^\mu V(p,s) ={}& 2 p^\mu
\;\;.
\end{split}
\end{equation}
We use helicity eigenstates, defined so that
\begin{equation}
\begin{split}
\gamma_5\s{s}U(p,\pm 1/2) &= \pm U(p,\pm 1/2)
\;\;,\\
\gamma_5\s{s}V(p,\pm 1/2) &= \pm V(p,\pm 1/2)
\;\;.
\end{split}
\end{equation}
Here the spin vector $s$ is
\begin{equation}
s = \frac{1}{m}\,p - \frac{m}{p\!\cdot\! n}\,n
\;\;,
\end{equation}
where $n$ is the auxiliary lightlike vector from Eq.~(\ref{eq:nldef}). Thus $s^2 = -1$ and $s\cdot p = 0$. 

A convenient definition that defines the phase of $U(p,-s)$ in terms of the phase of $U(p,s)$ is
\begin{equation}
U(p,-s) = \left(
1 + \frac{m}{p\!\cdot\!n}\,\s{n}
\right) V(p,s)
\;\;.
\end{equation}

For gluons, we need polarization vectors $\varepsilon^\mu(p,s;Q)$, representing a given helicity $s$ and defined with the aid of an auxiliary vector $Q$. The polarization vectors obey $p \cdot \varepsilon(p,s;Q) = 0$ and
\begin{equation}
Q\!\cdot\! \varepsilon(p,s;Q) = 0
\;\;.
\end{equation}
We can also write
\begin{equation}
\varepsilon(p,s;Q) = \varepsilon(p,s;n)
\;\;,
\end{equation}
where $n\cdot \varepsilon(p,s;n) = 0$ and $n$ is a lightlike vector defined in Eq.~(\ref{eq:nldef}). We define the phase by using the standard definition in terms of mass zero Dirac spinors \cite{helicity},
\begin{equation}
\varepsilon^{\mu}(p, \pm 1; n) = 
\pm\frac{\overline U({n,\pm 1/2})\gamma^{\mu}U({p,\pm 1/2})}
{\sqrt{2}\,\overline U({ p,\mp 1/2})U({n,\pm 1/2})}
\;\;.
\end{equation}

With this definition, there is a simple relation between polarization vectors defined with different auxiliary vectors $Q$ and thus different lightlike auxiliary vectors $n$ \cite{helicity},
\begin{equation}
\begin{split}
\varepsilon^{\mu}(p, \pm 1; n)-\varepsilon^{\mu}(p, \pm 1; n')
= \mp
\frac{\sqrt{2}\,\overline U({n,\pm 1/2}) U({n',\mp 1/2})}
{\overline U({n,\pm 1/2})U({p,\mp} 1/2)\ 
\overline U({p,\pm 1/2})U({n',\mp 1/2})}
\ p^{\mu}
\;\;.
\end{split}
\end{equation}
If we use these polarization vectors with the exact tree-level Feynman diagrams, gauge invariance of the matrix elements, together with the fact that $\varepsilon^{\mu}(p, s; n)$ differs from $\varepsilon^{\mu}(p, s; n')$ by a vector proportional to $p^\mu$, shows that the amplitude is independent of the choice of the auxiliary vector $Q$. Our matrix elements will be approximate and will be gauge invariant only to the extent that the splittings are close to the soft or collinear limits. Thus some dependence on the auxiliary vector used to define the polarization vectors will result.

\section{Splitting functions for the quantum states}
\label{sec:QuantumSpltting}

Consider for a moment a theory without spin, color, or flavors, say $\phi^3$ theory in six dimensions as in Sec.~\ref{sec:notation}. An $(m+1)$-parton scattering amplitude $M(\{\hat p\}_{m+1})$ is simple in the limit in which two of the partons are approximately collinear. Supposing that partons $m+1$ and $l$ are almost collinear, we have
\begin{equation}
M(\{\hat p\}_{m+1}) \approx v(\{\hat p\}_{m+1})\,M(\{p\}_{m})
\;\;,
\end{equation}
where $v(\{\hat p\}_{m+1}) = g/(2\hat p_l \cdot \hat p_{m+1})$. Here $\{p\}_{m}$ is determined from $\{\hat p\}_{m+1}$ by the (six-dimensional version of) the transformation described in Sec.~\ref{sec:mapping}. This factorization formula for the amplitude becomes exact in the limit that partons $m+1$ and $l$ become collinear. Away from the collinear limit, there is some freedom to choose the momentum mapping and the splitting amplitude $v(\{\hat p\}_{m+1})$. One has to make a definite choice based on ease of computation or conceptual simplicity. In the case of QCD, we have soft as well as collinear singularities, we have parton flavors (which are rather trivially treated) and we have color and spin, which are not so trivial. Let us see how to describe splitting in QCD. 

It has been known for a long time that QCD amplitudes factor in the soft and collinear limits \cite{sofcollfact}. Indeed, there are beautiful modern formulas for the factors \cite{helicity,DixonTASI,antenna}. We have adopted a more pedestrian approach that has at least the advantage of encompassing the soft and collinear limits at the same time and of including masses. A treatment of the squared amplitude that is rather similar to the approach of this paper, but at higher order, may be found in Ref.~\cite{CataniGrazzini}.

\subsection{Definition of the splitting functions $v_l$}
\label{sec:vl}

The QCD scattering amplitude for $m+1$ partons is a vector $\ket{\ME(\{\hat p, \hat f\}_{m+1})}$ in color\,$\otimes$\,spin space. In the limit that two partons, $l$ and $m+1$ are almost collinear, this amplitude takes a certain limiting form,
\begin{equation}
\label{eq:partonlsplits}
\ket{\ME(\{\hat p, \hat f\}_{m+1})} \sim
\ket{\ME_l(\{\hat p, \hat f\}_{m+1})}
\;\;,
\end{equation}
where $\ket{\ME_l(\{\hat p, \hat f\}_{m+1})}$ is to be defined precisely below. When $\hat p_{m+1}$ becomes soft, then all of the $\ket{\ME_l(\{\hat p, \hat f\}_{m+1})}$ amplitudes contribute to the limit,
\begin{equation}
\label{eq:partonissoft}
\ket{\ME(\{\hat p, \hat f\}_{m+1})} \sim
\sum_l\ket{\ME_l(\{\hat p, \hat f\}_{m+1})}
\;\;.
\end{equation}
We arrange the definition so that Eqs.~(\ref{eq:partonlsplits}) and (\ref{eq:partonissoft}) are exact in the collinear or soft limit respectively. We also arrange that $\ket{\ME_l(\{\hat p, \hat f\}_{m+1})}$ is defined for any $\{\hat p, \hat f\}_{m+1}$. Then these equations are approximate away from the limit. The amplitude $\ket{\ME_l(\{\hat p, \hat f\}_{m+1})}$ is then the contribution to the $(m+1)$-parton amplitude from the splitting of parton $l$ in the parton shower approximation.

We now need to define $\ket{\ME_l(\{\hat p, \hat f\}_{m+1})}$. This amplitude factors into a splitting operator times the $m$-parton matrix element evaluated at momenta and flavors $\{p, f\}_{m}$ determined from $\{\hat p, \hat f\}_{m+1}$ according to the transformation $Q_l(\{\hat p,\hat f\}_{m+1})$, Eq.~(\ref{eq:Qldef}) or Eq.~(\ref{eq:Qadef}),
\begin{equation}
\label{eq:splittingamplitudestructure}
\ket{\ME_l(\{\hat p, \hat f\}_{m+1})} = 
t^\dagger_l(f_l \to \hat f_l + \hat f_{m+1})\,
V^\dagger_l(\{\hat p, \hat f\}_{m+1})\, \ket{\ME(\{p, f\}_{m})}
\;\;.
\end{equation}
In Eq.~(\ref{eq:splittingamplitudestructure}), $V^\dagger(\{\hat p, \hat f\}_{m+1})$ is the analogue of $v(\{\hat p\}_{m+1})$ but is now an operator on the spin part of the color\,$\otimes$\,spin space. There is also an operator $t^\dagger_l(f_l \to \hat f_l + \hat f_{m+1})$ on the color part of the color\,$\otimes$\,spin space. This operator multiplies by the right color matrix. We will not comment further on it in this section, but will turn to the description of color in Sec.~\ref{sec:color}. The spin dependent splitting operator can be described in terms of its matrix elements,
\begin{equation}
\bra{\{\hat s\}_{m+1}}
V^\dagger_l(\{\hat p, \hat f\}_{m+1})\ket{\{s\}_m}
\;\;.
\end{equation}
This is a simple function of $\{\hat p, \hat f\}_{m+1}$, $\{\hat s\}_{m+1}$, and $\{s\}_m$. Furthermore, we can take it to be diagonal in the spectator spins,
\begin{equation}
\label{eq:Vtov}
\bra{\{\hat s\}_{m+1}}
V^\dagger_l(\{\hat p, \hat f\}_{m+1})\ket{\{s\}_m}
= 
\left(\prod_{j\notin\{l,m+1\}} \delta_{\hat s_j,s_j}\right)
v_l(\{\hat p, \hat f\}_{m+1},\hat s_{m+1},\hat s_{l},s_l)
\;\;.
\end{equation}
Our object in this section is to define the splitting functions $v$ from the QCD vertices. In writing formulas for $v_l(\{\hat p, \hat f\}_{m+1},\hat s_{m+1},\hat s_{l},s_l)$, we will use the momentum $p_l$. We understand that this is obtained from $\{\hat p, \hat f\}_{m+1}$ according to the transformation $Q_l(\{\hat p,\hat f\}_{m+1})$.

\subsection{Initial state $q \to q + {\rm g}$ splitting, quark scatters}
\label{sec:ISqqg1splitting}

\FIGURE{
\includegraphics[width = 10 cm]{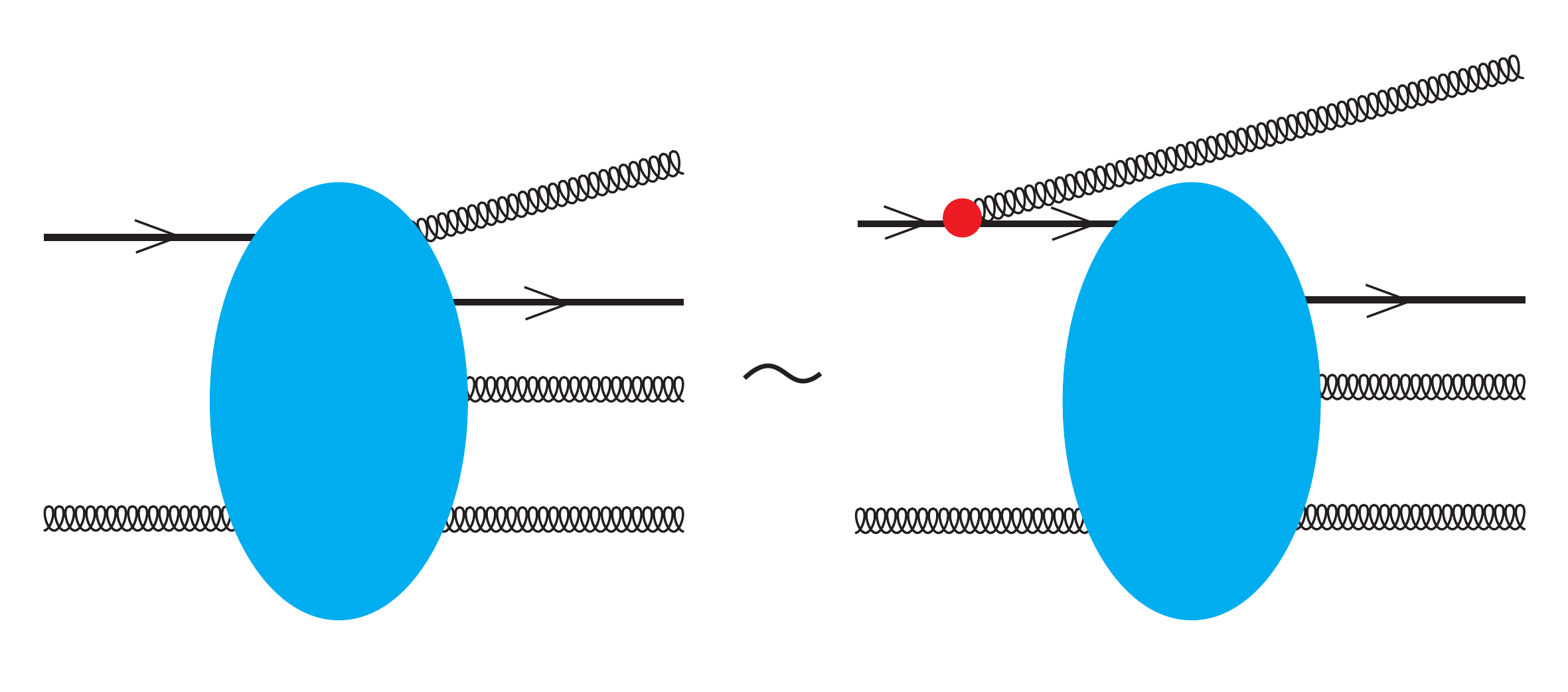}
\caption{Illustration of factorization at the amplitude level, leading to the definition of the splitting function in Eq.~(\ref{eq:qtoqgIspitting}). When the final state gluon becomes collinear with the initial state quark, the full amplitude is approximately the amplitude for one fewer parton, with the incoming quark after the gluon radiation approximated as being on shell, times a splitting function simply made from the QCD $qqg$ vertex and the singular quark propagator. If the final state gluon is soft, then this diagram is one of the possibilities. Then, the soft gluon could have been emitted from any of the external legs of the diagram and we must sum over all of the possible emissions. In this case, we can, however, use a simpler splitting function.
}
\label{fig:qqgIsplitting}
}

Consider an initial state $q \to q + {\rm g}$ splitting in which the gluon goes into the final state and the quark enters the hard scattering, as illustrated in Fig.~\ref{fig:qqgIsplitting}. We highlight this case because it exhibits some complications compared to final state $q \to q + {\rm g}$ splitting. The parton that splits could be either of the initial state partons. We examine the case that it is parton ``a.'' The kinematics were defined in Secs.~\ref{sec:ISsplittingkinematics} and \ref{sec:IScombinationkinematics}. In order to keep our notation for this subsection as simple as possible, we will write
\begin{equation}
\begin{aligned}
p &=  p_\La
\;\;,
\quad &
\eta &= \eta_\La
\;\;,
\\
\hat p &= \hat p_\La
\;\;,
&
\hat \eta &= \hat \eta_\La
\;\;,
\\
q & = \hat p_{m+1} 
\;\;, &
m & = m(f_\La) = m(\hat f_\La)
\;\;,
\\
\hat s & = \hat s_\La
\;\;,
&
\varepsilon_\mu & = \varepsilon_\mu(\hat p_{m+1},\hat s_{m+1};\hat Q)
\;\;.
\end{aligned}
\end{equation}
Here $\varepsilon$ is defined to be orthogonal to $\hat Q = \hat p_\La + \hat p_\Lb = \hat p_\La + p_\Lb$.

The Feynman rules for $\ket{M(\{\hat p,\hat f\}_{m+1})}$ give the following structure
\begin{equation}
\label{eq:split0}
M =
H
\frac{\s{P} + m}{P^2- m^2}\ 
g t^c \s{\varepsilon}^*
U(\hat p,\hat s)
\;\;.
\end{equation}
Here there are factors $U(\hat p,\hat s)$ for the initial state quark and $\varepsilon_\mu^*$ for the final state gluon. There is a vertex $g\gamma_\mu$ with a color matrix $t^c$. There is a propagator for the off-shell quark that carries momentum
\begin{equation}
P = \hat p - q
\;\;.
\end{equation}
The quark propagator has a denominator $P^2 - m^2$ and a numerator $\s{P} + m$. Finally, there is the rest of the diagram, $H$, which contains hard interactions. Thus $H$ carries a Dirac spinor index. We need to relate this to $\ket{\ME(\{p, f\}_{m})}$, where the momenta and flavors $\{p, f\}_{m}$ as well as the splitting variables $\{\zeta_{\rm p},\zeta_{\rm f}\}$ are given by the momentum and flavor mapping $Q_l(\{\hat p, \hat f\}_{m+1})$, Eq.~(\ref{eq:Qldef}).

We will be concerned with the behavior of $M$ in the soft and collinear limits. The soft limit is $q \to 0$ (for all four components of $q$). Since $P^2 - m^2 = -2 \hat p\cdot q$, the denominator is proportional to a single power of $q$, while the numerator stays finite in the $q \to 0$ limit. To describe the collinear limit, let $q_\perp$ be the part of $q$ orthogonal to $p_\LA$ and $p_\LB$.  The collinear limit is $q_\perp \to 0$ while $q\cdot p_\LB$ stays finite. Also, $m \to 0$ with $m^2 \lesssim |q_\perp^2|$. Then near the collinear limit, the denominator is
\begin{equation}
P^2 - m^2 = 
\left(\frac{\hat \eta}{\hat \eta - \eta}\ q_\perp^2
-\frac{\hat \eta - \eta}{\hat \eta}\ m^2
\right)(1 + {\cal O}( q_\perp^2, m^2))
\;\;.
\end{equation}
That is, the denominator has two powers of $q_\perp$. To analyze the numerator, we note that
\begin{equation}
\begin{split}
p ={}& \eta p_\LA + {\cal O}( q_\perp^2, m^2)
\;\;,
\\
\hat p ={}& \hat \eta p_\LA + {\cal O}( q_\perp^2, m^2)
\;\;,
\\
\hat q ={}& (\hat \eta - \eta)p_\LA + q_\perp + {\cal O}( q_\perp^2, m^2)
\;\;.
\end{split}
\end{equation}
The numerator has a factor $(\s{P} + m) \s{\varepsilon}^* U(\hat p,\hat s)$. Using $q\cdot \varepsilon = 0$ and $(\s{\hat p} - m))U(\hat p,\hat s) = 0$, one can rewrite this factor in the form
\begin{equation}
\label{eq:split1A}
(\s{P} + m)
\s{\varepsilon}^*
U(\hat p,\hat s)
= 2 \Bigl(\hat p - \frac{\hat \eta}{\hat \eta - \eta} q\Bigr)
 \cdot \varepsilon^*\
U(\hat p,\hat s)
+ \s{\varepsilon}^*
\left[\s{q} - \frac{\hat \eta - \eta}{\hat \eta}(\s{\hat p} - m)
\right]U(\hat p,\hat s)
\;\;.
\end{equation}
Looking at the numerator in this form, we see that it vanishes in the collinear limit proportionally to one power of $q_\perp$ or $m$. Thus $M$ is only half as singular in the collinear limit as it first appears.

We now develop an approximation for $M$. We can insert a factor 1 next to $H$ in Eq.~(\ref{eq:split0}) so that it reads
\begin{equation}
\label{eq:split1}
M =
H\,
\frac{\s{n} (\s{p} - m) + (\s{p} + m) \s{n}}{2 p\!\cdot\! n}\
\frac{\s{P} + m}{P^2- m^2}\
(g t^c \s{\varepsilon}^*)
U(\hat p,\hat s)
\;\;.
\end{equation}
Here $n$ is the lightlike vector $n = p_\LB$. 

We now notice that the contribution from the first term, namely 
\begin{equation}
\label{eq:split2}
M_{\rm ns} =
H\,
\frac{\s{n} (\s{p} - m) }{2 p\!\cdot\! n}
\frac{\s{P} + m}{P^2- m^2}\
(g t^c \s{\varepsilon}^*)
U(\hat p,\hat s)
\;\;,
\end{equation}
can be neglected because it does not have a collinear or soft singularity. To see this takes a little analysis. First, we write
\begin{equation}
p = P + (p + q - \hat p)
\;\;.
\end{equation}
Since $(\s{P} - m)(\s{P} + m) = P^2 - m^2$, we have
\begin{equation}
\label{eq:split3}
M_{\rm ns} =
H\,
\frac{\s{n}}{2 n\!\cdot\! p}\
(g t^a \s{\varepsilon}^*)
U(\hat p,\hat s)
+
H\,
\frac{\s{n} (\s{p} + \s{q} - \s{\hat p})}{2 p\!\cdot\! n}
\frac{\s{P} + m}{P^2- m^2}\
(g t^c \s{\varepsilon}^*)
U(\hat p,\hat s)
\;\;.
\end{equation}
The first term is non-singular because the denominator is cancelled. In the second term, the vector $(p + q - \hat p)$ vanishes in the collinear or soft limit. In the soft limit, it is proportional to one power of $q$, which cancels the single power of $q$ in the denominator. In the collinear limit, it is proportional to one power of $q_\perp$. As we have just seen, the rest of the numerator contains an additional factor of $q_\perp$. Together, these cancel the two powers of $q_\perp$ from the denominator. Thus no singularity remains.

We are left with
\begin{equation}
\label{eq:split4}
M_{\rm sing} =
H\,
\frac{(\s{p} + m) \s{n}}{2 p\!\cdot\! n}
\frac{\s{P} + m}{P^2- m^2}\
(g t^c \s{\varepsilon}^*)
U(\hat p,\hat s)
\;\;.
\end{equation}
In Eq.~(\ref{eq:split4}), the factor $\s{p} + m$ is 
\begin{equation}
\s{p} + m = \sum_{s} U(p,s) \overline U(p,s)
\;\;.
\end{equation}
The factor $U(p,s)$ is to be associated with $H$, giving the hard scattering amplitude for an incoming quark with spin $s$. The remaining factor, $\overline U(p,s)$, then becomes part of the splitting function.  This calculation leads us to define the splitting function as
\begin{equation}
\label{eq:qtoqgIspitting}
v_\La
=
-\frac{\sqrt{4\pi\as}}{(\hat p - q)^2 - m^2}\,
\varepsilon_\mu^*\,
\frac{\overline U({p,s})\s{n}(\s{\hat p} - \s{q} + m)
\gamma^\mu U({\hat p,\hat s})}
{2 p\!\cdot\! n}
\;\;.
\end{equation}
This does not include the color matrix and a factor $-1$, which will be included in the color operator $t^\dagger_l(f_l \to \hat f_l + \hat f_{m+1})$. As indicated by the derivation, Eq.~(\ref{eq:qtoqgIspitting}) is directly given by the factorized structure of QCD Feynman graphs in the soft and collinear limits. There is freedom to choose the form of the splitting function as one moves away from these limits. We have made a simple choice.

In the hard part of the diagram, we can make approximations that are valid for $q_\perp^2 \to 0$. In particular, we can adjust the momenta of the partons with indices other than ``a,'' replacing $\hat p_j$ by $p_j$ as defined by the momentum mapping $R_\La(\{p,f\}_m, \{\zeta_{\rm p},\zeta_{\rm f}\})$, Eq.~(\ref{eq:Radef}).

\subsection{Initial state $q \to q + {\rm g}$ splitting, gluon scatters}
\label{sec:ISqqg2splitting}

We consider next the process in which an initial state quark with label ``a'' splits to make a quark that goes into the final state (with label $m+1$) and a gluon that enters the hard scattering (with label ``a''). We simplify the notation as in the previous subsection by using
\begin{equation}
\begin{aligned}
p &=  p_\La
\;\;,
\quad &
\eta &= \eta_\La
\;\;,
\\
\hat p &= \hat p_\La
\;\;,
&
\hat \eta &= \hat \eta_\La
\;\;,
\\
q & = \hat p_{m+1} 
\;\;, &
m & = m(\hat f_\La) = m(\hat f_{m+1})
\;\;,
\\
\hat s & = \hat s_\La
\;\;,
&
s' &= \hat s_{m+1}
\;\;.
\end{aligned}
\end{equation}
The Feynman rules for $\ket{M(\{\hat p,\hat f\}_{m+1})}$ give the following structure
\begin{equation}
\label{eq:splitg0}
M =
H^\mu\,
\frac{D_{\mu\nu}(P;n)}{P^2}\
\overline U(q,s')
(g t^c \gamma^\nu)
U(\hat p,\hat s)
\;\;.
\end{equation}
Here $H$ is the hard part of the graph, now with a vector index, and there is a propagator for the off-shell gluon that carries momentum
\begin{equation}
P = \hat p - q
\;\;.
\end{equation}
We have chosen the axial gauge $n\cdot A = 0$, where $n$ is the lightlike vector $n = P_\LB$. The numerator of the gluon propagator is
\begin{equation}
D^{\mu\nu}(P;n) = - g^{\mu\nu} 
+ \frac{P^\mu n^\nu + n^\mu P^\nu}{P\!\cdot\! n}
\;\;.
\end{equation}
As in the previous subsection, $M$ is singular in the collinear limit, in which $q_\perp \to 0$ and $m \to 0$.\footnote{We do not need to be concerned with the soft limit, $q \to 0$ with $m \to 0$. Here, there is only a $1/\sqrt q$ singularity, which is too weak to create a logarithmically divergent integration over final states. Nevertheless, our approximation to $M$ also matches the behavior of $M$ in the soft limit.} As in the previous section, the numerator is proportional to $q_\perp$ and the denominator is proportional to $q_\perp^2$ in the collinear limit.

In order to find a suitable approximation for $M$, we insert 1 next to $H$, so that it reads
\begin{equation}
\label{eq:splitg1}
M =
H_\alpha\,
\left[
- D^{\alpha\mu}(p;n)
+ \frac{p^\alpha n^\mu + n^\alpha p^\mu}{p\!\cdot\! n}
\right]
\frac{D_{\mu\nu}(P;n)}{P^2}\
\overline U(q, s')
(g t^c \gamma^\nu)
U(\hat p, \hat s)
\;\;.
\end{equation}
We can drop the term  $(p^\alpha n^\mu + n^\alpha p^\mu)/(n\!\cdot\! p)$. To see this, we write
\begin{equation}
\frac{p^\alpha n^\mu + n^\alpha p^\mu}{p\!\cdot\! n}
= 
\frac{P^\alpha n^\mu + n^\alpha P^\mu}{P\!\cdot\! n}
+ D^{\alpha\mu}(p;n) - D^{\alpha\mu}(P;n)
\;\;.
\end{equation}
The term $P^\alpha n^\mu$ gives zero when contracted with $D_{\mu\nu}$. When we contract $n^\alpha P^\mu$ with $D_{\mu\nu}$, we get
\begin{equation}
\frac{n^\alpha P^\mu}{P\!\cdot\! n}\
\frac{D_{\mu\nu}(P;n)}{P^2}
= \frac{n^\alpha n_\nu}{(P\!\cdot\! n)^2}
\;\;,
\end{equation}
which does not have a collinear singularity. Finally, the difference $D^{\alpha\mu}(p;n) - D^{\alpha\mu}(P;n)$ is proportional to $q_\perp$ in the collinear limit because $P - p \propto q_\perp$ in this limit. The remaining numerator factor gives another factor $q_\perp$ in the collinear limit, so that the factor $q_\perp^2$ from the denominator is cancelled.

Thus we are left with $M \sim M_{\rm sing}$, where
\begin{equation}
\label{eq:splitg2}
M_{\rm sing} =
- H_\alpha\,
D^{\alpha\mu}(p;n)\
\frac{D_{\mu\nu}(P;n)}{P^2}\
\overline U(q, s')
(g t^c \gamma^\nu)
U(\hat p, \hat s)
\;\;.
\end{equation}
The factor $D^{\alpha\mu}(p;n)$ is
\begin{equation}
D^{\alpha\mu}(p;n) = \sum_{s}
\varepsilon^\alpha(p,s;n)\varepsilon^\mu(p,s;n)^*
= \sum_{s}
\varepsilon^\alpha(p,s;\hat Q)\varepsilon^\mu(p,s;\hat Q)^*
\;\;.
\end{equation}
It is equivalent to use $n = p_\LB$ or $\hat Q = \hat p_\La + \hat p_\Lb = \hat p_\La + p_\Lb$ to define the polarization vectors since they are orthogonal to both $p = p_\La$ and $p_\Lb$. The factor $\varepsilon^\alpha(p,s;\hat Q)$ is to be associated with $H$, giving the hard scattering amplitude for a gluon with spin $s$. The remaining factor, $\varepsilon^\mu(p,s;\hat Q)^*$, then becomes part of the splitting function.  

This calculation leads us to define the splitting function as
\begin{equation}
\begin{split}
\label{eq:qtoqgspitting}
v_\La
=
-\frac{\sqrt{4\pi\as}}{(\hat p - q)^2}\,
\varepsilon^\mu(p, s;\hat Q)^*\,
D_{\mu\nu}(\hat p - q ;n)\
\overline U(q, s')
\gamma^\nu
U(\hat p, \hat s)
\;\;.
\end{split}
\end{equation}
This does not include the color matrix, which is included in the color operator $t^\dagger_l(f_l \to \hat f_l + \hat f_{m+1})$.

In the hard part of the diagram, we can again make approximations that are valid for $q_\perp^2 \to 0$. In particular, we can adjust the momenta of the partons with indices other than ``a,'' replacing $\hat p_j$ by $p_j$ as defined by the momentum mapping $R_\La(\{p,f\}_m, \{\zeta_{\rm p},\zeta_{\rm f}\})$, Eq.~(\ref{eq:Radef}). We will also want to use $Q = p_\La + p_\Lb$ instead of $\hat Q = \hat p_\La + \hat p_\Lb$ to define the polarization vector for parton $l$ as it enters the hard scattering. This is just a change in notation rather than an approximation since $Q$ is in the $p$-$\hat Q$ plane.

\subsection{Other $q q g$ splittings}
\label{sec:Otherqqgsplitting}

For splittings involving a $qqg$ vertex, there are several other choices for the flavors $f_l$, $\hat f_l$ and $\hat f_{m+1}$ and for whether the index $l$ of the parton that splits is a final state index or an initial state index. The results for $v_l(\{\hat p, \hat f\}_{m+1},\hat s_{m+1},\hat s_{l},s_l)$ are listed in Table~\ref{tab:splitting}. In constructing this table, we keep track of two sign factors. First, there is a numerator sign that we compute as in the examples in the preceding subsections. This is a minus sign whenever a gluon or an antiquark enters the hard scattering from the initial state or leaves the hard scattering to the final state and a plus sign when a quark enters or leaves the hard scattering. Second, there is a color sign. There is always a color matrix $t^a$. We count a color ${\bf 3}$ line entering the hard scattering as a color $\bar{\bf 3}$ line leaving the hard scattering. Viewed this way, the color matrix is actually $(t^a)^T$. However, the generator of color rotations for the $\bar{\bf 3}$ representation of SU(3) is $-(t^a)^T$. Thus we include $-(t^a)^T$ as part of the color operator $t^\dagger_l(f_l \to \hat f_l + \hat f_{m+1})$ and include the minus sign as part of the splitting function $v_l$. This leaves a factor $-1$ in $v_l$ whenever a $\bar{\bf 3}$ line leaves the hard scattering. The sign included in $t^\dagger_l(f_l \to \hat f_l + \hat f_{m+1})$ is indicated in the last column of Table~\ref{tab:splitting}. The net sign of $v_l$ is shown in the table.

The construction for a final state splitting, as for an initial state splitting, makes use of a lightlike vector $n_l$ that is in the plane of $p_l$ and $\hat Q = \hat p_\La + \hat p_\Lb$, which is the same as the plane of $p_l$ and $Q = p_\La + p_\Lb$. The normalization of $n_l$ is not significant. Our choice for this vector was defined in Eq.~(\ref{eq:nldef}).

\TABLE{
\begin{tabular}{cccccc}
\hline
\\
$l$&$f_l$&$\hat f_l$&$\hat f_{m+1}$\!\!\!\!&
$\displaystyle{v_l \times \frac{1}{\sqrt{4\pi\as}}}$
&color \\
\\
$F$&
$q$&
$q$&
g&
$\displaystyle{
\varepsilon_\mu(\hat p_{m+1},\hat s_{m+1};\hat Q)^*\,
\frac{
\overline U({\hat p_l,\hat s_l})\gamma^\mu 
  [\s{\hat p}_l + \s{\hat p}_{m+1} + m(f_l)] \s{n_l} U({p_l,s_l})}
{2p_l\!\cdot\! n_l\ [(\hat p_l + \hat p_{m+1})^2 - m^2(f_l)]}
}
$&
$t^a$\\
&&&&&\\
$F$&
$\bar q$&
$\bar q$&
g&
$\displaystyle{
\varepsilon_\mu(\hat p_{m+1},\hat s_{m+1};\hat Q)^*\,
\frac{
\overline V({p_l,s_l})\s{n_l} 
  [\s{\hat p}_l + \s{\hat p}_{m+1} - m(f_l)]
  \gamma^\mu V({\hat p_l,\hat s_l})}
{2p_l\!\cdot\! n_l\ [(\hat p_l + \hat p_{m+1})^2 - m^2(f_l)]}
}
$&
$-t^a$\\
&&&&&\\
$I$&
$\bar q$&
$\bar q$&
g&
$\displaystyle{
-
\varepsilon_\mu(\hat p_{m+1},\hat s_{m+1};\hat Q)^*\,
\frac{\overline U({p_l,s_l})\s{n}_l
(\s{\hat p}_l - \s{\hat p}_{m+1} + m(f_l))
\gamma^\mu
 U({\hat p_l,\hat s_l})}
{2p_l\!\cdot\! n_l\ [(\hat p_l - \hat p_{m+1})^2 - m^2(f_l)]}
}
$&
$-t^a$\\
&&&&&\\
$I$&
$q$&
$q$&
g&
$\displaystyle{
-
\varepsilon_\mu(\hat p_{m+1},\hat s_{m+1};\hat Q)^*\,
\frac{\overline V({\hat p_l,\hat s_l})\gamma^\mu
(\s{\hat p}_l - \s{\hat p}_{m+1} - m(f_l))
   \s{n}_l V({p_l,s_l})}
{2p_l\!\cdot\! n_l\ [(\hat p_l - \hat p_{m+1})^2 - m^2(f_l)]}
}
$&
$t^a$\\
&&&&&\\
$F$&
g&
$q$&
$\bar q$&
$\displaystyle{
-
\varepsilon^\mu(p_{l},s_{l};\hat Q)
D_{\mu\nu}(\hat p_l + \hat p_{m+1},n_l)
\frac{
\overline U({\hat p_l,\hat s_l})\gamma^\nu V({\hat p_{m+1},\hat s_{m+1})}
}{(\hat p_l + \hat p_{m+1})^2}
}$&
$t^a$\\
&&&&&\\
$I$&
g&
$\bar q$&
$q$&
$\displaystyle{
-
\varepsilon^\mu(p_l,s_l;\hat Q)^*
D_{\mu\nu}(\hat p_l - \hat p_{m+1};n_l)
\frac{
\overline U({\hat p_{m+1},\hat s_{m+1}})\gamma^\nu 
U({\hat p_l,\hat s_l})
}{(\hat p_l - \hat p_{m+1})^2}
}
$&
$t^a$\\
&&&&&\\
$I$&
g&
$q$&
$\bar q$&
$\displaystyle{
-
\varepsilon^\mu(p_l,s_l;\hat Q)^*
D_{\mu\nu}(\hat p_l - \hat p_{m+1};n_l)
\frac{
\overline V({\hat p_l,\hat s_l})\gamma^\nu 
V({\hat p_{m+1},\hat s_{m+1}})
}{(\hat p_l - \hat p_{m+1})^2}
}
$&
$t^a$\\
&&&&&\\
$I$&
$q$&
g&
$q$&
$\displaystyle{
-
\varepsilon_\mu(\hat p_l,\hat s_l;\hat Q)\,
\frac{
\overline U({\hat p_{m+1},\hat s_{m+1}})\gamma^\mu 
[\s{\hat p}_l - \s{\hat p}_{m+1} - m(f_l)]\s{n_l}V({p_l,s_l})
}
{2p_l\!\cdot\! n_l\ [(\hat p_l - \hat p_{m+1})^2 - m^2(f_l)]}
}$&
$t^a$\\
&&&&&\\
$I$&
$\bar q$&
g&
$\bar q$&
$\displaystyle{
-
\varepsilon_\mu(\hat p_l,\hat s_l;\hat Q)\,
\frac{
\overline U({p_l,s_l})\s{n_l} 
[\s{\hat p}_l - \s{\hat p}_{m+1} + m(f_l)]
\gamma^\mu
V({\hat p_{m+1},\hat s_{m+1}})
}
{2p_l\!\cdot\! n_l\ [(\hat p_l - \hat p_{m+1})^2 - m^2(f_l)]}
}$&
$-t^a$
\\  [10 pt]
\hline
\end{tabular}
\caption{Splitting functions $v_l(\{\hat p, \hat f\}_{m+1},\hat s_{m+1},\hat s_{l},s_l)$ for splittings involving a $q\bar q g$ vertex, with a common factor $\sqrt{4\pi\as}$ removed. The values of $l$ are either in the set of initial state indices $I = \{\La,\Lb\}$ or in the set of final state indices $F = \{1,\dots,m\}$. The flavors $f_l$, $\hat f_l$, and $\hat f_{m+1}$ can be $f = {\rm g}$ or can be a quark index, ${q} \in \{{\rm u},{\rm d},\dots\}$, or an antiquark index $\bar {q} \in \{\bar{\rm u},\bar{\rm d},\dots\}$. Recall from Eq.~(\ref{eq:initialantiflavors}) that in the case of an initial state parton, $f_l$ and $\hat f_l$ denote the opposite of the incoming flavors of the parton. For a row in which $q$ denotes a quark index, $\bar q$ denotes the corresponding antiquark index. The next column gives the value of $v_l$ corresponding to the values of $l$, $f_l$, $\hat f_l$, and $\hat f_{m+1}$ indicated. The last column indicates the sign of the color matrix that is incorporated into the color operator $t_l^\dagger$. The lightlike vector $n_l$ is defined in Eq.~(\ref{eq:nldef}).}
\label{tab:splitting}
}

\subsection{Splitting with a $ggg$ vertex}
\label{sec:gggSplitting}

We construct the splitting function for a ${\rm g} \to {\rm g} + {\rm g}$ splitting in a similar fashion. In the case of a final state splitting, we use the ${\rm g}{\rm g}{\rm g}$ QCD vertex,
\begin{equation}
\label{eq:vgg}
v^{\alpha \beta \gamma}(p_a, p_b, p_c)
= g^{\alpha\beta} (p_a - p_b)^\gamma
+ g^{\beta\gamma} (p_b - p_c)^\alpha
+ g^{\gamma\alpha} (p_c - p_a)^\beta
\;\;,
\end{equation}
to define
\begin{equation}
\begin{split}
\label{eq:VggF}
v_l(\{\hat p, \hat f\}_{m+1},&\hat s_{m+1},\hat s_{l},s_l)
\\ & =
\frac{\sqrt{4\pi\as}}{2 \hat p_{m+1}\!\cdot\! \hat p_l}\, 
\varepsilon_{\alpha}(\hat p_{m+1}, \hat s_{m+1};\hat Q)^*
\varepsilon_{\beta}(\hat p_{l}, \hat s_l;\hat Q)^*
\varepsilon^{\nu}(p_{l}, s_l;\hat Q)
\\&\quad\times
v^{\alpha \beta \gamma}(\hat p_{m+1},\hat p_l,-\hat p_{m+1}-\hat p_l)\,
D_{\gamma\nu}(\hat p_l + \hat p_{m+1};n_l)
\;\;.
\end{split}
\end{equation}
For an initial state splitting, we have
\begin{equation}
\begin{split}
\label{eq:VggI}
v_l(\{\hat p, \hat f\}_{m+1},&\hat s_{m+1},\hat s_{l},s_l)
\\&=
-
\frac{\sqrt{4\pi\as}}{2 \hat p_{m+1}\!\cdot\! \hat p_l}\, 
\varepsilon_{\alpha}(\hat p_{m+1}, \hat s_{m+1};\hat  Q)^*
\varepsilon_{\beta}(\hat p_{l},\hat s_l; \hat Q)
\varepsilon^{\nu}(p_{l}, s_l; \hat Q)^*
\\&\quad\times
v^{\alpha \beta \gamma}(\hat p_{m+1}, -\hat p_l, \hat p_l - \hat p_{m+1})\,
D_{\gamma\nu}(\hat p_l - \hat p_{m+1};n_l)
\;\;.
\end{split}
\end{equation}
In each case, we have the exact QCD vertex and the exact propagator for the off-shell gluon in $n_l\cdot A = 0$ gauge followed by a projection onto the physical gluon degrees of freedom contained in the on-shell polarization vector.

\subsection{Soft splitting function}
\label{sec:softSplitting}

These splitting functions enable us to approximate the $(m+1)$-parton matrix element in the cases that $\hat p_{m+1}$ is collinear with $\hat p_l$ or else $\hat p_{m+1}$ is soft. In the special case that $\hat p_{m+1}$ is soft, or possibly soft and collinear with $\hat p_l$, a simpler splitting function can be used. When  $\hat p_{m+1}$ is soft, we have
\begin{equation}
\label{eq:partonissoftbis}
\ket{\ME(\{\hat p, \hat f\}_{m+1})} \sim
\sum_l\ket{\ME_l^{\rm soft}(\{\hat p, \hat f\}_{m+1})}
\;\;,
\end{equation}
where
\begin{equation}
\label{eq:splittingamplitudesoft}
\ket{\ME_l^{\rm soft}
(\{\hat p, \hat f\}_{m+1})} = t^\dagger_l(f_l \to \hat f_l + \hat f_{m+1})\,
V_l^{\dagger,{\rm soft}}(\{\hat p, \hat f\}_{m+1})\, \ket{\ME(\{p, f\}_{m})}
\;\;.
\end{equation}
The matrix elements of $V_l^{\dagger,{\rm soft}}$ are specified by a function $v_l^{\rm soft}$,
\begin{equation}
\label{eq:Vtovsoft}
\bra{\{\hat s\}_{m+1}}
V_l^{\dagger,{\rm soft}}(\{\hat p, \hat f\}_{m+1})\ket{\{s\}_m}
= 
\left(\prod_{j\notin\{l,m+1\}} \delta_{\hat s_j,s_j}\right)
v_l^{\rm soft}(\{\hat p, \hat f\}_{m+1},\hat s_{m+1},\hat s_{l},s_l)
\;\;.
\end{equation}
If parton $m+1$ is a quark or antiquark, $v_l^{\rm soft} = 0$. When parton $m+1$ is a gluon,  
\begin{equation}
\label{eq:VsoftF}
v_l^{\rm soft}(\{\hat p, \hat f\}_{m+1},\hat s_{m+1},\hat s_{l},s_l)
= 
\sqrt{4\pi\as}\,\delta_{\hat s_l, s_l}\,
\frac{
\varepsilon(\hat p_{m+1}, \hat s_{m+1};\hat Q)^*
\!\cdot\! \hat p_l}
{\hat p_{m+1}\!\cdot\! \hat p_l}\, 
\;\;.
\end{equation}
The functions $v_l^{\rm soft}$ are not as powerful as the functions $v_l$ because they provide good approximations only in the soft gluon limit. Nevertheless, we will have occasion to make use of them.
 
\section{Description of color}
\label{sec:color}

We will need a description of the quantum color state that is adapted to a
description of shower evolution. If we use an index notation, to each
parton with label $l$ there is associated a color index $a_l$, which
takes values $1,\dots,3$ for a quark or antiquark and takes values
$1,\dots,8$ for a gluon. There is also a spin index $\lambda_l$, which takes values $\pm 1/2$ for quark and $\pm 1$ for a gluon. We can expand $\ME$ in terms of color and spin basis vectors in the form
\begin{equation}
\begin{split}
\ME(\{p,f\}_{m})
^{a_\La,a_\Lb, a_1,\dots,a_m}
_{\lambda_\La, \lambda_\Lb, \lambda_1,\dots,\lambda_m}
 ={}& 
\sum_{\{c\}_{m}} 
\Psi(\{c\}_{m})^{a_\La,a_\Lb, a_1,\dots,a_m}
\sum_{\{s\}_{m}} 
\Xi(\{s\}_{m})_{\lambda_\La, \lambda_\Lb, \lambda_1,\dots,\lambda_m}
\\
&\times \ME(\{p,f,s,c\}_{m})
\;\;,
\end{split}
\end{equation}
where the $\Psi(\{c\}_{m})$ form a basis for the space of color singlet amplitudes with color labels $\{c\}_{m}$ and the $\Xi(\{s\}_{m})$ form a basis for the spin space with spin labels $\{s\}_{m}$. The quantities $\ME(\{p,f,s,c\})$ are the expansion coefficients. In a vector notation, this is
\begin{equation}
\label{eq:colorexpansion}
\ket{\ME(\{p,f\}_{m})}_{\rm c,s} = 
\sum_{\{c\}_{m}}\ket{\{c\}_{m}}_{\rm c}\otimes
\sum_{\{s\}_{m}}\ket{\{s\}_{m}}_{\rm s}\, 
\ME(\{p,f,s,c\}_{m})
\;\;.
\end{equation}
Here $\ket{\ME(\{p,f\}_{m})}_{\rm c,s}$ lies in the combined color-spin space while $\ket{\{c\}_{m}}_{\rm c}$ is a vector in color space and $\ket{\{s\}_{m}}_{\rm s}$ is a vector in spin space.

As discussed in Sec.~\ref{sec:spinstates}, we use a conventional treatment of spin. We assume that the spin labels $\lambda$ already represent parton helicities using suitable conventions for choosing corresponding Dirac spinors and polarization vectors. Then the basis vector labels can be simply $\{s\}_{m} = \{s_\La,s_\Lb, s_1,\dots,s_m\}$ and the basis vectors can be
\begin{equation}
\Xi(\{s\}_{m})_{\lambda_\La, \lambda_\Lb, \lambda_1,\dots,\lambda_m}
= 
\delta^{\lambda_\La}_{s_\La}
\delta^{\lambda_\Lb}_{s_\Lb}
\delta^{\lambda_1}_{s_1}\cdots
\delta^{\lambda_m}_{s_m}
\;\;.
\end{equation}
Then we have an orthonormal basis: $\brax{\{s'\}_{m}}\ket{\{s\}_{m}}$ is 1 if the spin labels are all the same and zero otherwise.

We use a treatment of color that is conventional but more subtle than the treatment for spin. We turn to this subject in this section.

\subsection{Color basis} 
\label{sec:ColorBasis}

We first note that as far as color is concerned, an initial state quark is equivalent to a final state antiquark and an initial state antiquark is equivalent to a final state quark. Thus, in the prose description in this section we use ``quark'' and ``$q$'' to refer to a final state quark or an initial state antiquark and we use ``antiquark'' and ``$\bar q$'' to refer to a final state antiquark or an initial state quark.\footnote{Recall from Eq.~(\ref{eq:initialantiflavors}) that in the case of an initial state parton, $f_l$ and $\hat f_l$ denote the opposite of the incoming flavors of the parton.}

We next note that the amplitude $\ket{\ME(\{p,f\}_{m})}$ us always invariant under an overall rotation of all of the parton colors. Thus what we really need is a basis for the space of color singlet amplitudes in the color space. There is a widely used and intuitively appealing way to do this that, furthermore, matches with the idea of color strings forming between outgoing partons \cite{colorbasis}.

The color basis vectors $\ket{\{c\}_{m}}_{\rm c}$ are labeled by a {\it
color string configurations} $\{c\}_{m}$. A color string configuration  can be described as a set $\{S_1,\dots,S_n\}$ of one or more {\it strings} $S$. There are two types of strings, {\it open strings} and {\it closed strings}. An open string is an ordered set of parton indices that we denote by $S = [l_1,l_2,\dots,l_{n-1},l_n]$. Here $l_1$ is the label of a quark,  $l_n$ is the label of an antiquark, and $l_2,\dots,l_{n-1}$ are labels of gluons. A  closed string is an ordered set of at least two parton indices that we denote by  $S = (l_1,l_2,\dots,l_{n-1},l_n)$. Here all of the indices label gluons and we treat sets that differ by a cyclic permutation of the indices as being the same. Thus a complete color string configuration for a quark, an antiquark, and five gluons might be $\{[1,6,3,7],(4,2,5)\}$. This is a notation for a possible $\{c\}_{5}$.

Now we can define the basis states. We take $\Psi(\{c\}_{m})$ to be a product
\begin{equation}
\Psi(\{c\}_{m})^{a_1,\dots,a_m} = 
\Psi(S_1)^{\{a\}_{[1]}}\,
\Psi(S_2)^{\{a\}_{[2]}}
\dots 
\Psi(S_K)^{\{a\}_{[K]}}
\;.
\end{equation}
Here we have denoted the set of color indices represented in string $k$ by 
\begin{equation}
\{a\}_{[k]} = \{a_{l_1},\dots,a_{l_n}\}
\end{equation}
if string $k$ is $[l_1,\dots, l_n]$ or $(l_1,\dots, l_n)$.

\FIGURE{
\includegraphics[width = 10 cm]{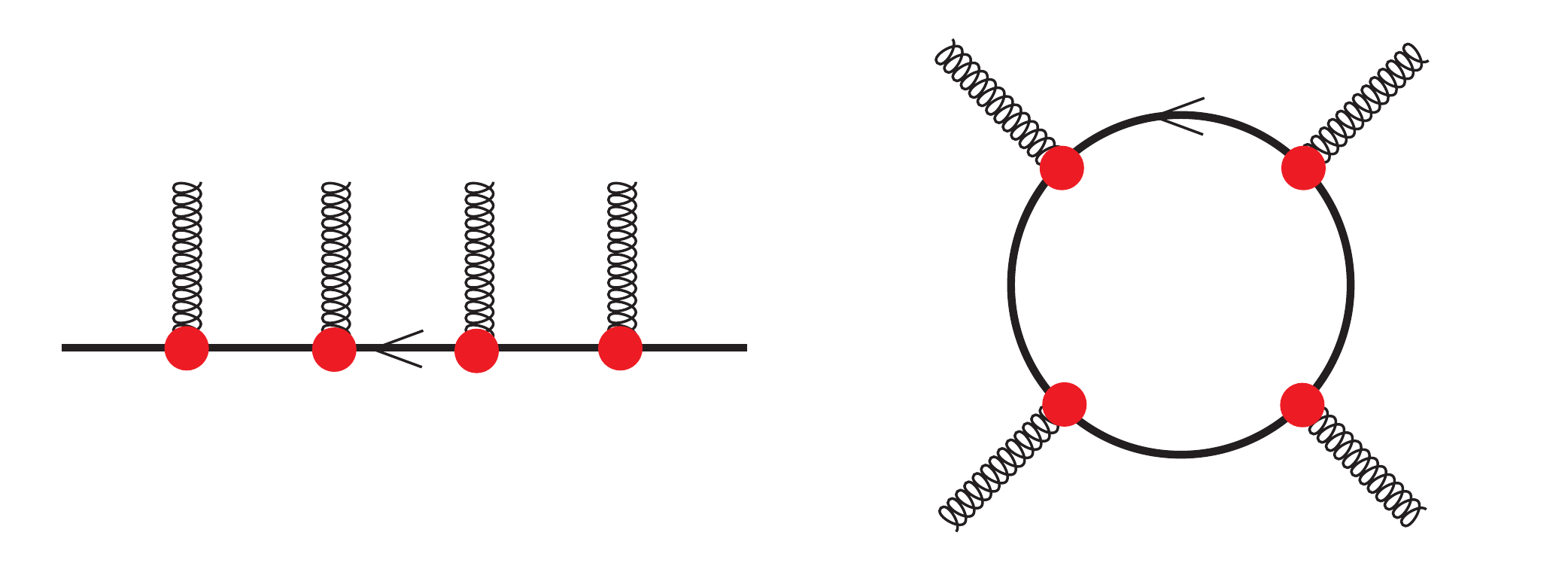}
\caption{Illustration an open string color basis state (left) and a closed sting color basis state (right).
}
\label{fig:strings}
}

We can now define the component factors $V(S)$. For notational convenience, we suppose that $l_i = i$ so that the partons along the string are numbered sequentially, $[1,2,\dots,n]$ or $(1,2,\dots,n)$.

We first consider an open string, as illustrated on the left in Fig.~\ref{fig:strings}. We define
\begin{equation}
\Psi(S)^{\{a\}}
= n(S)^{-1/2}
\left[
t^{a_2}t^{a_3}\cdots t^{a_{n-1}}
\right]_{a_1 a_n}
\;\;,
\end{equation}
where the $t^a$ are the SU(3) generator matrices for the fundamental
representation and we take the $a_1,a_n$ matrix element of the matrix product of the generator matrices (normalized to $T_{\rm R} = 1/2$). The normalization factor $n(S)$ is
\begin{equation}
\label{eq:nSopen}
n(S) = N_\Lc C_{\rm F}^{n-2}
\;\;.
\end{equation}
With this normalization,
\begin{equation}
\brax{S}\ket{S} \equiv 
\sum_{\{a\}}|\Psi(S)^{\{a\}}|^2
= 1
\;\;.
\label{eq:opennormalization}
\end{equation}

For a closed string with the same parton labels (now all gluons) we define, as illustrated on the right in Fig.~\ref{fig:strings},
\begin{equation}
\Psi(S)^{\{a\}}
= 
n(S)^{-1/2}\,
{\rm Tr}\left[
t^{a_1}t^{a_2}\cdots t^{a_{n}}
\right]
\;\;,
\end{equation}
where, again, the $t^a$ are the SU(3) generator matrices for the fundamental representation\footnote{One could use the adjoint representation here. However, an adjoint representation string is approximately equivalent to two fundamental representation strings and having two strings when one would do makes the description more complicated.} and where
\begin{equation}
n(S) = C_{\mathrm F}^n
\;\;.
\end{equation}
With this normalization,
\begin{equation}
\brax{S}\ket{S} \equiv 
\sum_{\{a\}}|\Psi(S)^{\{a\}}|^2
= 1 - \left(\frac{-1}{2N_\Lc C_{\rm F}}\right)^{n-1} 
\;\;.
\label{eq:closednormalization}
\end{equation}
This is approximately 1 in the limit of a large number of colors.

A general color basis state is a product of string states.\footnote{The building blocks used are the invariant matrices $\delta_{ij}$ to connect a quark and an antiquark and $t^a_{ij}$ to connect a quark, antiquark and gluon. One could also use the completely antisymmetric matrix $\epsilon_{ijk}$ to connect three quarks or three antiquarks. However, we don't need states made using $\epsilon_{ijk}$ because the amplitudes $M$ have net baryon number zero.} 
It includes a normalization factor $n(\{c\}_{m})^{-1/2}$,
\begin{equation}
n(\{c\}_{m}) = n(S_1)\,n(S_2)\cdots n(S_K)
\;\;.
\end{equation}
The normalization of the states is
\begin{equation}
\brax{ \{c\}_{m}}\ket{\{c\}_{m}} =
\prod_k \brax{S_k}\ket{S_k}
\;\;,
\end{equation}
where the factors are given in Eqs.~(\ref{eq:opennormalization}) and (\ref{eq:closednormalization}). Thus $\brax{ \{c\}_{m}}\ket{\{c\}_{m}} \approx 1$ in the large $N_\Lc$ limit.

The basis vectors $\ket{ \{c\}_{m}}$ are not exactly normalized and they are not orthogonal. However, the inner product between two different basis vectors is small in the limit of a large number of colors,
\begin{equation}
\label{eq:offdiagonalcolors}
\brax{\{c'\}_{m} }\ket{ \{c\}_{m} } = 
{\cal O}(1/N_\Lc^2)
\hskip 1 cm 
\{c'\}_{m} \ne \{c\}_{m}
\;\;.
\end{equation}
For instance, suppose $\{c\}_{m} =  [1,2,3,4]$ and suppose $\{c'\}_{m} = [1,3,2,4]$, with the positions of the two gluons reversed. Then $\brax{\{c'\}_{m}} \ket{\{c\}_{m}} = -1/(N_\Lc^2 - 1)$. The calculation is illustrated in Fig.~\ref{fig:stringnorms}.

\FIGURE{
\hskip 2 cm \includegraphics[width = 8 cm]{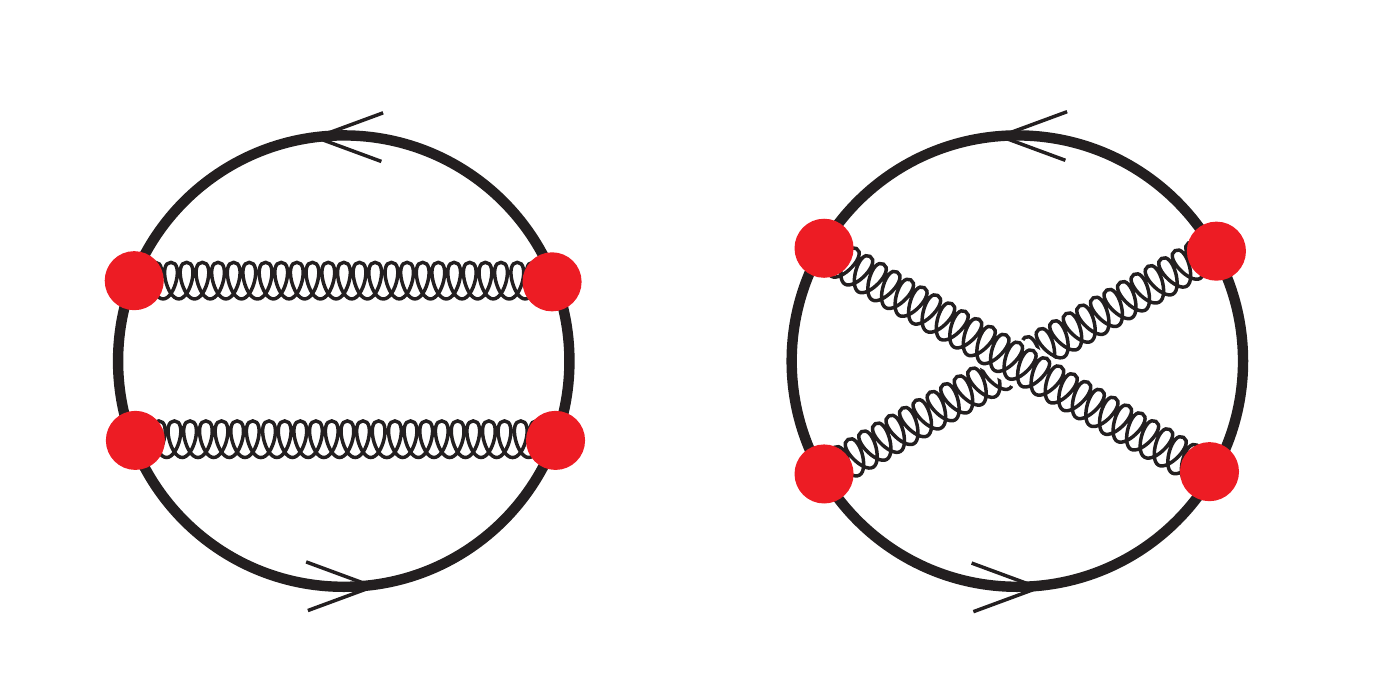}\hskip 2 cm
\caption{Inner products for color basis states. The left hand picture illustrates the inner product of the $[1,2,3,4]$ with itself. The color diagram shown gives $C_\LF^2 N_\Lc$, which is just canceled by the normalization factor $1/n(S)$ from Eq.~(\ref{eq:nSopen}). The right hand picture illustrates the inner product of the $[1,2,3,4]$ with $[1,3,2,4]$. The color diagram gives $-C_\LF/2$. Multiplying by $1/n(S)$ gives $-1/(N_\Lc^2 - 1)$.}
\label{fig:stringnorms}
}

\subsection{Parton insertion operators}
\label{sec:partoninsertion}

It will prove useful to define certain operators that act on an $m$ parton color state with partons with labels $\{\La,\Lb,1,\dots, m\}$ and add one parton with label $m+1$. We let $a^\dagger_+(l)$ insert a gluon just to the right of parton $l$ on whatever string contains parton $l$,
\begin{equation}
\begin{split}
\label{eq:aplusdef}
a^\dagger_+(l)\,\ket{\{\dots,[a_1,\dots, l,\dots, a_n],\dots\}}
={}& \ket{\{\dots,[a_1,\dots, l, m+1, \dots, a_n],\dots\}}
\;\;,
\\ 
a^\dagger_+(l)\,\ket{\{\dots,(a_1,\dots, l,\dots, a_n) ,\dots\}}
={}& \ket{\{\dots,(a_1,\dots,l,m+1,\dots a_n),\dots\}}
\;\;,
\\
a^\dagger_+(l)\,\ket{\{\dots,[l,\dots,a_n],\dots\}}
={}& \ket{\{\cdots [l, m+1,\dots, a_n],\dots\}}
\;\;,
\\
a^\dagger_+(l)\,\ket{\{\dots,[a_1,\dots,l],\dots\}}
={}& 0
\;\;.
\end{split}
\end{equation}
The first equation applies to the case that $l$ labels a gluon on an open string while the second equation applies to the case that $l$ labels a gluon on an closed string. The third equation applies when $l$ labels a quark at the end of an open string. The fourth equation applies when $l$ labels an antiquark at the end of an open string. In this case, there is no place to put the gluon, so the result is defined to be zero. Similarly, we define an operator $a^\dagger_-(l)$ that inserts a gluon just to the left of parton $l$ on whatever string contains parton $l$,
\begin{equation}
\begin{split}
\label{eq:aminusdef}
a^\dagger_-(l)\,\ket{\{\dots,[a_1,\dots, l,\dots, a_n],\dots\}}
={}& \ket{\{\dots,[a_1,\dots,m+1, l,\dots a_n],\dots\}}
\;\;,
\\
a^\dagger_-(l)\,\ket{\{\dots,(a_1,\dots, l,\dots, a_n),\dots\}}
={}& \ket{\{\dots,(a_1,\dots,m+1, l,\dots a_n),\dots\}}
\;\;,
\\
a^\dagger_-(l)\,\ket{\{\dots,[l,\dots,a_n],\dots\}}
={}& 0
\;\;,
\\
a^\dagger_-(l)\,\ket{\{\dots,[a_1,\dots,l],\dots\}}
={}& \ket{\{\dots,[a_1,\dots, m+1, l],\dots\}}
\;\;.
\end{split}
\end{equation}
We define an operator $a^\dagger_q(l)$ that breaks a string at the position of a gluon with label $l$, creating a quark that we take to inherit the label $l$ and an antiquark with the new label $m+1$. If $l$ is a gluon on an open string, this creates two open strings. If $l$ is a gluon on a closed string, this turns the closed string into an open string. If $l$ labels a quark or antiquark, we define the result to be zero. Thus
\begin{equation}
\begin{split}
\label{eq:aqdef}
a^\dagger_q(l)\,\ket{\{\dots,[a_1,\dots, l,\dots, a_n],\dots\}}
={}& \ket{\{\dots,[a_1,\dots, m+1],[l, \dots a_n],\dots\}}
\;\;,
\\ 
a^\dagger_q(l)\,\ket{\{\dots,(a_1,\dots, l,\dots, a_n),\dots\}}
={}& \ket{\{\dots,[l,\dots a_n,a_1,\dots,m+1],\dots\}}
\;\;,
\\
a^\dagger_q(l)\,\ket{\{\dots,[l,\dots,a_n],\dots\}}
={}& 0
\;\;,
\\
a^\dagger_q(l)\,\ket{\{\dots,[a_1,\dots,l],\dots\}}
={}& 0
\;\;.
\end{split}
\end{equation}
Finally, we define an operator $a^\dagger_s(l)$ that removes a gluon with label $l$ from its string, and creates a new open string consisting of just a quark that we take to inherit the label $l$ and an antiquark with the new label $m+1$. If $l$ labels a quark or antiquark, we define the result to be zero. Thus
\begin{equation}
\begin{split}
\label{eq:asdef}
a^\dagger_s(l)\,\ket{\{\dots,[a_1,\dots, l,\dots, a_n],\dots\}}
={}& \ket{\{\dots,[a_1, \dots, a_n],[l,m+1]\cdots}
\;\;,
\\ 
a^\dagger_s(l)\,\ket{\{\dots,(a_1,\dots, l,\dots, a_n),\dots\}}
={}& \ket{\{\dots,(a_1,\dots, a_n),[l,m+1],\dots\}}
\;\;,
\\
a^\dagger_s(l)\,\ket{\{\dots,[l,\dots,a_n],\dots\}}
={}& 0
\;\;,
\\
a^\dagger_s(l)\,\ket{\{\dots,[a_1,\dots,l],\dots\}}
={}& 0
\;\;.
\end{split}
\end{equation}
We will see the utility of these operators presently, when we study the color flow in parton splitting.

\subsection{Color evolution for the quantum states}
\label{sec:QuantumColor}

In this subsection, examine how the color vector changes with successive parton splittings. Consider starting with a color state $\ket{\psi}$ for $m$ partons and letting a final state parton with label $l$ and flavor $f_l$ emit a gluon with label $m+1$.\footnote{In \cite{CataniSeymour}, the final state labels are called $i$ and $j$. We identify $i$ with $l$, and $j$ with $m+1$. The labels for the initial $m$ partons are $\{\La,\Lb,1,\dots,m\}$ and for the subsequent state after splitting are $\{\La,\Lb,1,\dots,m,m+1\}$. The partons not involved in the splitting keep their labels.} After the gluon emission, parton $l$ retains its label $l$ and its flavor. This produces a new color state
\begin{equation}
\ket{\hat\psi} = t^\dagger_l(f_l \to  f_l + {\rm g}) \ket{\psi}
\;\;.
\end{equation}
We can define the color splitting operator $t^\dagger_l(f_l\to f_l + {\rm g})$ precisely by writing this out in component notation
\begin{equation}
\label{eq:splittingmatrix}
\hat\psi^{a_\La,a_\Lb, a_1,\dots, a_l,\dots,a_m, a_{m+1}} =
\sum_{\tilde a_l}
T(f_l)^{a_{m+1}}_{a_l,\tilde a_l}
\psi^{a_\La,a_\Lb, a_1,\dots,\tilde a_l,\dots,a_m}
\;\;.
\end{equation}
The matrix $T(f_l)^{a_{m+1}}$ here is the generator matrix in the {\bf 8} representation if $f_l = {\rm g}$, in the {\bf 3} representation if $f_l \in \{{\rm u},{\rm d},\dots\}$, and in the $\bar{\bf 3}$ representation if $f_l \in \{\bar{\rm u},\bar{\rm d},\dots\}$.\footnote{Specifically, $T(q)^a_{ij} = t^a_{ij}$, $T(\bar q)^a_{ij} = - t^a_{ji}$ and $T({\rm g})^a_{bc} = i f_{bac}$, where $q$ here represents any quark flavor and $\bar q$ represents any antiquark flavor.}

This notation applies for an initial state splitting too since, from Eq.~(\ref{eq:initialantiflavors}), in the case of an initial state parton, $f_l$ and $\hat f_l$ denote the opposite of the incoming flavors of the parton. In this case, the arrow in $f_l \to  f_l + {\rm g}$ and more generally in $f_l \to  \hat f_l + \hat f_{m+1}$ refers to backward evolution.

Now consider what happens when the splitting operator is applied to one of the color basis vectors, $\ket{\{c\}_m}$. For the case that $f_l = {\rm g}$ we have
\begin{equation}
\sum_{a_l} T({\rm g})^{\tilde a_{m+1}}_{\tilde a_l,a_l}\ \ t^{a_l}
= \sum_{a_l} i f_{\tilde a_l,\tilde a_{m+1},a_l}\ t^{a_l} 
= t^{\tilde a_l}t^{\tilde a_{m+1}} - t^{\tilde a_{m+1}}t^{\tilde a_l}
\;\;.
\end{equation}
Thus $t^\dagger_l({\rm g} \to {\rm g} + {\rm g})\ket{\{c\}_m}$ is a normalization factor times a difference of two basis vectors
\begin{equation}
\begin{split}
\left[\frac{n(\{\hat c_+\}_{m+1})}{n(\{c\}_{m})}\right]^{1/2}
\ket{\{\hat c_+\}_{m+1}} 
- \left[\frac{n(\{\hat c_-\}_{m+1})}{n(\{c\}_{m})}\right]^{1/2}
\ket{\{\hat c_-\}_{m+1}}
\;\;.
\end{split}
\end{equation}
In $\{\hat c_\pm\}_{m+1}$, gluon $l$ is replaced by gluons $l$ and $m+1$. Specifically, if in $\{c\}_{m}$ gluon $l$ appears in an open string as $[i_1,\dots,l,\dots,i_n]$ then in $\{\hat c_+\}_{m+1}$ this string is replaced by $[i_1,\dots,l,m+1,\dots,i_n]$ and in $\{\hat c_-\}_{m+1}$ this string is replaced by $[i_1,\dots,m+1,l,\dots,i_n]$. The normalization factor is
\begin{equation}
\frac{n(\{\hat c_\pm\}_{m+1})}{n(\{c\}_{m})} = {C_{\rm F}} \;\;.
\end{equation}
Analogous remarks apply for a closed string. We can state this result using the operators defined in Eqs.~(\ref{eq:aplusdef}) and (\ref{eq:aminusdef}),
\begin{equation}
\label{eq:coloropggg}
t^\dagger_l({\rm g} \to {\rm g} + {\rm g}) = 
\sqrt{C_{\rm F}}\, a^\dagger_+(l)
-\sqrt{C_{\rm F}}\, a^\dagger_-(l)
\;\;.
\end{equation}
This is illustrated in Fig.~\ref{fig:gggcolor}.

\FIGURE{
\includegraphics[width = 15 cm]{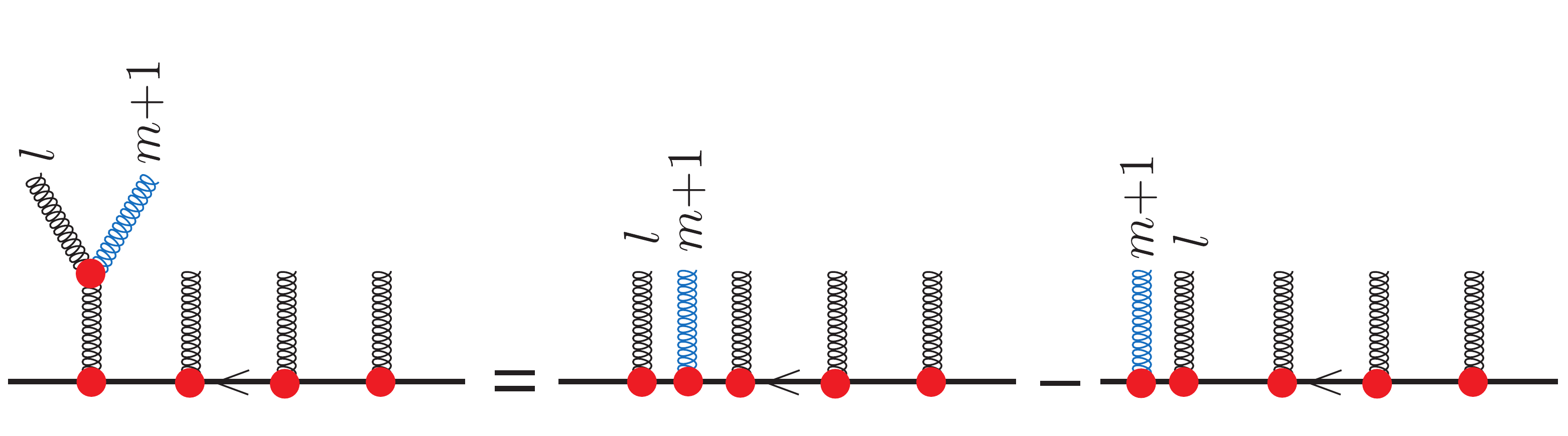}
\caption{Illustration of Eq.~(\ref{eq:coloropggg}). A gluon $l$ spits into gluon $l$ and gluon $m+1$. The result is to place the new gluon $m+1$ either to the right of gluon $l$, with insertion operator $a^\dagger_+(l)$ or to the left, with insertion operator $a^\dagger_-(l)$. The normalization factors $\sqrt C_{\rm F}$ are not indicated.}
\label{fig:gggcolor}
}

For a quark splitting to a quark plus a gluon we have a similar operator, one more generator matrix is inserted at the quark end of the string, so that
\begin{equation}
\label{eq:coloropqqg}
t^\dagger_l(q \to q + {\rm g}) = 
\sqrt{C_{\rm F}}\, a^\dagger_+(l)
\;\;.
\end{equation}
For an antiquark splitting to an antiquark plus a gluon, one more generator matrix is inserted at the antiquark end of the string, so that
\begin{equation}
\label{eq:coloropqbarqbarg}
t^\dagger_l(\bar q \to \bar q + {\rm g}) = 
-\sqrt{C_{\rm F}}\, a^\dagger_-(l)
\;\;.
\end{equation}
In these equations, $q$ represents any quark flavor and $\bar q$ represents any antiquark flavor. 

The operators for gluon emission obey an important identity. The matrix element $M$ and our approximations to it are always color singlets,
\begin{equation}
\label{eq:colorsinglet}
\sum_l\sum_{\tilde a_l}
T(f_l)^{a_{m+1}}_{a_l,\tilde a_l}
\psi^{a_\La,a_\Lb, a_1,\dots,\tilde a_l,\dots,a_m}
= 0
\;\;.
\end{equation}
Thus
\begin{equation}
\label{eq:tlidentity}
\sum_l t^\dagger_l(f_l\to f_l + {\rm g}) \ket{\psi} = 0
\;\;.
\end{equation}
We will use this identity in the following sections to rearrange the splitting formulas.

The last possibility for parton splitting is that of a gluon splitting into a quark-antiquark pair with flavors $q$ and $\bar q$. The color splitting operator is defined by
\begin{equation}
\ket{\hat\psi} = t^\dagger_l({\rm g}  \to q  + \bar q) \ket{\psi}
\;\;,
\end{equation}
where
\begin{equation}
\hat\psi^{a_\La,a_\Lb, a_1,\dots,a_m,a_{m+1}} =
\sum_{\tilde a_l}
t^{\tilde a_l}_{a_l,a_{m+1}}
\psi^{a_\La,a_\Lb, a_1,\dots,\tilde a_l,\dots,a_m}
\;\;.
\end{equation}
Our convention here is that the quark inherits the gluon label, $l$, and the antiquark gets the new label, $m+1$. To analyze this, we use
\begin{equation}
t^{\tilde a_l}_{a_l,a_{m+1}}\ t^{\tilde a_l}_{c,d}
= \frac{1}{2}\,\delta_{c,a_{m+1}}\delta_{a_l,d}
-\frac{1}{2N_\Lc}\,\delta_{a_l,a_{m+1}}\delta_{c,d}
\;\;.
\end{equation}
Thus $t_l(g \to q + \bar q) \ket{\{c\}_{m}}$ is a combination of two basis vectors,
\begin{equation}
\begin{split}
\frac{1}{2}
\left[\frac{n(\{\hat c_1\}_{m+1})}{n(\{c\}_{m})}\right]^{1/2}
\ket{\{\hat c_1\}_{m+1}}
- \frac{1}{2N_\Lc}
\left[\frac{n(\{\hat c_2\}_{m+1})}{n(\{c\}_{m})}\right]^{1/2}
\ket{\{\hat c_2\}_{m+1}}
\;\;.
\end{split}
\label{eq:gqqbarcolor}
\end{equation}
In $\{\hat c_1\}_{m+1}$, the string in which gluon $l$ resides is split.  Specifically, if in $\{c\}_{m}$, gluon $l$ appears in an open string as $[i_1,\dots,i_A,l,i_B\dots i_{n}]$, then in $\{\hat c_1\}_{m+1}$ this string is replaced by two strings, $[i_1,\dots,i_A,m+1][l,i_B,\cdots,i_n]$. On the other hand, if in $\{c\}_{m}$, gluon $l$ appears in a closed string as $(i_1,\dots,i_A,l,i_B,\dots i_{n})$, then in $\{\hat c_1\}_{m+1}$ this string is replaced by an open string, $[l,i_B,\dots,i_{n},i_1\dots,i_A,m+1]$. In either case,
\begin{equation}
\frac{n(\{\hat c_1\}_{m+1})}{n(\{c\}_{m})}
= \frac{N_\Lc}{C_{\rm F}}
\;\;.
\end{equation}

In the second color state in Eq.~(\ref{eq:gqqbarcolor}), $\ket{\{\hat c_2\}_{m+1}}$, gluon $l$ is simply removed from its string and a new trivial $q\bar q$ string is created. Specifically, if in $\{c\}_{m}$ gluon $l$ appears in an open string as $[i_1,\dots,i_A,l,i_B\dots i_{n}]$, then in $\{\hat c_2\}_{m+1}$ this string is replaced by two strings, $[i_1,\dots,i_A,i_B,\cdots,i_n]$ and $[l,m+1]$. On the other hand, if in $\{c\}_{m}$ gluon $l$ appears in an closed string as $(i_1,\dots,i_A,l,i_B,\dots i_{n})$, then in $\{\hat c_2\}_{m+1}$ this string is replaced by a closed string $(i_1,\dots,i_A,i_B,\dots i_{n})$ and the open string $[l,m+1]$. In either case,
\begin{equation}
\frac{n(\{\hat c_2\}_{m+1})}{n(\{c\}_{m})}
= \frac{N_\Lc}{C_{\rm F}}
\;\;.
\end{equation}

We can state this result using the operators defined in Eqs.~(\ref{eq:aqdef}) and (\ref{eq:asdef}),
\begin{equation}
\label{eq:coloropgqqbar}
t^\dagger_l({\rm g} \to q + \bar q) = 
\sqrt{\frac{N_\Lc}{4C_{\rm F}}}\  a^\dagger_q(l)
-\sqrt{\frac{1}{4N_\Lc C_{\rm F}}}\  a^\dagger_s(l)
\;\;.
\end{equation}

We can now see the advantage of this choice of color basis, beyond the fact that it is widely used for NLO calculations. In this basis, the description of parton splitting is very simple. For splittings $q \to q {\rm g}$ and $\bar q \to \bar q {\rm g}$, we simply add a gluon to a string. For splittings ${\rm g} \to {\rm g}{\rm g}$, we replace one gluon on a string by two. There are two terms corresponding to the two possible orders for the two gluons on the string. For a splitting $g \to \bar q  q$, we split a string or we remove the gluon from its string and create a new $q  \bar q$ string. The normalization factors may be considered to be just a matter of bookkeeping. However, one should note that the splittings $q \to q {\rm g}$, $\bar q \to \bar q {\rm g}$. and ${\rm g} \to {\rm g}{\rm g}$ come with numerical factors $\sqrt{C_{\rm F}}$ in the amplitude. These factors are large for large $N_\Lc$. The splitting ${\rm g} \to q  \bar q$ comes with a numerical factor for the first term, $[{N_\Lc}/(4 C_{\rm F})]^{1/2}$ that is {\it not} large in the large $N_\Lc$ limit. (The numerical factor for the second term is small in the large $N_\Lc$ limit.) Thus the color factor for the splitting ${\rm g} \to q  \bar q$ is smaller than the color factors for the other splittings. This makes ${\rm g} \to q  \bar q$ splitting somewhat disfavored even though there are several $q  \bar q$ flavor choices available.

\section{Evolution for the statistical states}
\label{sec:ClassicalEvolution}

We have seen how parton splitting works at the quantum amplitude level. We now need to use these results to formulate the effect of splitting on the density operator $\rho$ and thus the statistical state $\sket\rho$. Suppose that we have the function
\begin{equation}
\sbrax{\{p,f,s',c',s,c\}_m}
\sket{\rho} 
\end{equation}
describing $m$ final state partons plus the two initial state partons. After one splitting, we will have $m+1$ final state partons, described by 
\begin{equation}
\sbrax{\{\hat p,\hat f,\hat s',\hat c',\hat s,\hat c\}_{m+1}}\sket{\rho}
\;\;.
\end{equation}
We need this function within the soft and collinear splitting approximations that we have used for the amplitudes.

Any of the partons can split. In particular, parton $i$ in $\ket{\ME(\{p, f\}_{m})}$ and parton $j$ in $\bra{\ME(\{p, f\}_{m})}$ can split. In the simplest case (and the only case incorporated into typical parton shower Monte Carlo event generators), $i = j$. However, one can have $i \ne j$ and still get the same momenta and flavors $\{\hat p,\hat f\}_{m+1}$ from both splittings and still get a logarithmic divergence if one were to integrate over $\{\hat p\}_{m+1}$. This happens when parton $i$ emits a gluon $m+1$ in $\ket{\ME(\{p, f\}_{m})}$ and parton $j$ emits gluon $m+1$ in $\bra{\ME(\{p, f\}_{m})}$. An interference graph of this sort has no collinear divergence from $\hat p_{m+1}$ being collinear to $\hat p_i$ or $\hat p_j$. However it does have a soft divergence from $\hat p_{m+1} \to 0$. For this reason, we need to include the case $i \ne j$.

For each splitting, we relate $\{\hat p,\hat f\}_{m+1}$ to the starting momenta and flavors $\{p,f\}_m$ and splitting variables $\{\zeta_{\rm p},\zeta_{\rm f}\}$ using one of the mappings
\begin{equation}
\label{eq:Rldefagain}
\{\hat p, \hat f\}_{m+1} = R_l(\{p,f\}_m, \{\zeta_{\rm p},\zeta_{\rm f}\})
\;\;,
\end{equation}
as specified in Sec.~\ref{sec:mapping}. In the case $i \ne j$, we can use either the mapping with $l = i$ or the mapping with $l = j$.\footnote{Either mapping suffices because in either case the map becomes the identity map in the soft limit $\hat p_{m+1} \to 0$.} We find it most useful to average over these two possibilities. We use the mapping with $l = i$ with a weight $A_{ij}$ and the mapping with $l = j$ with weight $A_{ji}$. In general, the weights can depend on the momenta $\{\hat p\}_{m+1}$. In that case, the momentum mapping $R_l$ with $l=i$ is used for $A_{ij}$ and the momentum mapping $R_l$ with $l=j$ is used for $A_{ji}$ . In this paper, the default value is $A_{ij} = 1/2$, but our notation allows for other choices. In the case $i = j$, we use the mapping with $l = i = j$.

We are thus led to write the density operator $\hat \rho$ after splitting as a sum of contributions $\hat \rho^{(l)}_{ij}$, where the superscript indicates the treatment of the kinematics and the subscripts indicate which partons split,
\begin{equation}
\begin{split}
\label{eq:hatrho}
\hat \rho(\{\hat p, \hat f\}_{m+1}) ={}& \sum_{l} 
\hat \rho^{(l)}_{ll}(\{\hat p, \hat f\}_{m+1})
\\&
+\sum_{\substack{i,j \\ i\ne j}}\bigl\{
A_{ij}(\{\hat p\}_{m+1})\,
\hat \rho^{(i)}_{ij}(\{\hat p, \hat f\}_{m+1})
+
A_{ji}(\{\hat p\}_{m+1})\,
\hat \rho^{(j)}_{ij}(\{\hat p, \hat f\}_{m+1})
\bigr\}
\\
={}& \sum_{l} 
\Bigl\{
\hat \rho^{(l)}_{ll}(\{\hat p, \hat f\}_{m+1})
\\&
+\sum_{k\ne l}
A_{lk}(\{\hat p\}_{m+1})
\left[
\hat \rho^{(l)}_{lk}(\{\hat p, \hat f\}_{m+1})
+
\hat \rho^{(l)}_{kl}(\{\hat p, \hat f\}_{m+1})
\right]
\Bigr\}
\;\;.
\end{split}
\end{equation}
Here the sums run over the set $\{\La,\Lb,1,\dots,m\}$. We use Eq.~(\ref{eq:splittingamplitudestructure}) for the contributions $\hat \rho^{(l)}_{ij}$ in the case $i = j$,
\begin{equation}
\begin{split}
\label{eq:rholdef}
\rho^{(l)}_{ll}(\{\hat p, \hat f\}_{m+1}) ={}& 
t^\dagger_l(f_l \to \hat f_l + \hat f_{m+1})\,
V^\dagger_l(\{\hat p, \hat f\}_{m+1})\
\rho(\{p, f\}_{m})
\\ & \times
V_l(\{\hat p, \hat f\}_{m+1})\,
t_l(f_l \to \hat f_l + \hat f_{m+1})
\\ & \times
S_l(\{\hat f\}_{m+1})
\;\;.
\end{split}
\end{equation}
Here $S_l(\{\hat f\}_{m+1})$ is a counting factor that is determined by the conventions we have used to label the final state partons,
\begin{equation}
\label{eq:Sldef}
S_l(\{\hat f\}_{m+1}) = \left\{
\begin{array}{cl}
1/2\;, &  l \in \{1,\dots,m\},\ \hat f_l = \hat f_{m+1} = {\rm g}  \\
  1\;, &  l \in \{1,\dots,m\},\ \hat f_l \ne {\rm g}, \hat f_{m+1} = {\rm g}  \\
  0\;, &  l \in \{1,\dots,m\},\ \hat f_l = {\rm g}, \hat f_{m+1} \ne {\rm g}  \\
  1\;, &  l \in \{1,\dots,m\},\ \hat f_l = q, \hat f_{m+1} = \bar q  \\
  0\;, &  l \in \{1,\dots,m\},\ \hat f_l = \bar q, \hat f_{m+1} = q  \\
  1\;, &  l \in \{\La,\Lb\}
\end{array}
\right.
\;\;.
\end{equation}
This factor is zero for $\hat f_l = {\rm g}$ and $\hat f_{m+1} = q$ or $\hat f_{m+1} = \bar q$, even though flavor conservation allows this combination in a $q \to q + {\rm g}$ or $\bar q \to \bar q + {\rm g}$ splitting, because we have chosen to label the final daughter gluon in these cases as parton $m+1$. Similarly, this factor is zero for $\hat f_l = \bar q$ and $\hat f_{m+1} = q$, even though flavor conservation allows this combination in a ${\rm g} \to q + \bar q$ splitting, because we have chosen to label the final daughter antiquark in this case as parton $m+1$. In the case of a final state ${\rm g} \to {\rm g} + {\rm g}$ splitting on line $l$, the splitting probability is symmetric under interchange of the labels $l$ and $m+1$, so that integrating over $\hat p_l$ and $\hat p_{m+1}$ would count the same physical configuration twice. The factor 1/2 corrects for this symmetry. This issue is discussed in some detail in Appendix \ref{sec:symmetryfactors}.

For $i \ne j$, we can use the simpler splitting operator $V^{{\rm soft}}$ as in Eq.~(\ref{eq:partonissoftbis}),
\begin{equation}
\begin{split}
\label{eq:rhoijdef}
\hat\rho^{(l)}_{ij}(\{\hat p, \hat f\}_{m+1}) ={}& 
t^\dagger_i(f_i \to \hat f_i + \hat f_{m+1})\,
V^{\dagger,{\rm soft}}_i(\{\hat p, \hat f\}_{m+1})\
\rho(\{p, f\}_{m})
\\ & \times
V^{\rm soft}_j(\{\hat p, \hat f\}_{m+1})\,
t_j(f_j \to \hat f_j + \hat f_{m+1})
\;\;.
\end{split}
\end{equation}
Note that $V^{\dagger,{\rm soft}}_i$ and $V^{\rm soft}_j$ both vanish if parton $m+1$ is not a gluon. Thus $\hat\rho^{(l)}_{ij} = 0$ for $i \ne j$ unless parton $m+1$ is a gluon.

This definition for $\hat\rho^{(l)}_{ij}$ defines a mapping of the statistical states in which $\sket{\rho}$ becomes 
\begin{equation}
\sket{\hat\rho^{(l)}_{ij}} = {\cal S}^{(l)}_{ij}\sket{\rho}
\;\;,
\end{equation}
after splitting. Then
\begin{equation}
\begin{split}
\sbra{\{\hat p,\hat f,\hat s',\hat c',\hat s,\hat c\}_{m+1}}
{\cal S}^{(l)}_{ij}\sket{\rho} ={}& 
\frac{1}{m!}
\int\big[d\{p,f,s',c',s,c\}_m\big]
\\&\times
\sbra{\{\hat p,\hat f,\hat s',\hat c',\hat s,\hat c\}_{m+1}}
{\cal S}^{(l)}_{ij}
\sket{\{p,f,s',c',s,c\}_m}
\\&\times
 \sbrax{\{p,f,s',c',s,c\}_m}
\sket{\rho}
\;\;.
\end{split}
\end{equation}
The splitting operator ${\cal S}^{(l)}_{ij}$ is defined by giving its matrix elements,
\begin{equation}
\begin{split}
\label{eq:Sijdef}
\sbra{\{\hat p,\hat f,\hat s',\hat c',\hat s,\hat c\}_{m+1}}&
{\cal S}^{(l)}_{ij}\sket{\{p,f,s',c',s,c\}_m} =
\\ &
\sbra{\{\hat c',\hat c\}_{m+1}}
{\cal G}(i,j;\{\hat f\}_{m+1})
\sket{\{c',c\}_m}
\\&\times
\sbra{\{\hat s',\hat s\}_{m+1}}
{\cal W}(i,j;\{\hat f,\hat p\}_{m+1})
\sket{\{s',s\}_m}
\\&\times
\sbra{\{\hat p,\hat f\}_{m+1}}{\cal P}_{l}\sket{\{p,f\}_m}
\\&\times
(m+1)\,
\frac
{n_\Lc(a) n_\Lc(b)\,\eta_{\La}\eta_{\Lb}}
{n_\Lc(\hat a) n_\Lc(\hat b)\,
 \hat \eta_{\La}\hat \eta_{\Lb}}\,
\frac{
f_{\hat a/A}(\hat \eta_{\La},\mu^{2}_{F})
f_{\hat b/B}(\hat \eta_{\Lb},\mu^{2}_{F})}
{f_{a/A}(\eta_{\La},\mu^{2}_{F})
f_{b/B}(\eta_{\Lb},\mu^{2}_{F})}
\;\;.
\end{split}
\end{equation}
We discuss each factor in turn. 

The first factor, ${\cal G}(i,j;\{\hat f\}_{m+1})$, describes how the splittings change the parton colors, 
\begin{equation}
\begin{split}
\label{eq:Gijdef}
\sbra{\{\hat c',\hat c\}_{m+1}}
{\cal G}(i,j;\{\hat f\}_{m+1})
\sket{\{c',c\}_m}
={}&
\dualL\bra{\{\hat c\}_{m+1}}
t^\dagger_i(f_i \to \hat f_i + \hat f_{m+1})
\ket{\{c\}_{m}}
\\ &\times
\bra{\{c'\}_{m}}
t_j(f_j \to \hat f_j + \hat f_{m+1})
\ket{\{\hat c'\}_{m+1}}\dualR
\;\;.
\end{split}
\end{equation}

In the next factor, ${\cal W}(i,j;\{\hat f,\hat p\}_{m+1})$, contains the spin-dependent splitting functions. The simplest case occurs for $i = j$ with anything other than a final state ${\rm g} \to {\rm g} + {\rm g}$ splitting. For that case
\begin{equation*}
\begin{split}
\sbra{\{\hat s',\hat s\}_{m+1}}
{\cal W}{}&(l,l;\{\hat f,\hat p\}_{m+1})
\sket{\{s',s\}_m}
\\={}&
S_l(\{\hat f\}_{m+1})\,
\bra{\{\hat s\}_{m+1}}
V^{\dagger}_l(\{\hat p, \hat f\}_{m+1})
\ket{\{s\}_{m}}
\bra{\{s'\}_{m}}V_l(\{\hat p, \hat f\}_{m+1})\ket{\{\hat s'\}_{m+1}}
\;\;.
\end{split}
\end{equation*}
For $i = j$ in general we define 
\begin{equation}
\begin{split}
\label{eq:Wlldef}
\sbra{\{\hat s',\hat s\}_{m+1}}&
{\cal W}{}(l,l;\{\hat f,\hat p\}_{m+1})
\sket{\{s',s\}_m}
\\={}&
S_l(\{\hat f\}_{m+1})\bigg\{
\bra{\{\hat s\}_{m+1}}
V^{\dagger}_l(\{\hat p, \hat f\}_{m+1})
\ket{\{s\}_{m}}
\bra{\{s'\}_{m}}V_l(\{\hat p, \hat f\}_{m+1})\ket{\{\hat s'\}_{m+1}}
\\ &+
\theta(l \in \{1,\dots,m\},\, \hat f_l = \hat f_{m+1} = {\rm g})\,
\sbra{\{\hat s',\hat s\}_{m+1}}\widetilde{\cal W}(l,l;\{\hat p\}_{m+1})
\sket{\{s',s\}_m}
\bigg\}
\;\;.
\end{split}
\end{equation}
The second term is included in the special case of a final state ${\rm g} \to {\rm g} + {\rm g}$ splitting. In this case, the first term is symmetric under an interchange $l\leftrightarrow m+1$ between the two final state gluons. It gives a leading singularity when gluon $m+1$ is soft but also gives a leading singularity when gluon $l$ is soft. We seek to use the freedom to assign labels to ensure that gluon $m+1$ can be soft but not gluon $l$. The operator $\widetilde{\cal W}$ in the added term is antisymmetric under $l\leftrightarrow m+1$, so that it gives zero contribution after integration over the final state momenta. The simplest choice for $\widetilde{\cal W}$ would be the first term times $[2\,\theta(\hat p_{m+1}\cdot Q < \hat p_{l}\cdot Q) - 1]$ which is antisymmetric under $l\leftrightarrow m+1$. Then the total ${\cal W}$ would be just the first term times $2\,\theta(\hat p_{m+1}\cdot Q < \hat p_{l}\cdot Q)$. Clearly this eliminates the singularity when gluon $l$ is soft. However, we adopt a slightly more subtle procedure. Write the tensor that defines the three gluon vertex as $v = v_1 + v_2 + v_3$,
\begin{equation}
\begin{split}
\label{eq:vgg123}
v_1^{\alpha \beta \gamma}(p_a, p_b, p_c)
={}& g^{\alpha\beta} (p_a - p_b)^\gamma
\;\;,
\\
v_2^{\alpha \beta \gamma}(p_a, p_b, p_c)
={}& g^{\beta\gamma} (p_b - p_c)^\alpha
\;\;,
\\
v_3^{\alpha \beta \gamma}(p_a, p_b, p_c)
={}& g^{\gamma\alpha} (p_c - p_a)^\beta
\;\;.
\end{split}
\end{equation}
Then we can define partial vertex functions analogous to those in Eq.~(\ref{eq:VggF}) by
\begin{equation}
\begin{split}
\label{eq:Vgg123}
v_{J,l}(\{\hat p, \hat f\}_{m+1},&\hat s_{m+1},\hat s_{l},s_l)
\\ & =
\frac{\sqrt{4\pi\as}}{2 \hat p_{m+1}\!\cdot\! \hat p_l}\, 
\varepsilon_{\alpha}(\hat p_{m+1}, \hat s_{m+1};\hat Q)^*
\varepsilon_{\beta}(\hat p_{l}, \hat s_l;\hat Q)^*
\varepsilon^{\nu}(p_{l}, s_l;\hat Q)
\\&\quad\times
v^{\alpha \beta \gamma}_J(\hat p_{m+1},\hat p_l,-\hat p_{m+1}-\hat p_l)\,
D_{\gamma\nu}(\hat p_l + \hat p_{m+1};n_l)
\;\;.
\end{split}
\end{equation}
for $J \in \{1,2,3\}$. Finally we define partial operators $V$ by the analogue of Eq.~(\ref{eq:Vtov}),
\begin{equation}
\label{eq:Vtov123}
\bra{\{\hat s\}_{m+1}}
V^{\dagger}_{J,l}(\{\hat p, \hat f\}_{m+1})\ket{\{s\}_m}
= 
\left(\prod_{j\notin\{l,m+1\}} \delta_{\hat s_j,s_j}\right)
v_{J,l}(\{\hat p, \hat f\}_{m+1},\hat s_{m+1},\hat s_{l},s_l)
\;\;.
\end{equation}
Now the first term in ${\cal W}$ comes from $\sum_{J,K=1}^3 v_J \times v_K^*$. The term with a leading singularity when gluon $m+1$ is soft is $v_2 \times v_2^*$. The term with a leading singularity when gluon $l$ is soft is $v_3 \times v_3^*$. Therefore we need to get rid of $v_3 \times v_3^*$ and double $v_2 \times v_2^*$ to make up for the factor $1/2$ in $S_l$. We thus define
\begin{equation}
\begin{split}
\label{eq:tildeWdef}
\big(\{{}\hat s',\hat s\}_{m+1}\big|
\widetilde{\cal W}&(l,l;\{\hat p\}_{m+1})
\sket{\{s',s\}_m} = 
\\&
\bra{\{\hat s\}_{m+1}}
V^{\dagger}_{2,l}(\{\hat p, \hat f\}_{m+1})
\ket{\{s\}_{m}}
\bra{\{s'\}_{m}}
V_{2,l}(\{\hat p, \hat f\}_{m+1})
\ket{\{\hat s'\}_{m+1}}
\\& - 
\bra{\{\hat s\}_{m+1}}
V^{\dagger}_{3,l}(\{\hat p, \hat f\}_{m+1})
\ket{\{s\}_{m}}
\bra{\{s'\}_{m}}
V_{3,l}(\{\hat p, \hat f\}_{m+1})
\ket{\{\hat s'\}_{m+1}}
\;\;.
\end{split}
\end{equation}
This is computationally very simple even if it takes some time to explain.

For $i \ne j$, we use the simpler splitting operators $V^{\rm soft}$,
\begin{equation}
\begin{split}
\label{eq:Wijdef}
\sbra{\{\hat s',\hat s\}_{m+1}}
{\cal W}(i,j;\{\hat f,\hat p\}_{m+1})
\sket{\{s',s\}_m}
={}&
\bra{\{\hat s\}_{m+1}}
V^{\dagger,{\rm soft}}_i(\{\hat p, \hat f\}_{m+1})
\ket{\{s\}_{m}}
\\ &\times
\bra{\{s'\}_{m}}
V^{\rm soft}_j(\{\hat p, \hat f\}_{m+1})
\ket{\{\hat s'\}_{m+1}}
\;\;.
\end{split}
\end{equation}

In the next factor, the matrix element $\sbra{\{\hat p,\hat f\}_{m+1}}{\cal P}_{l}\sket{\{p,f\}_m}$ is a delta function that enforces the requirement that the momenta and flavors $\{p,f\}_{m}$ are related to the momenta and flavors after the splitting by the mapping $Q_l$. The definition is
\begin{equation}
\label{eq:calPdef1}
\frac{1}{m!}
\int\big[d\{p,f\}_{m}\big]\,
\sbra{\{\hat p,\hat f\}_{m+1}}{\cal P}_{l}\sket{\{p,f\}_m}\,
h\big(\{p, f\}_{m}\big)
= h\big(\{p', f'\}_{m}\big)
\;\;,
\end{equation}
where $h(\{p, f\}_{m})$ is an arbitrary test function and
\begin{equation}
\label{eq:Qldefagainandagain}
\{\{p',f'\}_m, \{\zeta_{\rm p},\zeta_{\rm f}\}\}
= Q_l(\{\hat p,\hat f\}_{m+1})
\;\;
\end{equation}
as specified in Sec.~\ref{sec:mapping}. Another useful identity for ${\cal P}_l$ is\footnote{To derive Eq.~(\ref{eq:calPdef2}) from Eq.~(\ref{eq:calPdef1}), we add one more integration to Eq.~(\ref{eq:calPdef1}), in the form $\int\big[d\{\hat p,\hat f\}_{m+1}\big]\, g\big(\{\hat p, \hat f\}_{m+1}\big)$.
Then on the right hand side, we change integration variables from $\{\hat p',\hat f'\}_{m+1}$ to $\{\{p,f\}_m, \{\zeta_{\rm p},\zeta_{\rm f}\}\}$ according to Eqs.~(\ref{eq:jacobianFSdef}) and (\ref{eq:jacobianISdef}). This gives a result that is equivalent to Eq.~(\ref{eq:calPdef2}).}
\begin{equation}
\begin{split}
\label{eq:calPdef2}
\frac{1}{(m+1)!}
\int &\big[d\{\hat p, \hat f\}_{m+1}\big]\ 
g\big(\{\hat p, \hat f\}_{m+1}\big)\,
\sbra{\{\hat p,\hat f\}_{m+1}}{\cal P}_{l}\sket{\{p,f\}_m}\,
\\={}& 
\frac{1}{(m+1)}
\sum_{\zeta_{\rm f}\in \Phi_{l}(f_{l})}\,
\int\! d\zeta_{\rm p}\
\theta(\zeta_{\rm p} \in \varGamma_{l}(\{p\}_{m},\zeta_{\rm f}))\,
g\big(\{\hat p', \hat f'\}_{m+1}\big)\big)\, 
\;\;,
\end{split}
\end{equation}
where $g\big(\{\hat p, \hat f\}_{m+1}\big)$ is an arbitrary test function and
\begin{equation}
\label{eq:Rldefagainbis}
\{\hat p',\hat f'\}_{m+1}
= R_l(\{\{p,f\}_m, \{\zeta_{\rm p},\zeta_{\rm f}\}\})
\end{equation}
is the inverse transformation to $Q_l$, as specified in Sec.~\ref{sec:mapping}. In this form, we display an integration over the splitting variables that would occur in an implementation of this formalism as a computer program.

The counting factor $(m+1)$ is the ratio of the factor $(m+1)!$ in the normalization integral for $\sbra{\{\hat p,\hat f,\hat s',\hat c',\hat s,\hat c\}_{m+1}} {\cal S}^{(l)}_{ij} \sket{\rho}$ to the factor $m!$ in the normalization integral for $\sbrax{\{p,f,s',c',s,c\}_{m}}\sket{\rho}$. This factor is derived in Appendix \ref{sec:symmetryfactors}. The factor with parton distributions comes from Eqs.~(\ref{eq:rhodef1}) and (\ref{eq:rhodef2}). In the case of a final state splitting, this factor is 1.

We can now assemble our result. From Eq.~(\ref{eq:hatrho}) we have
\begin{equation}
\sket{\hat\rho} = {\cal S} \sket{\rho}
\;\;,
\end{equation}
where the total splitting operator is
\begin{equation}
\label{eq:calSl}
{\cal S} = \sum_l{\cal S}_l\;\;,\quad
{\cal S}_l = {\cal S}^{(l)}_{ll}
+\sum_{k\ne l}
{\cal A}_{lk}
\left\{ 
{\cal S}^{(l)}_{lk}
+ {\cal S}^{(l)}_{kl}
\right\}
\;\;.
\end{equation}
Here the sums run over the set $\{\La,\Lb,1,\dots,m\}$ and ${\cal A}_{lk}$ is the operator on the space of statistical states that multiplies a basis vector $\sket{\{\hat p,\hat f,\hat s',\hat c',\hat s,\hat c\}_{m+1}}$ by the corresponding function $A_{lk}(\{\hat p\}_{m+1})$.

We manipulate the result a bit. Because the quantum amplitudes are color singlets, as reflected in Eq.~(\ref{eq:tlidentity}), when $\hat f_{m+1} = {\rm g}$ the gluon emission operators ${\cal G}(l,k;\{\hat f\}_{m+1})$ obey
\begin{equation}
\label{eq:calGidentities}
\sum_k {\cal G}(l,k;\{\hat f\}_{m+1}) = 0
\qquad,
\qquad 
\sum_k {\cal G}(k,l;\{\hat f\}_{m+1}) = 0
\;\;.
\end{equation}
Then the ${\cal G}(l,k;\{\hat f\}_{m+1})$ are not independent. For this reason, there can be color coherence cancellations that are always present but are not evident if we use all of the possible ${\cal G}(l,k;\{\hat f\}_{m+1})$ operators. Accordingly, when $\hat f_{m+1} = {\rm g}$, we eliminate ${\cal G}(l,l;\{\hat f\}_{m+1})$ with the replacement
\begin{equation}
\label{eq:Gllreplacement}
{\cal G}(l,l;\{\hat f\}_{m+1}) = 
- \frac{1}{2}\sum_{k\ne l} {\cal G}(l,k;\{\hat f\}_{m+1})
- \frac{1}{2}\sum_{k\ne l} {\cal G}(k,l;\{\hat f\}_{m+1})
\;\;.
\end{equation}
We will see in Sec.~\ref{sec:softcoherence} how this allows cancellations to occur at a low level of a calculation. With this replacement, the total splitting operator is given for the case $\hat f_{m+1} = {\rm g}$ by
\begin{equation}
\begin{split}
\label{eq:Salldefg}
\big(\{\hat p,\hat f,{}&\hat s',\hat c',\hat s,\hat c\}_{m+1}\big|
{\cal S}_{l}\sket{\{p,f,s',c',s,c\}_{m}}
\\={}&
(m+1)
\sbra{\{\hat p,\hat f\}_{m+1}}{\cal P}_{l}\sket{\{p,f\}_m}\,
\frac
{n_\Lc(a) n_\Lc(b)\,\eta_{\La}\eta_{\Lb}}
{n_\Lc(\hat a) n_\Lc(\hat b)\,
 \hat \eta_{\La}\hat \eta_{\Lb}}\,
\frac{
f_{\hat a/A}(\hat \eta_{\La},\mu^{2}_{F})
f_{\hat b/B}(\hat \eta_{\Lb},\mu^{2}_{F})}
{f_{a/A}(\eta_{\La},\mu^{2}_{F})
f_{b/B}(\eta_{\Lb},\mu^{2}_{F})}
\\&\times
\sum_{\substack{k\in \{\La,\Lb,1,\dots,m\}\\ k\ne l}}
\bigg\{
\sbra{\{\hat c',\hat c\}_{m+1}}
{\cal G}(l,k;\{\hat f\}_{m+1})
\sket{\{c',c\}_m}
\\&\hskip 2.5 cm
\times
\Big[ 
A_{lk}(\{\hat p\}_{m+1})
\sbra{\{\hat s',\hat s\}_{m+1}}
{\cal W}(l,k;\{\hat f,\hat p\}_{m+1})
\sket{\{s',s\}_m}
\\&\hskip 3 cm
- 
\frac{1}{2}
\sbra{\{\hat s',\hat s\}_{m+1}}
{\cal W}(l,l;\{\hat f,\hat p\}_{m+1})
\sket{\{s',s\}_m}
\Big]
\\&\hskip 2 cm
+
\sbra{\{\hat c',\hat c\}_{m+1}}
{\cal G}(k,l;\{\hat f\}_{m+1})
\sket{\{c',c\}_m}
\\&\hskip 2.5 cm
\times
\Big[ 
A_{lk}(\{\hat p\}_{m+1})
\sbra{\{\hat s',\hat s\}_{m+1}}
{\cal W}(k,l;\{\hat f,\hat p\}_{m+1})
\sket{\{s',s\}_m}
\\&\hskip 3 cm
- 
\frac{1}{2}
\sbra{\{\hat s',\hat s\}_{m+1}}
{\cal W}(l,l;\{\hat f,\hat p\}_{m+1})
\sket{\{s',s\}_m}
\Big]
\bigg\}
\;\;.
\end{split}
\end{equation}
In the case that $\hat f_{m+1} \ne {\rm g}$, that is $\{\hat f_l,\hat f_{m+1}\} = \{q,\bar q\}$, this becomes
\begin{equation}
\begin{split}
\label{eq:Salldefnotg}
\big(\{\hat p,\hat f,\hat s',\hat c',\hat s,\hat c\}_{m+1}|
{}&{\cal S}_{l}\sket{\{p,f,s',c',s,c\}_{m}} = (m+1)\,
\sbra{\{\hat p,\hat f\}_{m+1}}{\cal P}_{l}\sket{\{p,f\}_m}\,
\\&\times
\frac
{n_\Lc(a) n_\Lc(b)\,\eta_{\La}\eta_{\Lb}}
{n_\Lc(\hat a) n_\Lc(\hat b)\,
 \hat \eta_{\La}\hat \eta_{\Lb}}\,
\frac{
f_{\hat a/A}(\hat \eta_{\La},\mu^{2}_{F})
f_{\hat b/B}(\hat \eta_{\Lb},\mu^{2}_{F})}
{f_{a/A}(\eta_{\La},\mu^{2}_{F})
f_{b/B}(\eta_{\Lb},\mu^{2}_{F})}
\\&\times
\sbra{\{\hat c',\hat c\}_{m+1}}
{\cal G}(l,l;\{\hat f\}_{m+1})
\sket{\{c',c\}_m}
\\&\times
\sbra{\{\hat s',\hat s\}_{m+1}}
{\cal W}(l,l;\{\hat f,\hat p\}_{m+1})
\sket{\{s',s\}_m}
\;\;.
\end{split}
\end{equation}

In the following section, we will use ${\cal S}$ to define the splitting operator at shower time $t$, ${\cal H}_{\rm I}(t)$. Then in Sec.~\ref{sec:colorstructure} we will study the evolution of the color structure that is expressed in Eqs.~(\ref{eq:Salldefg}) and (\ref{eq:Salldefnotg}). In Sec.~\ref{sec:softcoherence}, we will examine the how quantum coherence for soft gluon emission is contained in Eq.~(\ref{eq:Salldefg}).

\section{The operator ${\cal H}_{\rm I}(t)$}
\label{sec:HIoft}

It remains to define the operator ${\cal H}_{\rm I}(t)$. The integral of this operator over $t$ gives the total probability for a splitting at any scale,
\begin{equation}
\int _0^\infty dt\ {\cal H}_{\rm I}(t) = {\cal S}
\;\;.
\end{equation}
To get the probability for a splitting at the scale corresponding to shower time $t$, we simply need to insert a delta function that defines $t$. Our default choice is
\begin{equation}
\label{eq:tdef}
t = \log\left(\frac{Q_0^2}
{|(\hat p_l 
+(-1)^{\delta_{l,\La} + \delta_{l,\Lb}} \hat p_{m+1})^2 - m^2(f_l)|}\right)
\;\;.
\end{equation}
Here the virtuality is defined using $\hat p_l + \hat p_{m+1}$ for a final state splitting and $\hat p_l - \hat p_{m+1}$ for an initial state splitting. We take $Q_0^2$ to be the hardness scale of the initial hard scattering that starts the parton shower, so that the initial value of $t$ is zero. One could take $Q_0^2$ to be the minimum of the values $2 p_i\cdot p_j$ for final state particles $i,j$ from the initial hard scattering. Here we can neglect all quark masses compared to $Q_0$. Other definitions of $t$ are possible. For instance, many authors use a measure of the transverse momentum in a parton splitting.

When we use the definition (\ref{eq:tdef}) of $t$, we obtain the corresponding definition of ${\cal H}_{\rm I}(t)$. 
\begin{equation}
\begin{split}
\label{eq:Hdef}
\big(\{\hat p,\hat f,\hat s',\hat c',\hat s,\hat c\}_{m+1}\big|
{\cal H}_{\rm I}(t){}&\sket{\{p,f,s',c',s,c\}_{m}}
\\
=\sum_{l\in \{\La,\Lb,1,\dots,m\}}
&\big(\{\hat p,\hat f,\hat s',\hat c',\hat s,\hat c\}_{m+1}\big|
{\cal S}_{l}\sket{\{p,f,s',c',s,c\}_{m}}
\\&\times
\delta\!\left(
t - \log\left(\frac{Q_0^2}{|(\hat p_l +(-1)^{\delta_{l,\La} + \delta_{l,\Lb}}
\hat p_{m+1})^2 - m^2(f_l)|}\right)
\right)\;\;,
\end{split}
\end{equation}
where  ${\cal S}_{l}$ is defined in Eqs.~(\ref{eq:Salldefg}) and (\ref{eq:Salldefnotg}). In these equations there are parton distribution functions evaluated at a factorization scale $\mu_{\LF}$, which we define according to Eq.~(\ref{eq:muFdef}) in terms of $t$. In addition, there is a factor $\alpha_{\rm s}$, which needs to be evaluated at a scale $\mu_{\rm R}$ that has not been made explicit in the notation. The argument presented in Sec.~(\ref{sec:resolution}) indicates that the momentum scale at the splitting is just the resolution scale that we use for $\mu_\LF$. Thus we take
\begin{equation}
\label{eq:muFRdef}
\mu_{\LF}^2 = \mu_{\rm R}^2 = Q_0^2\, e^{-t}
\;\;.
\end{equation}

\section{Color evolution of the statistical states}
\label{sec:colorstructure}

We are now in a position to say something about the color structure of the statistical states and the evolution of this structure. Notice that in general $\sbrax{\{p,f,s',c',s,c\}_{m}}\sket{\rho}$ can be non-zero for $\{c'\}_{m} \ne \{c\}_{m}$. Even if we start with $\{c'\}_{n} = \{c\}_{n}$, at some early stage of evolution, splitting will generate states with $\{c'\}_{m} \ne \{c\}_{m}$ at later stages. However, in the end we measure something that is color independent. This means taking the color trace of $\rho$. That is, we multiply $\sbrax{\{p,f,s',c',s,c\}_{m}} \sket{\rho}$ by $\brax{\{c'\}_{m}}\ket{\{c\}_{m}}$ and sum over the color configurations $\{c'\}_{m}$ and $\{c\}_{m}$, as in Eq.~(\ref{eq:sigmaFsimple}). Recall that  when $\{c'\}_{m} \ne \{c\}_{m}$, the inner product $\brax{\{c'\}_{m}} \ket{\{c\}_{m}}$ is of order $1/N_\Lc^2$ to some power and is thus small in the large $N_\Lc^2$ limit and numerically small for $N_\Lc = 3$. Thus configurations with $\{c'\}_{m} \ne \{c\}_{m}$ are not very important. Furthermore, if we start with a state with $\{c'\}_{m} \ne \{c\}_{m}$, splitting cannot generate $\{\hat c'\}_{m+1} = \{\hat c\}_{m+1}$ at the next stage. Thus one can speak of a ``leading color approximation'' in which contributions with $\{c'\}_{m} \ne \{c\}_{m}$ are always dropped. This is what happens in most parton shower Monte Carlo programs. We do not drop terms, but we should understand what happens in the leading color approximation.

In order to understand splitting in the leading color configuration, suppose that we apply ${\cal H}_{\rm I}(t)$ to a state $\sket{\{p,f,s',c',s,c\}_{m}}$ in which $\{c'\} = \{c\}$. Thus we consider the action of the color splitting operator ${\cal G}(i,j;\{\hat f\}_{m+1})$, on a state $\sket{\{c',c\}_m}$ with $\{c'\} = \{c\}$,
\begin{equation}
\begin{split}
\label{eq:Gcolor}
\sbra{\{\hat c',\hat c\}_{m+1}}
{\cal G}(i,j;\{\hat f\}_{m+1})
\sket{\{c,c\}_m}
={}&
\dualL\bra{\{\hat c\}_{m+1}}
t^\dagger_i(f_i \to \hat f_i + \hat f_{m+1})
\ket{\{c\}_{m}}
\\ &\times
\dualL\bra{\{\hat c'\}_{m+1}}
t^\dagger_j(f_j \to \hat f_j + \hat f_{m+1})
\ket{\{c\}_{m}}^*
\;\;.
\end{split}
\end{equation}
We first consider gluon emission. That is $\hat f_{m+1} = {\rm g}$.
Then, after using Eq.~(\ref{eq:Gllreplacement}), all of the contributions to ${\cal H}_{\rm I}(t)\sket{\rho}$ come from ${\cal G}(i,j;\{\hat f\}_{m+1})$ with $i \ne j$. In order to have a definite case in mind, let us suppose that $i$ and $j$ are gluons. Then, using Eq.~(\ref{eq:coloropggg}) to represent the $t^\dagger$ operators in terms of the operators $a^\dagger$ that insert the added gluon in particular places in the string basis states, our matrix element becomes
\begin{equation}
\begin{split}
\label{eq:leadingcolor}
C_{\rm F} &
\bra{\{\hat c\}_{m+1}}a^\dagger_+(i)\ket{\{c\}_{m}}\
\bra{\{\hat c'\}_{m+1}}a^\dagger_+(j)\ket{\{c\}_{m}}^*
\\ +\
C_{\rm F}&
\bra{\{\hat c\}_{m+1}}a^\dagger_-(i)\ket{\{c\}_{m}}\
\bra{\{\hat c'\}_{m+1}}a^\dagger_-(j)\ket{\{c\}_{m}}^*
\\ -\
 C_{\rm F}&
\bra{\{\hat c\}_{m+1}}a^\dagger_+(i)\ket{\{c\}_{m}}\
\bra{\{\hat c'\}_{m+1}}a^\dagger_-(j)\ket{\{c\}_{m}}^*
\\ -\
C_{\rm F}&
\bra{\{\hat c\}_{m+1}}a^\dagger_-(i)\ket{\{c\}_{m}}\
\bra{\{\hat c'\}_{m+1}}a^\dagger_+(j)\ket{\{c\}_{m}}^*
\;\;.
\end{split}
\end{equation}

Consider the case that gluon $j$ is just to the right of gluon $i$ along a color string. Then if the new gluon is inserted to the right of gluon $i$, we get the same state as when the new gluon is inserted to the left of gluon $j$. That is, the third term,  with $a^\dagger_+(i)$ and $a^\dagger_-(j)$, gives a non-zero matrix element when $\{\hat c'\}_{m+1} = \{\hat c\}_{m+1}$. This contribution is thus kept in the leading color approximation. The other three contributions would be thrown away in this approximation. This is illustrated in Fig.~\ref{fig:ijsplitting}.

\FIGURE{
\includegraphics[width = 4 cm]{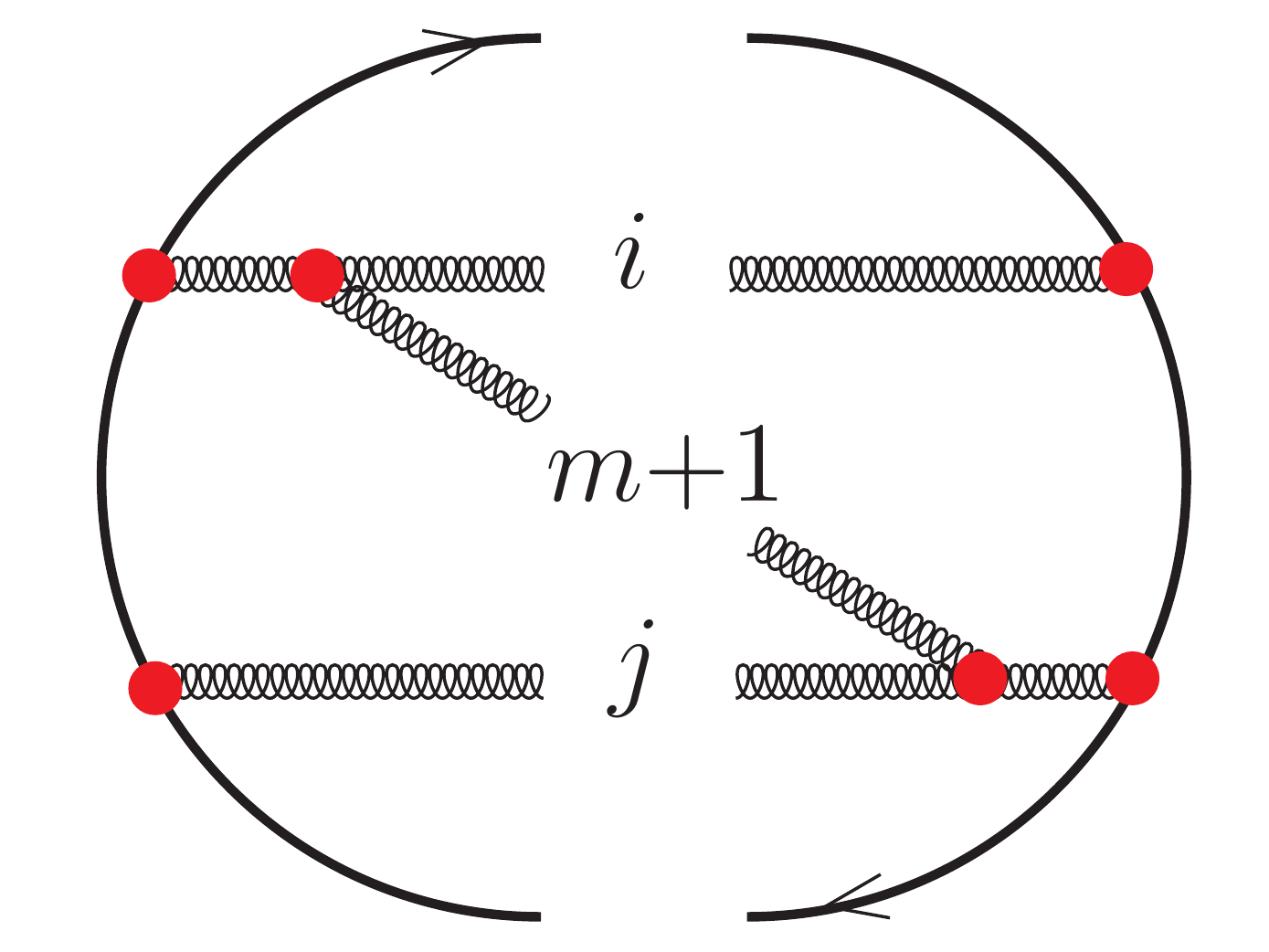}
\includegraphics[width = 4 cm]{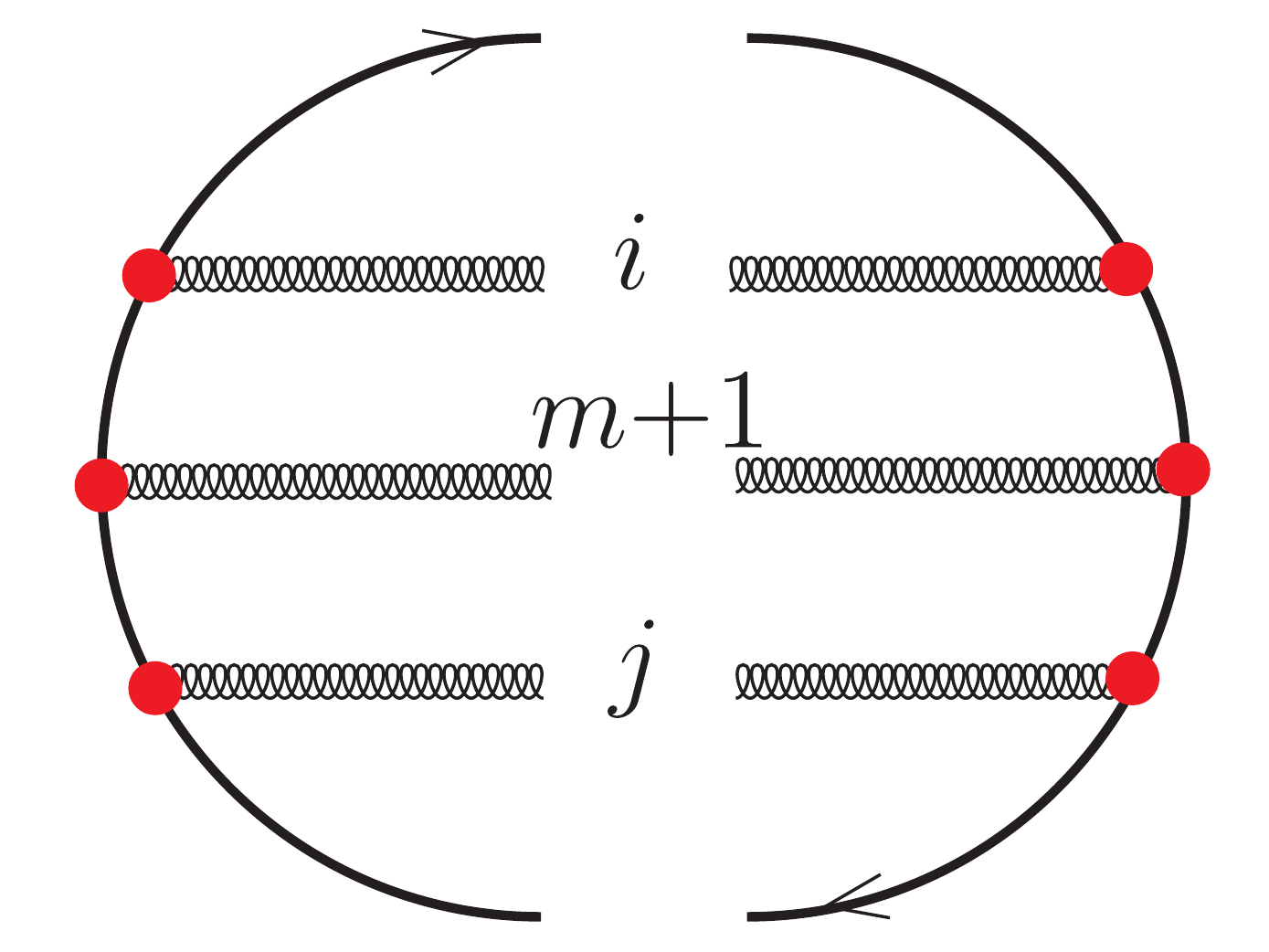}
\includegraphics[width = 4 cm]{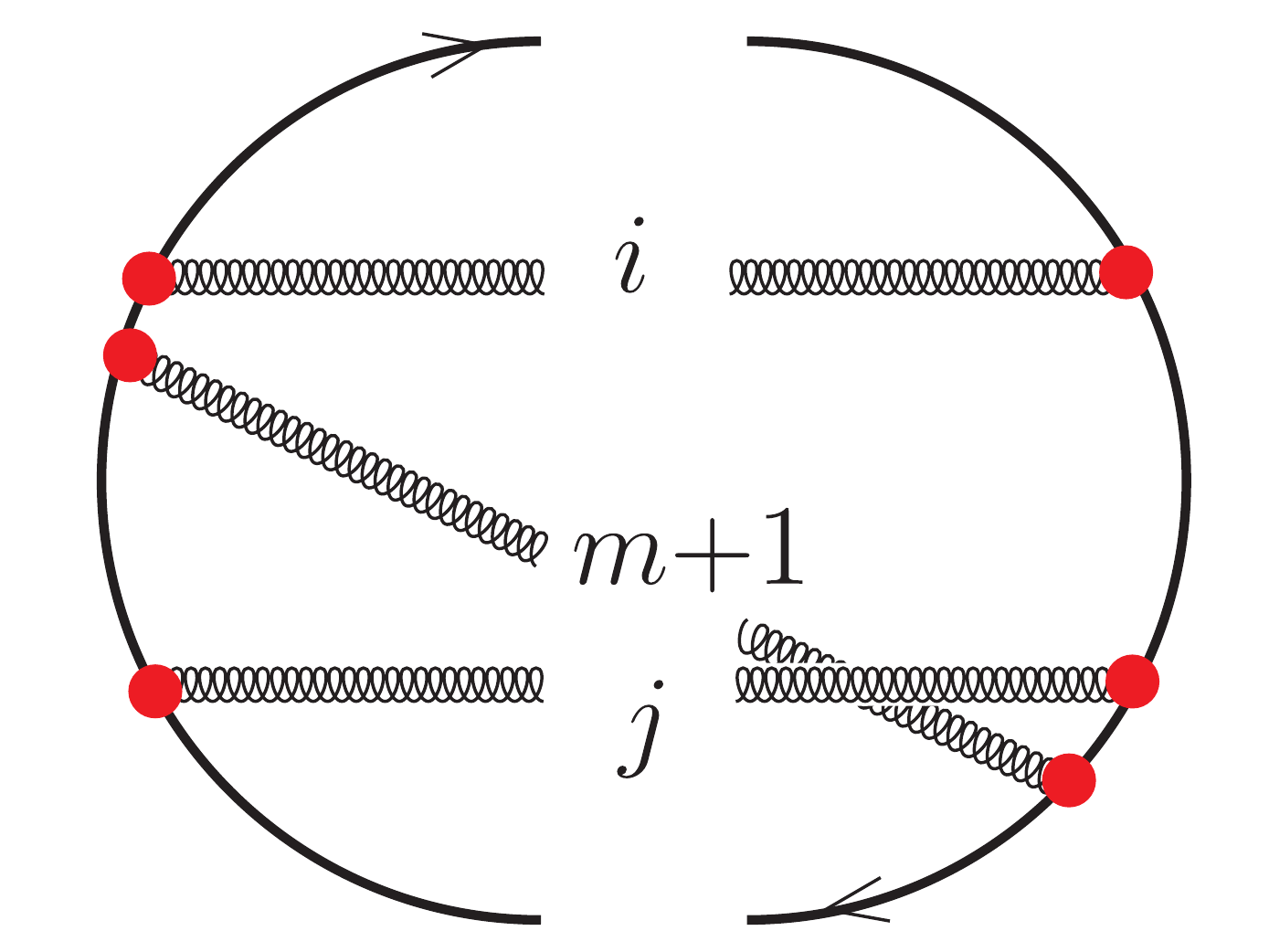}
\caption{Illustration of Eq.~(\ref{eq:leadingcolor}). The first diagram shows the color configuration corresponding to the operator ${\cal G}(i,j;\{\hat f\}_{m+1})$: the gluon's color is emitted from parton $i$ and absorbed on gluon $j$. Using Eq.~(\ref{eq:coloropggg}) to write this in terms of color basis states, one contribution is shown in the middle diagram. This is the leading color contribution. Another contribution is shown in the right-hand diagram. This contribution is subleading. If $i$ and $j$ had not been color connected to each other, there would have been no contribution that survives in the leading color approximation.
}
\label{fig:ijsplitting}
}

Similarly, if gluon $j$ is just to the left of gluon $i$ along a color string, then the contribution with $a^\dagger_-(i)$ and $a^\dagger_+(j)$ makes a leading color contribution, while the other three contributions would be thrown away in the leading color approximation.

Suppose now that gluon $j$ is {\it not} next to gluon $i$ along a color string. Then all four terms would be thrown away in the leading color approximation.

We can summarize this by saying that there is a term that is kept in the leading color approximation when gluons $i$ and $j$ are ``color connected'': next to each other along a color string. An analogous analysis leads to the same conclusion if one or both of partons $i$ and $j$ are quarks.

One may say that in the leading color approximation, the operating units are color dipoles, consisting of partons that are next to each other along a single color string. Within this approximation, a gluon may be considered to carry the ${\bf 3}\times \bar{\bf 3}$ representation of $SU(3)$ instead of the $\bf 8$ representation. Then, for instance, the $\bar{\bf 3}$ half of a gluon forms a dipole with the ${\bf 3}$ part of the neighboring gluon or with a neighboring quark. A dipole can emit a gluon. But there is no interference between diagrams in which the gluon is emitted by different dipoles. This is illustrated in Fig.~\ref{fig:colordipole}. The color dipole picture was introduced as the basis of the parton shower program \textsc{Ariadne} \cite{ariadne}.

\FIGURE{
\includegraphics[width = 5 cm]{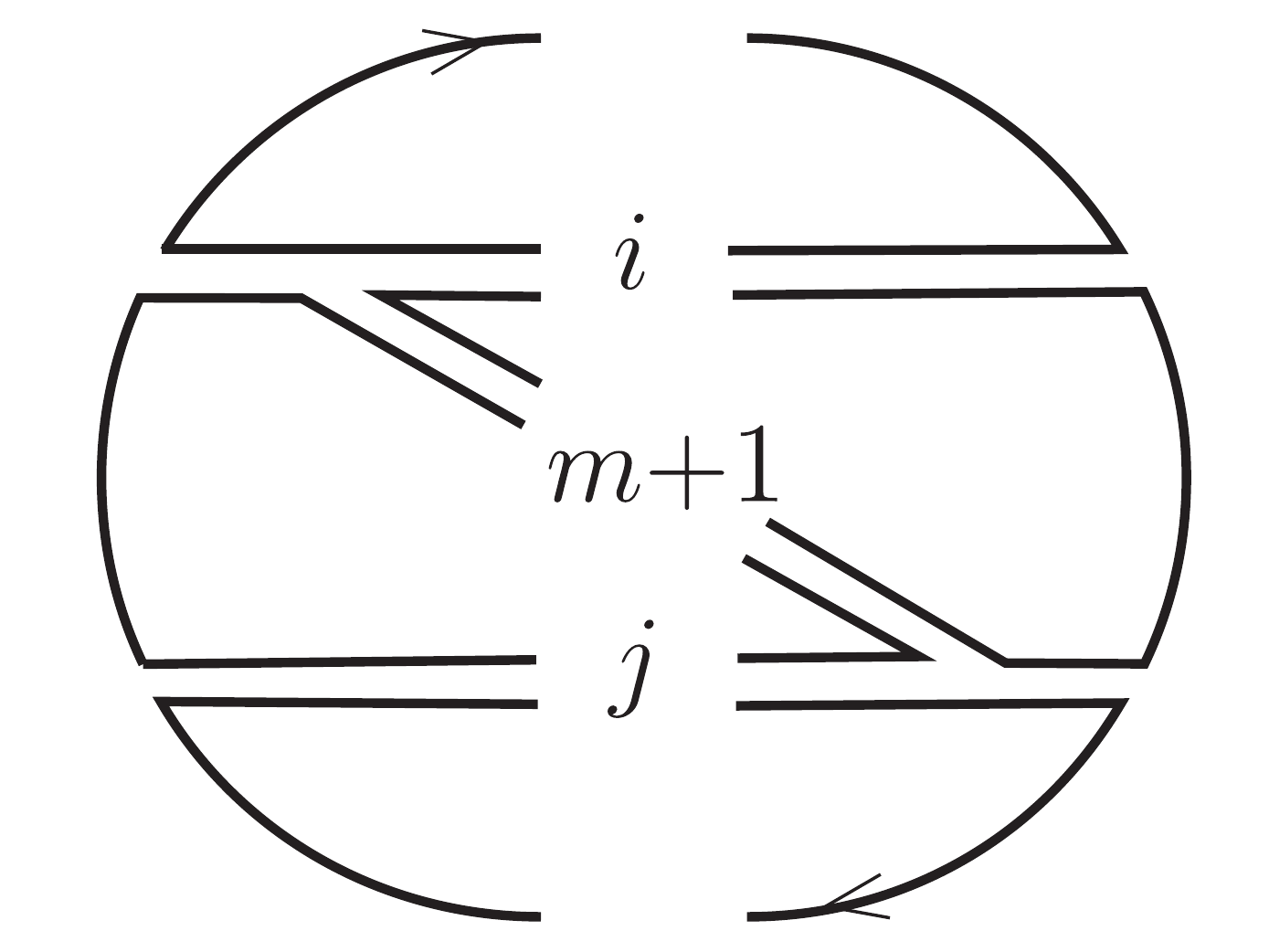}
\caption{The starting diagram from Fig.~\ref{fig:ijsplitting} in the leading color approximation, in which gluons are emitted from color dipoles. The $i$-$j$ color dipole is a unit that emits a gluon and absorbs it on the other side of the diagram. At the next stage there will be an $i$-$(m+1)$ color dipole and an $(m+1)$-$j$ color dipole.
}
\label{fig:colordipole}
}

With the formalism presented in this paper, one can easily implement this approximation, but one can also keep corrections to it.

We have considered gluon emission. The case of a ${\rm g} \to q + \bar q$ splitting is rather different. Then the contributing terms are from ${\cal G}(l,l;\{\hat f\}_{m+1})$. For the operators $t^\dagger_l({\rm g} \to q + \bar q)$ in Eq.~(\ref{eq:Gcolor}), we use Eq.~(\ref{eq:coloropgqqbar}). In the leading term, with operator $a^\dagger_q(l)$, the color string containing the gluon breaks at the position of the gluon, with the creation of two new string ends. There is a second term, with operator $a^\dagger_s(l)$, in which gluon $l$ simply disappears from its string and a new $q$-$\bar q$ string is created. However, the coefficient of $a^\dagger_s(l)$ is small in the $N_\Lc \to \infty$ limit and would be dropped in the leading color approximation. In fact, the coefficient of the leading $a^\dagger_q(l)$ is finite in the $N_\Lc \to \infty$ limit, instead of growing with $N_\Lc$. Thus ${\rm g} \to q + \bar q$ splitting is color suppressed compared to ${\rm g} \to {\rm g} + {\rm g}$ splitting. Some of the color suppression is cancelled by the number of available $q \bar q$ flavors. Parton shower Monte Carlo programs normally include the leading term in $g \to q + \bar q$ splitting. With the formalism presented in this paper, we can keep both terms in ${\rm g} \to q + \bar q$ splitting.

\section{Soft gluon coherence}
\label{sec:softcoherence}

When a soft gluon is emitted from the partons involved in a hard scattering, the angular distribution of the emitted gluon reflects the structure of the emitting partons as a whole. In particular, a soft gluon emitted from a pair of colored partons at an angle that is bigger than the angle between the partons effectively sees just one parton with the total color charge of the pair. Since the splitting operator ${\cal H}_{\rm I}(t)$ coherently sums the leading soft gluon singularities for emissions from a hard partonic system, including all of the interference diagrams, soft gluon coherence is automatically included. It is of interest to see how this happens and what form it takes in the color basis used in this paper.

Consider the matrix element of the splitting operator ${\cal S}_l$ for the case $\hat f_{m+1} = {\rm g}$, as given in Eq.~(\ref{eq:Salldefg}) with the default choice $A_{lk} = 1/2$. There is a sum over indices $k\in \{\La,\Lb,1,\dots,m\}$ with $k\ne l$. There are two terms with different color factors, $\sbra{\{\hat c',\hat c\}_{m+1}} {\cal G}(l,k;\{\hat f\}_{m+1})\sket{\{c',c\}_m}$ and $\sbra{\{\hat c',\hat c\}_{m+1}} {\cal G}(k, l;\{\hat f\}_{m+1})\sket{\{c',c\}_m}$. These have analogous structures, so it suffices to analyze one of them. The coefficient of the color factor $-\sbra{\{\hat c',\hat c\}_{m+1}} {\cal G}(l,k;\{\hat f\}_{m+1})\sket{\{c',c\}_m}$ is
\begin{equation}
\begin{split}
\sbra{\{\hat s',\hat s\}_{m+1}}
{\cal W}_{\rm tot}(l,k;&\{\hat f,\hat p\}_{m+1})
\sket{\{s',s\}_m} \equiv
\\
&  
\sbra{\{\hat s',\hat s\}_{m+1}}
{\cal W}(l,l;\{\hat f,\hat p\}_{m+1})
\sket{\{s',s\}_m}
\\&
- 
\sbra{\{\hat s',\hat s\}_{m+1}}
{\cal W}(l,k;\{\hat f,\hat p\}_{m+1})
\sket{\{s',s\}_m}
\;\;.
\end{split}
\end{equation}
The first term is from the square of the amplitude for emitting the gluon from parton $l$. The second term represents the interference between the emission of gluon $m+1$ from parton $l$ and the emission of this gluon from parton $k$.  Let us evaluate this in the soft-gluon approximation, $V \to V^{\rm soft}$, as defined in Eqs.~(\ref{eq:Vtovsoft}) and (\ref{eq:VsoftF}). We can also approximate $\hat p_l \sim p_l$ and $\hat p_k \sim p_k$. With these replacements, we have
\begin{equation}
\begin{split}
\sbra{\{\hat s',\hat s\}_{m+1}}
{\cal W}_{\rm tot}&(l,k;\{\hat f,\hat p\}_{m+1})
\sket{\{s',s\}_m} \sim
\\&
\left(\prod_{j \ne m+1} \delta_{\hat s_j,s_j}\delta_{\hat s'_j,s'_j}\right)
(4\pi\as)\ \frac{
\varepsilon(\hat p_{m+1}, \hat s_{m+1};Q)^*
\!\cdot\! p_l}
{\hat p_{m+1}\!\cdot\! p_l}
\\&\times
\left\{
\frac{
\varepsilon(\hat p_{m+1}, \hat s_{m+1};Q)
\!\cdot\! p_l}
{\hat p_{m+1}\!\cdot\! p_l}
-
\frac{
\varepsilon(\hat p_{m+1}, \hat s_{m+1};Q)
\!\cdot\! p_k}
{\hat p_{m+1}\!\cdot\! p_k}
\right\}
\;\;.
\end{split}
\end{equation}
There are singularities when $\hat p_{m+1}$ is collinear with $p_{l}$ and when $\hat p_{m+1}$ collinear with $p_{k}$. The singularity for $\hat p_{m+1}$ collinear with $p_{k}$ is not strong enough to give a logarithmic divergence when we integrate over the direction of $\hat p_{m+1}$. The singularity for $\hat p_{m+1}$ collinear with $p_{l}$ {\em is} strong enough to give a logarithmic divergence. Here, the leading singularity comes from the first term, while the second term, representing the interference graph, gives only an integrable collinear singularity. 

When $\hat p_{m+1}$ is not close to being collinear with $p_{l}$ or $p_{k}$, both terms are important. However, suppose that the angle between $p_{l}$ and $p_{k}$ is small and that $\hat p_{m+1}$ makes an angle with either of them that is substantially greater than this angle. Then the two contributions approximately cancel each other. That is, soft gluon radiation associated with this color factor is approximately confined to a cone about the directions of partons $l$ and $k$ with opening angle on the order of the angle between $p_{l}$ and $p_{k}$. 

Thus there is approximate angular ordering of soft gluon emissions. One can also make an exact statement about angular ordering \cite{angleorder}. If we sum over the spins of the soft gluon, we can use
\begin{equation}
\sum_{\hat s_{m+1}}
\varepsilon^\mu(\hat p_{m+1}, \hat s_{m+1};Q)
\varepsilon^\nu(\hat p_{m+1}, \hat s_{m+1};Q)^*
= - g^{\mu\nu}
+ \frac{\hat p_{m+1}^\mu Q^\nu + Q^\mu\hat p_{m+1}^\nu}{\hat p_{m+1}\!\cdot\! Q}
- \frac{Q^2\,\hat p_{m+1}^\mu \hat p_{m+1}^\nu}{(\hat p_{m+1}\!\cdot\! Q)^2}
\;\;.
\end{equation}
Then we can integrate over the azimuthal angle $\phi$ specifying the rotation of $\hat p_{m+1}$ about $p_l$ in the rest frame of $Q$. This gives
\begin{equation}
\begin{split}
\int \frac{d\phi}{2\pi} \sum_{s_{m+1},\hat s_{m+1}} 
\delta_{s_{m+1},\hat s_{m+1}}&
\sbra{\{\hat s',\hat s\}_{m+1}}
{\cal W}_{\rm tot}(l,k;\{\hat f,\hat p\}_{m+1})
\sket{\{s',s\}_m}
= 
\\&
\left(\prod_{j \ne m+1} \delta_{\hat s_j,s_j}\delta_{\hat s'_j,s'_j}\right)
(4\pi\as)\ I_{\rm soft}
\;\;,
\end{split}
\end{equation}
where
\begin{equation}
\begin{split}
I_{\rm soft} ={}& \int \frac{d\phi}{2\pi}\
\frac{p_l^\mu}
{\hat p_{m+1}\!\cdot\! p_l}
\left\{
\frac{p_l^\nu}
{\hat p_{m+1}\!\cdot\! p_l}
-
\frac{
p_k^\nu}
{\hat p_{m+1}\!\cdot\! p_k}
\right\}
\\&\times
\left\{
-g^{\mu\nu}
+ \frac{\hat p_{m+1}^\mu Q^\nu + Q^\mu\hat p_{m+1}^\nu}{\hat p_{m+1}\!\cdot\! Q}
- \frac{Q^2\,\hat p_{m+1}^\mu \hat p_{m+1}^\nu}{(\hat p_{m+1}\!\cdot\! Q)^2}
\right\}
\;\;.
\end{split}
\end{equation}
One can multiply this out and perform the integral using
\begin{equation}
\int \frac{d\phi}{2\pi}\ \frac{1}{A^2 + B^2 + 2AB \cos \phi}
= \frac{1}{|A^2 - B^2|}
\;\;.
\end{equation}

This gives
\begin{equation}
\label{eq:Isoftresult}
\begin{split}
I_{\rm soft} = {}&
\frac{Q \!\cdot\! p_l}{p_l\!\cdot\! \hat p_{m+1}\,Q\!\cdot\! \hat p_{m+1}}
\left\{
\frac{\hat p_{m+1}\!\cdot\! \omega\!\cdot\! p_k\, \beta_{l}}
{\sqrt{(\hat p_{m+1}\!\cdot\!\omega\!\cdot\! p_k)^2 
	+ m^{2}(f_{k})\, (\omega\!\cdot \!\hat p_{m+1})^2}}
+1
\right\}
- \frac{m^{2}(f_{l})}{(p_{l}\!\cdot\!\hat p_{m+1})^{2}}
\;\;.
\end{split}
\end{equation}
Here $\beta$ is the velocity of parton $l$, 
\begin{equation}
\beta_{l} = \frac{\sqrt{(p_{l}\!\cdot\!Q)^{2} - m^{2}(f_{l})Q^{2}}}{p_{l}\!\cdot\!Q}\;\;.
\end{equation}
We have written $\omega\cdot v$ for the product $\omega^{\mu}_{\,\alpha}v^\alpha$ of a vector $v$ with the tensor $\omega$ defined by
\begin{equation}
\omega^{\mu\nu} = \frac{Q^\mu p_l^\nu - p_l^\mu Q^\nu}{Q\!\cdot\! p_l}
\;\;.
\end{equation}
The tensor $\beta_l^{-1}\omega^{\mu\nu}$ is the unit antisymmetric tensor in the $p_l$-$Q$ plane and projects onto this plane. It obeys $\beta_l^{-2}\omega^{\mu}_{\ \alpha}\omega^{\alpha}_{\ \nu} = P^\mu_\nu$, where $P^\mu_\nu$ is the projection operator onto the $p_l$-$Q$ plane. 

In the case that all of the masses are zero, Eq.~(\ref{eq:Isoftresult}) becomes
\begin{equation}
\label{eq:Isoftresult0}
\begin{split}
I_{\rm soft} = {}&
\frac{Q \!\cdot\! p_l}{p_l\!\cdot\! \hat p_{m+1}\,Q\!\cdot\! \hat p_{m+1}}
\left\{
\frac{\hat p_{m+1}\!\cdot\! \omega\!\cdot\! p_k}
{\sqrt{(\hat p_{m+1}\!\cdot\!\omega\!\cdot\! p_k)^2 }}
+1
\right\}
\qquad\text{ (masses vanish) }
\;\;.
\end{split}
\end{equation}
The ratio in the first term in Eq.~(\ref{eq:Isoftresult0}) is either $+1$ or $-1$. It is $+1$ if
\begin{equation}
\frac{\hat p_{m+1}\!\cdot\! Q}{\hat p_{m+1}\!\cdot\! p_l} >
\frac{p_k\!\cdot\! Q}{p_k\!\cdot\! p_l}
\;\;.
\end{equation}
That is, the ratio is $+1$ if, in the rest frame of $Q$, $\hat p_{m+1}$ makes a smaller angle with $p_l$ than does $p_k$. In this small angle region, we have
\begin{equation}
\label{eq:Isoftresultnarrow}
I_{\rm soft} =
\frac{2Q \!\cdot\! p_l}{p_l\!\cdot\! \hat p_{m+1}\ Q\!\cdot\! \hat p_{m+1}}
\qquad\text{(masses vanish, small angle)}
\;\;.
\end{equation}
In the wide angle region, where $\hat p_{m+1}$ makes a larger angle with $p_l$ than does $p_k$, we have
\begin{equation}
\label{eq:Isoftresultwide}
I_{\rm soft} = 0
\qquad\text{(masses vanish, wide angle)}
\;\;.
\end{equation}
Thus, in the massless case, after a sum over spins and an average over the azimuthal angle, the soft radiation outside a cone centered on $p_l$ and extending out to $p_k$ cancels exactly. This phenomenon is known as angular ordering. It can be used to define the angular ordering approximation, in which one simply neglects the radiation outside of this cone even for a fixed set of spins and a fixed azimuthal angle. This angular ordering approximation is commonly used in parton shower Monte Carlo event generators. With the formalism of this paper, one can make this approximation if desired, but it is not required.

In the massive case, it is no longer true that soft radiation outside a cone centered on $p_l$ and extending out to $p_k$ cancels after integrating over the azimuthal angle.

\section{Inclusive evolution}
\label{sec:inclusiveH}

As described in Sec.~\ref{sec:evolution}, the Sudakov exponent is constructed from the operator ${\cal V}(t)$. In turn, ${\cal V}(t)$ is constructed from $\sbra{1}{\cal H}_{\rm I}(t)\sket{\{p,f,s',c',s,c\}_{m}}$, which tells the total probability for the state $\sket{\{p,f,s',c',s,c\}_{m}}$ to split at time $t$. In this section, we determine the structure of $\sbra{1}{\cal H}_{\rm I}(t)\sket{\{p,f,s',c',s,c\}_{m}}$.

We know the matrix elements of ${\cal H}_{\rm I}(t)$, so we simply insert the completeness relation (\ref{eq:completeness}) for the basis states and use Eq.~(\ref{eq:braF1def}) for the inner product of $\sbra{1}$ with a basis state. Thus
\begin{equation}
\begin{split}
\label{eq:HI1start}
\sbra{1}{\cal H}_{\rm I}(t)
\sket{\{p,f,s',c',s,c\}_{m}}
={}& \frac{1}{(m+1)!}\int
\big[d\{\hat p,\hat f,\hat s',\hat c',\hat s,\hat c\}_{m+1}\big]
\\ &\times
\brax{\{\hat s'\}_{m+1}}\ket{\{\hat s\}_{m+1}}\,
\brax{\{\hat c'\}_{m+1}}\ket{\{\hat c\}_{m+1}}
\\ &\times
\sbra{\{\hat p,\hat f,\hat s',\hat c',\hat s,\hat c\}_{m+1}}
{\cal H}_{\rm I}(t)\sket{\{p,f,s',c',s,c\}_{m}}
\;\;.
\end{split}
\end{equation}
We use Eq.~(\ref{eq:Hdef}) to express ${\cal H}_{\rm I}(t)$ in terms of a sum of splitting operators ${\cal S}_l$ and then use Eqs.~(\ref{eq:Salldefg}) and (\ref{eq:Salldefnotg}) for the matrix elements of ${\cal S}_l$. We encounter
\begin{equation}
\begin{split}
\label{eq:HI1integral}
\frac{1}{(m+1)!}
\int \big[d\{\hat p,\hat f&\}_{m+1}\big]
\sbra{\{\hat p,\hat f\}_{m+1}}{\cal P}_{l}\sket{\{p,f\}_m}
\cdots
\\&
= 
\frac{1}{(m+1)}
\sum_{\zeta_{\rm f}\in \Phi_{l}(f_l)}
\int d\zeta_{\rm p}\
\theta(\zeta_{\rm p} \in \varGamma_{l}(\{p\}_{m},\zeta_{\rm f}))\ 
\cdots
\;\;,
\end{split}
\end{equation}
where we have used Eq.~(\ref{eq:calPdef2}). Thus we are really integrating over the splitting variables. Inside the integral and the sum over $l$, the variables $\{\hat p,\hat f\}_{m+1}$ are determined from $\{p,f\}_m$ and the splitting variables $\{\zeta_{\rm f},\zeta_{\rm f}\}$ by the transformation $R_l$, Eqs.~(\ref{eq:Rldef}) and (\ref{eq:Radef}). Then
\begin{equation}
\begin{split}
\label{eq:HI1together}
\sbra{1}&{\cal H}_{\rm I}(t)
\sket{\{p,f,s',c',s,c\}_{m}}
=
\\&
\sum_l
\sum_{\zeta_{\rm f}\in \Phi_{l}(f_l)}
\int d\zeta_{\rm p}\
\theta(\zeta_{\rm p} \in \varGamma_{l}(\{p\}_{m},\zeta_{\rm f}))
\\&\times
\delta\!\left(
t - \log\left(\frac{Q_0^2}
{|(\hat p_l 
+(-1)^{\delta_{l,\La} + \delta_{l,\Lb}}
\hat p_{m+1})^2 - m^2(f_l)|}\right)
\right)
\\&\times
\frac
{n_\Lc(a) n_\Lc(b)\,\eta_{\La}\eta_{\Lb}}
{n_\Lc(\hat a) n_\Lc(\hat b)\,
 \hat \eta_{\La}\hat \eta_{\Lb}}\,
\frac{
f_{\hat a/A}(\hat \eta_{\La},\mu^{2}_{F})
f_{\hat b/B}(\hat \eta_{\Lb},\mu^{2}_{F})}
{f_{a/A}(\eta_{\La},\mu^{2}_{F})
f_{b/B}(\eta_{\Lb},\mu^{2}_{F})}
\\ &\times
\biggl\{
\theta(\hat f_{m+1}\ne {\rm g})\
\bra{\{c'\}_{m}}
g_{ll}(\{\hat f\}_{m+1})
\ket{\{c\}_{m}}
\bra{\{s'\}_{m}}
w_{ll}(\{\hat f,\hat p\}_{m+1})
\ket{\{s\}_{m}}
\\ & +
\theta(\hat f_{m+1} = {\rm g})\sum_{k \ne l}
\bigg[
\bra{\{c'\}_{m}}
g_{lk}(\{\hat f\}_{m+1})
\ket{\{c\}_{m}}
\\&\quad\quad
\times 
\bra{\{s'\}_{m}}
A_{lk}(\{\hat p\}_{m+1})
w_{lk}(\{\hat f,\hat p\}_{m+1})
-
\frac{1}{2}\,
w_{ll}(\{\hat f,\hat p\}_{m+1})
\ket{\{s\}_{m}}
\\& \quad\ +
\bra{\{c'\}_{m}}
g_{kl}(\{\hat f\}_{m+1})
\ket{\{c\}_{m}}
\\&\quad\quad\times 
\bra{\{s'\}_{m}}
A_{lk}(\{\hat p\}_{m+1})
w_{kl}(\{\hat f,\hat p\}_{m+1})
-
\frac{1}{2}\,
w_{ll}(\{\hat f,\hat p\}_{m+1})
\ket{\{s\}_{m}}
\bigg]
\bigg\}
\;\;.
\end{split}
\end{equation}
Here the color dependent function is
\begin{equation}
\begin{split}
\bra{\{c'\}_{m}}
g_{ij}(\{\hat f\}_{m+1})
\ket{\{c\}_{m}}
={}&
\sum_{\{\hat c',\hat c\}_{m+1}}\brax{\{\hat c'\}_{m+1}}\ket{\{\hat c\}_{m+1}}
\\&\times
\sbra{\{\hat c',\hat c\}_{m+1}}
{\cal G}(i,j;\{\hat f\}_{m+1})
\sket{\{c',c\}_m}
\end{split}
\end{equation}
and the spin dependent function is
\begin{equation}
\begin{split}
\bra{\{s'\}_{m}}
w_{ij}(\{\hat f,\hat p\}_{m+1})
\ket{\{s\}_{m}}
={}&
\sum_{\{\hat s',\hat s\}_{m+1}}
\brax{\{\hat s'\}_{m+1}}\ket{\{\hat s\}_{m+1}}
\\&\times
\sbra{\{\hat s',\hat s\}_{m+1}}
{\cal W}(i,j;\{\hat f,\hat p\}_{m+1})
\sket{\{s',s\}_m}
\;\;.
\end{split}
\end{equation}

Let us look at the color factor in Eq.~(\ref{eq:HI1together}) first. Using the definition (\ref{eq:Gijdef}) of ${\cal G}(l,k;\{\hat f\}_{m+1})$, we have
\begin{equation}
\begin{split}
\label{eq:gijdef0}
\bra{\{c'\}_{m}}
g_{ij}(\{\hat f\}_{m+1})
\ket{\{c\}_{m}}
={}&
\sum_{\{\hat c',\hat c\}_{m+1}}
\bra{\{c'\}_{m}}
t_j(f_j \to \hat f_j + \hat f_{m+1})
\ket{\{\hat c'\}_{m+1}}\dualR
\\ &\times
\brax{\{\hat c'\}_{m+1}}\ket{\{\hat c\}_{m+1}}\,
\dualL\bra{\{\hat c\}_{m+1}}
t^\dagger_i(f_i \to \hat f_i + \hat f_{m+1})
\ket{\{c\}_{m}}
\;\;.
\end{split}
\end{equation}
Then using the completeness relations (\ref{eq:dualcompleteness1}) and (\ref{eq:dualcompleteness2}) we find
\begin{equation}
\label{eq:gijdef}
\bra{\{c'\}_{m}}
g_{ij}(\{\hat f\}_{m+1})
\ket{\{c\}_{m}}
=
\bra{\{c'\}_{m}}
t_j(f_j \to \hat f_j + \hat f_{m+1})\,
t^\dagger_i(f_i \to \hat f_i + \hat f_{m+1})
\ket{\{c\}_{m}}
\;\;.
\end{equation}
The operator $t_j t^\dagger_i$ is written as $\bm{T}_j\cdot \bm{T}_i$ in the work of Catani and Seymour on the dipole subtraction scheme for next-to-leading order calculations \cite{CataniSeymour}. In that work, there is a sum over final states, while in this paper we follow the evolution of the exclusive final state that comes between the operators $t_j$ and $t^\dagger_i$. We get back to the inclusive case when we form $\sbra{1}{\cal H}_{\rm I}(t)
\sket{\{p,f,s',c',s,c\}_{m}}$. The operators $g_{ij}$ have some simple properties. From the definition Eq.~(\ref{eq:splittingmatrix}), we see that
\begin{equation}
\label{gijsymmetry}
\bra{\{c'\}_{m}}
g_{ij}(\{\hat f\}_{m+1})
\ket{\{c\}_{m}}
=
\bra{\{c'\}_{m}}
g_{ji}(\{\hat f\}_{m+1})
\ket{\{c\}_{m}}
\;\;.
\end{equation}
Furthermore, when $i=j$, the operators are proportional to the unit operator,
\begin{equation}
\bra{\{c'\}_{m}}
g_{ll}(\{\hat f\}_{m+1})
\ket{\{c\}_{m}} = 
\brax{\{c'\}_{m}}
\ket{\{c\}_{m}}\times
\left\{
\begin{array}{ll}
C_{\rm F}\;, & 
  \{\hat f_l,\hat f_{m+1}\} = \{q,{\rm g}\} \ {\rm or}\ \{\bar q,{\rm g}\}\\
C_{\rm A}\;, & 
  \{\hat f_l,\hat f_{m+1}\} = \{{\rm g},{\rm g}\} \\
T_{\rm R}\;, & \{\hat f_l,\hat f_{m+1}\} = \{q,\bar q\}
\end{array}
\right.
\;\;.
\end{equation}
The last of these cases occurs in Eq.~(\ref{eq:HI1together}).

Let us look next at the spin dependent factor in Eq.~(\ref{eq:HI1together}). In order to do this, we need to introduce the possibility of averaging over the azimuthal angle of parton splitting. For the splitting of parton $l$, we define the transverse part, $q_\perp$ of $\hat p_{m+1}$ by
\begin{equation}
\label{eq:qTdef}
\hat p_{m+1} = a p_l + b n_l + q_\perp
\;\;,
\end{equation}
where $n_l$ is the lightlike vector defined in Eq.~(\ref{eq:nldef}) and $q_\perp \cdot p_l = q_\perp \cdot n_l = 0$. We let $\phi$ be the angle of $q_\perp$ as measured in any convenient coordinate system, so that $\int d\phi$ means integrating over $q_\perp$ at fixed $|q_\perp|$. Thus integrating over $\phi$ is part of integrating over the splitting variables $\zeta_{\rm p}$.

Consider first the spin dependent factors for the case $i\ne j$, which arises from interference diagrams. Using the definition (\ref{eq:Wijdef}) of ${\cal W}(i,j;\{\hat f,\hat p\}_{m+1})$ and the orthogonality of the spin basis vectors, this is
\begin{equation}
\begin{split}
\bra{\{s'\}_{m}}
w_{ij}(\{\hat f,\hat p\}_{m+1})
\ket{\{s\}_{m}}
={}&
\sum_{\{\hat s\}_{m+1}}
\bra{\{s'\}_{m}}
V^{\rm soft}_j(\{\hat p, \hat f\}_{m+1})
\ket{\{\hat s\}_{m+1}}
\\&\times
\bra{\{\hat s\}_{m+1}}
V^{\dagger,{\rm soft}}_i(\{\hat p, \hat f\}_{m+1})
\ket{\{s\}_{m}}
\;\;.
\end{split}
\end{equation}
Using the definition (\ref{eq:Vtovsoft}) and (\ref{eq:VsoftF}) of $V^{{\rm soft}}$, this is
\begin{equation}
\begin{split}
\bra{\{s'\}_{m}}&
w_{ij}(\{\hat f,\hat p\}_{m+1})
\ket{\{s\}_{m}}
\\
&=
\sum_{\{\hat s\}_{m+1}}
\Bigg(\prod_{n \in \{\La,\Lb,1,\dots,m\}} \delta_{\hat s_n,s_n}
\delta_{\hat s_n,s'_n}\Bigg)
(4\pi \as)
\\&\quad\times
\frac{
\varepsilon(\hat p_{m+1}, \hat s_{m+1};Q)^*
\!\cdot\! \hat p_i}
{\hat p_{m+1}\!\cdot\! \hat p_i}\
\frac{
\varepsilon(\hat p_{m+1}, \hat s_{m+1};Q)
\!\cdot\! \hat p_j}
{\hat p_{m+1}\!\cdot\! \hat p_j}
\\&=
\Bigg(\prod_{n \in \{\La,\Lb,1,\dots,m\}} \delta_{s_n,s'_n}\Bigg)
(4\pi \as)\sum_{\hat s_{m+1}}
\frac{
\varepsilon(\hat p_{m+1}, \hat s_{m+1};Q)^*
\!\cdot\! \hat p_i}
{\hat p_{m+1}\!\cdot\! \hat p_i}\
\frac{
\varepsilon(\hat p_{m+1}, \hat s_{m+1};Q)
\!\cdot\! \hat p_j}
{\hat p_{m+1}\!\cdot\! \hat p_j}
\\&=
\brax{\{s'\}_m}\ket{\{s\}_m}
(4\pi \as)\sum_{\hat s_{m+1}}
\frac{
\varepsilon(\hat p_{m+1}, \hat s_{m+1};Q)^*
\!\cdot\! \hat p_i}
{\hat p_{m+1}\!\cdot\! \hat p_i}\
\frac{
\varepsilon(\hat p_{m+1}, \hat s_{m+1};Q)
\!\cdot\! \hat p_j}
{\hat p_{m+1}\!\cdot\! \hat p_j}
\;\;.
\end{split}
\end{equation}
Notice that $\bra{\{s'\}_{m}} w_{ij}(\{\hat f,\hat p\}_{m+1})\ket{\{s\}_{m}}$ is proportional to a unit matrix $\brax{\{s'\}_m}\ket{\{s\}_m}$ in the spin indices and that it is symmetric under $i \leftrightarrow j$. The coefficient of $\brax{\{s'\}_m}\ket{\{s\}_m}$ is a product of splitting functions $v_i^{\rm soft}$ and $v_j^{{\rm soft}\,*}$ as defined in Eq.~(\ref{eq:VsoftF}), summed over $\hat s_i$ and $\hat s_j$ and averaged over $s_i$ and $s_j$. That is 
\begin{equation}
\bra{\{s'\}_{m}}
w_{ij}(\{\hat f,\hat p\}_{m+1})
\ket{\{s\}_{m}}
=
\brax{\{s'\}_m}\ket{\{s\}_m}\,
\overline w_{ij}(\{\hat f,\hat p\}_{m+1})
\;\;,
\end{equation}
where $\overline w_{ij}$ is
\begin{equation}
\begin{split}
\label{eq:wijef}
\overline w_{ij}&(\{\hat f,\hat p\}_{m+1})
= \overline w_{ji}(\{\hat f,\hat p\}_{m+1})
\\&=
\frac{1}{4}\,\sum_{\hat s_{m+1}}\sum_{\hat s_{i}}\sum_{s_{i}}
\sum_{\hat s_{j}}\sum_{s_{j}}
v_j^{\rm soft}(\{\hat p, \hat f\}_{m+1},\hat s_{m+1},\hat s_{j},s_j)^*\,
v_i^{\rm soft}(\{\hat p, \hat f\}_{m+1},\hat s_{m+1},\hat s_{i},s_i)
\;\;.
\end{split}
\end{equation}

Consider next the case $i = j = l \in \{1,\dots,m\}$ with $\hat f_l \ne {\rm g}$ or with $l = {\rm a}$ or ${\rm b}$ and any $\hat f_l$. Using the definition (\ref{eq:Wlldef}) of ${\cal W}(l,l;\{\hat f,\hat p\}_{m+1})$ and orthogonality for the spin basis vectors, this is
\begin{equation}
\begin{split}
\label{eq:wll}
\bra{\{s'\}_{m}}
w_{ll}(\{\hat f,\hat p\}_{m+1})
\ket{\{s\}_{m}}
={}&
\sum_{\{\hat s\}_{m+1}}
S_l(\{\hat f\}_{m+1})\
\bra{\{s'\}_{m}}
V_l(\{\hat p, \hat f\}_{m+1})
\ket{\{\hat s\}_{m+1}}
\\&\times
\bra{\{\hat s\}_{m+1}}
V^{\dagger}_l(\{\hat p, \hat f\}_{m+1})
\ket{\{s\}_{m}}
\;\;.
\end{split}
\end{equation}
Using the definition of $V$ as given in Eq.~(\ref{eq:Vtov}), this is
\begin{equation}
\begin{split}
\bra{\{s'\}_{m}}
w_{ll}(\{\hat f,\hat p\}_{m+1})
\ket{\{s\}_{m}}
={}&
\sum_{\{\hat s\}_{m+1}}
\Biggl(\,\prod_{n\notin\{l,m+1\}} \delta_{\hat s_n,s_n}
\delta_{\hat s_n,s'_n}\Biggr)
S_l(\{\hat f\}_{m+1})
\\&\times
v_l^*(\{\hat p, \hat f\}_{m+1},\hat s_{m+1},\hat s_{l},s'_l)\,
v_l(\{\hat p, \hat f\}_{m+1},\hat s_{m+1},\hat s_{l},s_l)
\\={}&
\sum_{\hat s_{m+1}}\sum_{\hat s_{l}}
\Biggl(\,\prod_{n\notin\{l,m+1\}} \delta_{s'_n,s_n}
\Biggr)
S_l(\{\hat f\}_{m+1})
\\&\times
v_l^*(\{\hat p, \hat f\}_{m+1},\hat s_{m+1},\hat s_{l},s'_l)\,
v_l(\{\hat p, \hat f\}_{m+1},\hat s_{m+1},\hat s_{l},s_l)
\;\;.
\end{split}
\end{equation}
This is proportional to the unit matrix in the spin indices $s_n,s'_n$ for all $n$ except possibly for $s_l,s'_l$, the indices that appear in the functions $v_l$. However, in Eq.~(\ref{eq:HI1together}) there is an integration over the azimuthal angle $\phi$ as part of the integration over $\zeta_{\rm p}$. Once we sum over the final state spin indices $\hat s_l, \hat s_{m+1}$ and integrate over $\phi$, the result is invariant under rotations about the $p_l$ axis in the rest frame of $Q$. Thus the result must vanish for $s_l \ne s'_l$. In addition, the parity invariance of the splitting vertices implies that the result is invariant under a reflection through a plane containing $p_l$, $Q$, and any vector transverse to $p_l$ and $Q$. Under this transformation, $s_l \leftrightarrow - s_l$ and $s'_l \leftrightarrow - s'_l$. Thus the result is proportional to $\delta_{s_l,s'_l}$. The coefficient of the $\delta_{s_l,s'_l}$ can be obtained by setting $s'_l \to s_l$, summing over $s_l$, and multiplying by 1/2. Thus
\begin{equation}
\begin{split}
\int\frac{d\phi}{2\pi}\
\bra{\{s'\}_{m}}&
w_{ll}(\{\hat f,\hat p\}_{m+1})
\ket{\{s\}_{m}}
\\={}&
\int\frac{d\phi}{2\pi}
\left(\prod_{n=1}^m \delta_{s'_n,s_n}
\right)
\frac{1}{2}\,\sum_{\hat s_{m+1}}\sum_{\hat s_{l}}\sum_{\bar s_{l}}
S_l(\{\hat f\}_{m+1})
\\&\times
v_l^*(\{\hat p, \hat f\}_{m+1},\hat s_{m+1},\hat s_{l},\bar s_l)\,
v_l(\{\hat p, \hat f\}_{m+1},\hat s_{m+1},\hat s_{l},\bar s_l)
\\={}&
\brax{\{s'\}_m}\ket{\{s\}_m}\
S_l(\{\hat f\}_{m+1})
\\&\times
\frac{1}{2}\,\sum_{\hat s_{m+1}}\sum_{\hat s_{l}}\sum_{\bar s_{l}}
\big|v_l(\{\hat p, \hat f\}_{m+1},\hat s_{m+1},\hat s_{l},\bar s_l)\big|^2
\;\;.
\end{split}
\end{equation}
In the last line, we have not written the average over $\phi$ because, once we have summed over all of the spins, the result is independent of $\phi$.

For the special case $i = j = l \in \{1,\dots,m\}$ with $\hat f_l = {\rm g}$, there is an extra term in the definition (\ref{eq:Wlldef}) of ${\cal W}(l,l;\{\hat f,\hat p\}_{m+1})$. The extra term, involving $\widetilde {\cal W}$ defined in Eq.~(\ref{eq:tildeWdef}), is built from some of the separate terms in the three gluon vertex. Their treatment is essentially the same as the treatment just given for the other vertex functions. The result for $i = j = l$ in general is
\begin{equation}
\int\frac{d\phi}{2\pi}\ \bra{\{s'\}_{m}}
w_{ll}(\{\hat f,\hat p\}_{m+1})
\ket{\{s\}_{m}}
=
\brax{\{s'\}_m}\ket{\{s\}_m}\,
\overline w_{ll}(\{\hat f,\hat p\}_{m+1})
\;\;,
\end{equation}
where
\begin{equation}
\begin{split}
\label{eq:wlldef}
\overline w_{ll}(\{\hat f,\hat p\}_{m+1})={}&
S_l(\{\hat f\}_{m+1})\
\frac{1}{2}\,\sum_{\hat s_{m+1}}\sum_{\hat s_{l}}\sum_{s_{l}}
\\&\times
\bigg\{
\big|v_l(\{\hat p, \hat f\}_{m+1},\hat s_{m+1},\hat s_{l},s_l)\big|^2
\\ & + 
\theta(l \in \{1,\dots,m\},\hat f_l = \hat f_{m+1} = {\rm g})\
\\&\times
\Big[
|v_{2,l}(\{\hat p, \hat f\}_{m+1},\hat s_{m+1},\hat s_{l},s_l)|^2
-
|v_{3,l}(\{\hat p, \hat f\}_{m+1},\hat s_{m+1},\hat s_{l},s_l)|^2\,
\Big]\bigg\}
\;\;.
\end{split}
\end{equation}
Here $v_{2,l}$ and $v_{3,l}$ are defined in Eq.~(\ref{eq:Vgg123}).

We conclude that $\sbra{1}{\cal H}_\LI(t)\sket{\{p,f,s',c',s,c\}_{m}}$ has the form given in Eq.~(\ref{eq:1H}),
\begin{equation}
\label{eq:1Hencore}
\sbra{1}{\cal H}_\LI(t)\sket{\{p,f,s',c',s,c\}_{m}}
= 2\,\brax{\{s'\}_{m}}\ket{\{s\}_{m}}\,
\bra{\{c'\}_{m}}h(t,\{p,f\}_{m})\ket{\{c\}_{m}}
\;\;.
\end{equation}
It is proportional to the unit matrix in spin but is not proportional to the unit matrix, or even diagonal, in color. The matrix $\bra{\{c'\}_{m}}h(t,\{p,f\}_{m})\ket{\{c\}_{m}}$ is
\begin{equation}
\begin{split}
\label{eq:hdef}
\bra{\{c'\}_{m}}&h(t,\{p,f\}_{m})\ket{\{c\}_{m}}
=
\\&
\frac{1}{2}
\sum_l
\sum_{\zeta_{\rm f}\in \Phi_{l}(f_l)}
\int d\zeta_{\rm p}\
\theta(\zeta_{\rm p} \in \varGamma_{l}(\{p\}_{m},\zeta_{\rm f}))
\\&\times
\delta\!\left(
t - \log\left(\frac{Q_0^2}
{|(\hat p_l 
+(-1)^{\delta_{l,\La} + \delta_{l,\Lb}}
 \hat p_{m+1})^2 - m^2(f_l)|}\right)
\right)
\\&\times
\frac
{n_\Lc(a) n_\Lc(b)\,\eta_{\La}\eta_{\Lb}}
{n_\Lc(\hat a) n_\Lc(\hat b)\,
 \hat \eta_{\La}\hat \eta_{\Lb}}\,
\frac{
f_{\hat a/A}(\hat \eta_{\La},\mu^{2}_{F})
f_{\hat b/B}(\hat \eta_{\Lb},\mu^{2}_{F})}
{f_{a/A}(\eta_{\La},\mu^{2}_{F})
f_{b/B}(\eta_{\Lb},\mu^{2}_{F})}
\\ &\times
\biggl\{
\theta(\hat f_{m+1}\ne {\rm g})\
\brax{\{c'\}_{m}}\ket{\{c\}_{m}}\,
T_{\rm R}\
\overline w_{ll}(\{\hat f,\hat p\}_{m+1})
\\ &\ \ +
\theta(\hat f_{m+1} = {\rm g})\sum_{k \ne l}
\bra{\{c'\}_{m}}
g_{lk}(\{\hat f\}_{m+1})
\ket{\{c\}_{m}}
\\& \quad\quad\times
\big[
2\,A_{lk}(\{\hat p\}_{m+1})\,
\overline w_{lk}(\{\hat f,\hat p\}_{m+1})
-
\overline w_{ll}(\{\hat f,\hat p\}_{m+1})
\big]
\bigg\}
\;\;.
\end{split}
\end{equation}
Here $\bra{\{c'\}_{m}} g_{ij}(\{\hat f\}_{m+1}) \ket{\{c\}_{m}}$ is given in Eq.~(\ref{eq:gijdef}) while $\overline w_{ij}(\{\hat f,\hat p\}_{m+1})$ for $i\ne j$ is given in Eq.~(\ref{eq:wijef}) and $\overline w_{ll}(\{\hat f,\hat p\}_{m+1})$ is given in Eq.~(\ref{eq:wlldef}). The scales $\mu_\LF$ used in the parton distribution functions and $\mu_{\rm R}$ used in $\alpha_{\rm s}$ are given by Eq.~(\ref{eq:muFRdef}). Inside the integral and the sum over $l$, the variables $\{\hat p,\hat f\}_{m+1}$ are determined from $\{p,f\}_m$ and the splitting variables $\{\zeta_{\rm p},\zeta_{\rm f}\}$ by the transformation $R_l$, Eqs.~(\ref{eq:Rldef}) and (\ref{eq:Radef}).

Notice that the matrix $\bra{\{c'\}_{m}} h(t,\{p,f\}_{m})\ket{\{c\}_{m}}$ is not diagonal in color. However, the matrix elements with $\{c'\}_{m} \ne \{c'\}_{m}$ are suppressed by powers of $1/N_\Lc$. For the term with a factor $\brax{\{c'\}_{m}}\ket{\{c\}_{m}}$, this was already noted in Eq.~(\ref{eq:offdiagonalcolors}). To see this for the term involving $g_{lk}$, we should write the matrix element of $g_{lk}$ in the form of Eq.~(\ref{eq:gijdef0}),
\begin{equation}
\begin{split}
\label{eq:gijdef1}
\bra{\{c'\}_{m}}
g_{lk}(\{\hat f\}_{m+1})
\ket{\{c\}_{m}}
={}&
\sum_{\{\hat c',\hat c\}_{m+1}}
\dualL\bra{\{\hat c'\}_{m+1}}
t_k^\dagger(f_k \to \hat f_k + \hat f_{m+1})
\ket{\{c'\}_{m}}
\\ &\times
\brax{\{\hat c'\}_{m+1}}\ket{\{\hat c\}_{m+1}}\,
\dualL\bra{\{\hat c\}_{m+1}}
t^\dagger_l(f_l \to \hat f_l + \hat f_{m+1})
\ket{\{c\}_{m}}
\;\;.
\end{split}
\end{equation}
The leading contribution for $1/N_\Lc \to 0$ comes when $\{c'\}_{m} = \{c\}_{m}$ and partons $l$ and $k$ are color connected: they lie next to each other on a string. As discussed in Sec.~\ref{sec:colorstructure}, we can use the representations (\ref{eq:coloropggg}), (\ref{eq:coloropqqg}), (\ref{eq:coloropqbarqbarg}), and (\ref{eq:coloropgqqbar}) for the operators $t_l^\dagger$ and $t_k^\dagger$ to see that the action of $t_l^\dagger$ and $t_k^\dagger$ can produce $\{\hat c'\}_{m+1} = \{\hat c\}_{m+1}$. If $\{c'\}_{m} \ne \{c\}_{m}$, then the action of $t_l^\dagger$ and $t_k^\dagger$ always produces $\{\hat c'\}_{m+1} \ne \{\hat c\}_{m+1}$, so that we get a color suppressed inner product $\brax{\{\hat c'\}_{m+1}}\ket{\{\hat c\}_{m+1}}$.

\section{End of the shower}
\label{sec:ShowerEnd}

As the shower progresses toward smaller and smaller resolution scales $\mu^2 = Q_0^2 e^{-t}$, there must come a point at which the perturbative basis of the evolution equation is no longer valid. Then, at some evolution time $t_{\rm f}$, the shower evolution should be stopped. In the event that the resolution scale of the desired measurement function is larger than $\mu_{\rm f}^2 = Q_0^2 e^{-t_{\rm f}}$, whatever happens beyond that is not seen by the measurement. Then one could simply apply the measurement function, calculating
\begin{equation}
\label{eq:Ftimesrho}
\sbrax{F}\sket{\rho(t_{\rm f})}
\;\;.
\end{equation}
Let us suppose that the measurement function does not see the spins or colors of the final state partons. Then, as in Eq.~(\ref{eq:sigmaFsimple}), we need
\begin{equation}
\label{eq:sigmaFsimple1}
\begin{split}
\sbrax{F}\sket{\rho(t_{\rm f})} ={}&
\sum_m\frac{1}{m!}
\int \big[d\{p,f,s',c',s,c\}_{m}\big]
F(\{p,f\}_{m})\
\brax{\{s'\}_{m}}\ket{\{s\}_{m}}
\brax{\{c'\}_{m}}\ket{\{c\}_{m}}
\\&\times
\sbrax{\{p,f,s',c',s,c\}_{m}}\sket{\rho(t_{\rm f})}
\;\;.
\end{split}
\end{equation}

It is significant that, although the possibility of $\{c'\}_{m} \ne \{c\}_{m}$ is included in the shower evolution presented in this paper, the matrix $\brax{\{c'\}_{m}} \ket{\{c\}_{m}}$ is almost diagonal, with off-diagonal matrix elements being suppressed by factors of $1/N_\Lc^2$. The situations with respect to spins and colors are different. In the end, we must have $\{s'\}_{m} = \{s\}_{m}$ exactly. However, the spins get shuffled at each stage of shower evolution and there is no reason that a state with $\{s'\}_{m} \ne \{s\}_{m}$ at an earlier stage of evolution cannot evolve into a state with $\{s'\}_{m} = \{s\}_{m}$ at the end. On the other hand, color differences between $\{c'\}_{m}$ and $\{c\}_{m}$ are a little like entropy. Once $\{c'\}_{m} \ne \{c\}_{m}$ at an early stage of evolution, we can never get $\{c'\}_{m} = \{c\}_{m}$ at the end. For this reason, the most important part of color evolution is the part that maintains $\{c'\}_{m} = \{c\}_{m}$ throughout. This is evolution in the leading color dipole approximation.

Now, what if we wish to use a measurement function with a resolution scale smaller than $\mu_{\rm f}^2$. Then we need a model for what happens at smaller resolution scales (or later and earlier proper times than given by $x^2 \sim 1/\mu_{\rm f}^2$). Our model should certainly include hadronization. We can easily extend the formalism presented here to encompass hadronization. We have only to replace $\sbrax{F}\sket{\rho(t_{\rm f})}$ by 
\begin{equation}
\label{eq:Ftimesrho1}
\sbra{F_{\rm h}}{\cal U}(\infty,t_{\rm f})\sket{\rho(t_{\rm f})}
\;\;.
\end{equation}
Here ${\cal U}(\infty,t_{\rm f})$ represents a model for what happens after Monte Carlo time $t_{\rm f}$. It starts with partonic states and maps them into the space of hadronic states. Typically the hadronic states are labeled by momenta and hadronic flavors but not spins. Then $\sbra{F_{\rm h}}$ represents the measurement function in the space of hadronic states. For purposes of discussing the partonic shower, we can denote
\begin{equation}
\label{eq:Feff}
\sbra{F_{\rm h}}{\cal U}(\infty,t_{\rm f}) =
\sbra{F_{\rm eff}}
\;\;.
\end{equation}
Thus $F_{\rm eff}$ is the true hadronic measurement function translated back to the partonic level.

Assuming that the hadronization model does not use color or spin information, the measured cross section then takes the form
\begin{equation}
\label{eq:sigmaFsimple2}
\begin{split}
\sbra{F_{\rm h}}{\cal U}(\infty,t_{\rm f})\sket{\rho(t_{\rm f})}
 ={}& \sbrax{F_{\rm eff}}\sket{\rho(t_{\rm f})}
 \\
  ={}&
\sum_m\frac{1}{m!}
\int \big[d\{p,f,s',c',s,c\}_{m}\big]
F_{\rm eff}(\{p,f\}_{m})
\\&\times
\brax{\{s'\}_{m}}\ket{\{s\}_{m}}
\brax{\{c'\}_{m}}\ket{\{c\}_{m}}
\sbrax{\{p,f,s',c',s,c\}_{m}}\sket{\rho(t_{\rm f})}
\;\;.
\end{split}
\end{equation}

Typically, hadronization models {\em do} use color information. The color field interacting with the outgoing partons is represented as a classical color string. The string then fragments into hadrons. This applies directly in \textsc{Pythia} \cite{Pythia} and in a different way in \textsc{Herwig} \cite{Herwig}, where the color strings fragment into color singlet parton clusters immediately. The formalism of this paper is set up with the color string picture in mind. The color states $\{c\}_{m}$ exactly map onto string configurations, as explained in Sec.~\ref{sec:ColorBasis}. Thus for terms in Eq.~(\ref{eq:sigmaFsimple2}) with $\{c'\}_{m} = \{c\}_{m}$, one can use $\{c\}_{m}$ as the input to the hadronization model. The string model does not tell us what to do with $\{c'\}_{m} \ne \{c\}_{m}$. A reasonable suggestion would be to use $\{c'\}_{m}$ half the time and $\{c\}_{m}$ half the time. This could be represented as 
\begin{equation}
\label{eq:sigmaFsimple3}
\begin{split}
\sbra{F_{\rm h}}{\cal U}(\infty,t_{\rm f})\sket{\rho(t_{\rm f})} ={}&
\sum_m\frac{1}{m!}
\int \big[d\{p,f,s',c',s,c\}_{m}\big]
\\&\times
\frac{1}{2}
\Big[
F_{\rm eff}(\{p,f,c\}_{m})
+\
F_{\rm eff}(\{p,f,c'\}_{m})\
\Big]
\\&\times
\brax{\{s'\}_{m}}\ket{\{s\}_{m}}
\brax{\{c'\}_{m}}\ket{\{c\}_{m}}
\sbrax{\{p,f,s',c',s,c\}_{m}}\sket{\rho(t_{\rm f})}
\;\;.
\end{split}
\end{equation}

If the hadronization model is based on strings, one needs to do something with the string ends that connect to the initial state partons. Consider, for example, the case that the hard collision at scale $\mu_{\rm f}^2$ is a quark-quark collision and concentrate on one of the initial state quarks. From the point of view of backwards evolution from the scale $Q_0^2$, this initial state quark appears as a color $\overline {\bf 3}$ line. The initial state quark is part of a colorless hadron, but the net color $\overline {\bf 3}$ is carried by the spectator quarks from this hadron. Thus in a color string model, the color strings from the hard interactions should connect to the spectator partons. Of course, this is not a completely simple problem. In a realistic model, the spectator partons from the two hadrons have many interactions with each other, possibly followed by their own (not very hard) parton showers. After all of these interactions, the spectator partons left over from a quark-quark collision must have $\overline {\bf 3} \otimes \overline {\bf 3}$ color, but its internal color state can be quite complicated. Modeling the spectator interactions is well beyond the scope of this paper but is addressed by Sj\"ostrand and Skands in Ref.~\cite{SjostrandSkands}.
 
\section{Conclusions}
\label{sec:conclusions}

We have presented a formulation of parton showering for hadron-hadron collisions. The prediction for a cross section corresponding to an observable $F$ is given by $\sbrax{F}\sket{\rho(t_{\rm f})}$ or one of the other formulas in Sec.~\ref{sec:ShowerEnd}, depending on the treatment of hadronization, which is not covered in this paper. The dynamics of the quantum density $\sket{\rho(t)}$ is given by the evolution operator ${\cal U}(t,t')$, so that $\sket{\rho(t_{\rm f})} = {\cal U}(t_{\rm f},0)\sket{\rho(0)}$, where $\sket{\rho(0)}$ is determined from the hard matrix element that starts the shower. Thus the shower dynamics represented in ${\cal U}(t,t')$ is based on factorization of soft and collinear singularities from hard scattering. The basic formula is Eq.~(\ref{eq:evolution}), which we can rewrite as
\begin{equation}
\label{eq:evolutionbis}
{\cal U}(t_{\rm f},t') = {\cal N}(t_{\rm f},t')
+ \int_{t'}^{t_{\rm f}}\! d\tau\ 
{\cal U}(t_{\rm f},\tau)\,
[{\cal H}_{\LI}(\tau)
- {\cal V}_{\LS}(\tau)]
\,{\cal N}(\tau,t')
\;\;.
\end{equation}
Here ${\cal H}_\LI(t)$ is a parton splitting operator, as defined in the preceding sections and ${\cal V}_\LS(t)$ represents a virtual interaction that interchanges colors. The operator ${\cal N}(\tau,t')$ generates the standard sort of Sudakov exponential that gives the probability not to have an interaction between shower times $t'$ and $\tau$. We provide Table~\ref{tab:KeyFormulas} to indicate where the various functions needed to compute ${\cal U}$ can be found.

\TABLE{
\begin{tabular}{lll}
\hline
quantity & equation & uses
\\ [3 pt]
${\cal U}(t,t')$ &
Eq.~(\ref{eq:evolution}) &
${\cal H}_{\LI}$,\ 
${\cal N}$,\
${\cal V}_{\LS}$
\\
${\cal H}_{\LI}(\tau)$ &
Eq.~(\ref{eq:Hdef}) &
${\cal S}_{l}$
\\
${\cal V}(t,\{p,f\}_{m})$ &
Eq.~(\ref{eq:calVdef}) &
$h$
\\
${\cal V}_{\LS}(t,\{p,f\}_{m})$ &
Eq.~(\ref{eq:VS}) &
${\cal V}$,\ ${\cal V}_{\LE}$
\\
${\cal V}_{\LE}(t,\{p,f\}_{m})$ &
Eq.~(\ref{eq:VEdef}) &
$h$
\\
${\cal N}(t, t')$ &
Eq.~(\ref{eq:calN}) &
${\cal V}_{\LE}$
\\
$h(t,\{p,f\}_{m})$ &
Eq.~(\ref{eq:hdef}) &
$g_{ij}$,\
$\overline w_{lk}$,
$A_{lk}$
\\ $g_{ij}(\{\hat f\}_{m+1})$ &
Eq.~(\ref{eq:gijdef0}) &
$t^\dagger_l$,\
color basis
\\
$\overline w_{ij}(\{\hat f,\hat p\}_{m+1})$ &
Eqs.~(\ref{eq:wijef},\,\ref{eq:wlldef}) &
$v^{\rm soft}_i$, $v_l$, $v_{J,l}$
\\
$t^\dagger_l(f_j \to \hat f_j + \hat f_{m+1})$ &
Eqs.~(\ref{eq:coloropggg}-\ref{eq:coloropqbarqbarg},\,\ref{eq:coloropgqqbar}) &
$a_+^\dagger$,\
$a_-^\dagger$,\
$a^\dagger_q$,\
$a^\dagger_s$
\\
$a^\dagger_+$,\ $a^\dagger_-$,\ $a^\dagger_q$,\ $a^\dagger_s$ &
Eq.~(\ref{eq:aplusdef}-\ref{eq:asdef}) &
color basis
\\
color basis &
Sec.~\ref{sec:ColorBasis} &
\
\\
${\cal S}_l$ &
Eq.~(\ref{eq:calSl}) &
${\cal S}^{(l)}_{ij}$, $A_{lk}$\
\\
${\cal S}^{(l)}_{ij}$ &
Eq.~(\ref{eq:Sijdef}) &
${\cal G}$,\
${\cal W}$,\
${\cal P}_{l}$
\\
${\cal G}(i,j;\{\hat f\}_{m+1})$ &
Eq.~(\ref{eq:Gijdef}) &
$t^\dagger_i$,\
color basis
\\
${\cal W}{}(l,l;\{\hat f,\hat p\}_{m+1})$ &
Eq.~(\ref{eq:Wlldef}) &
$S_l$,\
$V^{\dagger}_l$,\
$\widetilde{\cal W}$
\\
${\cal W}(i,j;\{\hat f,\hat p\}_{m+1})$ &
Eq.~(\ref{eq:Wijdef}) &
$V^{\dagger,{\rm soft}}_i$
\\
$\widetilde{\cal W}(l,l;\{\hat p\}_{m+1})$ &
Eq.~(\ref{eq:tildeWdef}) &
$V_{J,l}$
\\
$V^{\dagger}_{J,l}(\{\hat p, \hat f\}_{m+1})$ &
Eq.~(\ref{eq:Vtov123}) &
$v_{J,l}$
\\
$v_{J,l}(\{\hat p, \hat f\}_{m+1},\hat s_{m+1},\hat s_{l},s_l)$ &
Eq.~(\ref{eq:Vgg123}) &
\
\\
$S_l(\{\hat f\}_{m+1})$ &
Eq.~(\ref{eq:Sldef}) &
\
\\
$V^\dagger_l(\{\hat p, \hat f\}_{m+1})$ &
Eq.~(\ref{eq:Vtov}) &
$v_l$
\\
$v_l(\{\hat p, \hat f\}_{m+1},\hat s_{m+1},\hat s_{l},s_l)$ &
Table~\ref{tab:splitting}&
\
\\
 &
Eqs.~(\ref{eq:VggF},\,\ref{eq:VggI}) &
\
\\
$V_l^{\dagger,{\rm soft}}(\{\hat p, \hat f\}_{m+1})$ &
Eq.~(\ref{eq:Vtovsoft}) &
$v_l^{\rm soft}$
\\
$v_l^{\rm soft}(\{\hat p, \hat f\}_{m+1},\hat s_{m+1},\hat s_{l},s_l)$ &
Eq.~(\ref{eq:VsoftF}) &
\\
${\cal P}_{l}$ &
Eq.~(\ref{eq:calPdef1}) &
$Q_l$
\\
$Q_l(\{\hat p,\hat f\}_{m+1})$ &
Sec.~\ref{sec:mapping} &
\
\\
$\int \big[d\{p,f\}_{m}\big]$ &
Eq.~(\ref{eq:pfmeasure}) &
\\
$A_{lk}(\{p\}_{m+1})$ &
Sec.~\ref{sec:ClassicalEvolution} &
\
\\ [3 pt]
\hline
\end{tabular}
\caption{Key formulas.}
\label{tab:KeyFormulas}
}

Eq.~(\ref{eq:evolutionbis}) has the proper form to conveniently generate a parton shower. Starting with a state at time $t'$, one would use ${\cal N}(\tau,t')$ to determine the time $\tau$ for a parton splitting (or color rearrangement). Possibly there is no splitting before the cutoff time $t_{\rm f}$, as represented by the first term. Otherwise, the operators ${\cal H}_{\LI}(\tau)$ and ${\cal V}_{\LS}(\tau)$ give a new partonic state at time $\tau$.  Now we operate with ${\cal U}(t_{\rm f},\tau)$, which is to say that we apply this procedure again.

We offer here concluding remarks under several headings.

\paragraph{Implementation.}
Equation~(\ref{eq:evolutionbis}), together with the formulas in Sec.~\ref{sec:ShowerEnd} and the definitions given throughout this paper, represents a certain approximation for the cross section $\sigma[F]$ corresponding to a given observable $F$. When Eq.~(\ref{eq:evolutionbis}) is iterated, the result is expressed in the form of certain integrals and sums. It will be a significant challenge to find ways to implement Eq.~(\ref{eq:evolutionbis}) in a manner that allows an efficient calculation of $\sigma[F]$. We expect that there is more than one way to attack this problem. The choice affects the efficiency of calculation, but not the result, $\sigma[F]$.  We leave implementation issues to later work. 

\paragraph{Evolution variable.} 
We have chosen the evolution variable $t$ to be proportional to the virtuality in the splitting. An alternative would be the transverse momentum in the splitting. The \textsc{Herwig} choice of the splitting angle does not work well with the formalism presented here since a parton emitted at a fixed angle can be arbitrarily soft, necessitating introducing the final hardness cutoff $\mu_{\rm f}^2$ at each splitting.

\paragraph{Momentum mapping.}
It is not kinematically possible for an on-shell parton to split into two on-shell partons. However, it is useful to approximate the mother parton as being exactly on-shell in calculating the (relatively) hard scattering in which the mother parton participates. To make this approximation, we need to take the needed momentum from somewhere else. Sometimes this is done by taking momentum from the mother's sister in the previous splitting. In the Catani-Seymour subtraction scheme for doing next-to-leading order calculations, there is a ``spectator parton'' that donates the required momentum \cite{CataniSeymour}. Rather than taking momentum from a single parton that might not have much to give, we have chosen to take a little momentum from each final state parton, with each donating according to how much momentum it has. Note that we keep momentum in balance at each step, rather than waiting until the end of the shower to make adjustments.

\paragraph{Common evolution.}
In Eq.~(\ref{eq:evolutionbis}), each parton has a chance to split or interchange colors between shower times $t$ and $t+dt$. In the very simplest form of a parton shower, each parton could evolve independently, at least if one ignores momentum conservation and adjusts the momenta only at the end. Then the complete evolution of each parton $l$ could be traced out without keeping track of the other partons. Effectively, there could be a separate time $t_l$ for each parton. However, independent evolution could still be implemented using a common evolution variable $t$, giving each parton its chance to split between times $t$ and $t + dt$. The physical distinction that characterizes independent evolution is that the evolution is independent if the various functions involved in the splitting of parton $l$ do not involve the states of the other partons. With the definitions of the operators used in Eq.~(\ref{eq:evolutionbis}), the use of a common shower time variable is {\em required}, since each splitting changes the whole partonic state and affects the probabilities for other partons to split at later shower times.

\paragraph{Other choices.} 
Within Eq.~(\ref{eq:evolutionbis}), there are quite a number of other choices required. For instance, the splitting functions must have a particular form in the limits of soft and collinear splittings. However, away from these limits there is a certain freedom to choose. Where choices like this were needed, we exercised the freedom to choose based mostly on conceptual simplicity. Other authors might choose differently.

\paragraph{Interference and angular ordering.} 
In Eq.~(\ref{eq:evolutionbis}), quantum interference between emissions of a gluon from different partons is treated exactly in the soft gluon limit. Suppression of wide angle emissions is a result. Typically, parton shower generators make an ``angular ordering'' approximation to this result. Once one has an implementation of Eq.~(\ref{eq:evolutionbis}), it will be interesting to make the standard angular ordering approximation and see how good an approximation it is.

\paragraph{Spin.} 
Partons produced in a hard scattering carry spin, which can affect the angular distribution of their subsequent splittings. Eq.~(\ref{eq:evolutionbis}) includes the full spin information. Typically, parton shower generators average over spins, thus discarding this information. Once one has an implementation of Eq.~(\ref{eq:evolutionbis}), it will be interesting to insert spin averages everywhere and see how good an approximation it is.

\paragraph{Color.} 
Partons produced in a hard scattering carry color, which can affect the pattern of future splittings. The formalism presented here includes the full color information. Typically, parton shower generators make use of a leading color approximation that amounts to taking the first term in an expansion about $1/N_{\rm c} = 0$. Once one has an implementation of Eq.~(\ref{eq:evolutionbis}), it will be interesting to make the leading color approximation everywhere and see how good an approximation it is.

\paragraph{Improvements needed.} 
We leave for future work the question of how one could match the parton shower to the exact matrix elements for $2 \to n$ scattering instead of simply starting with $2 \to 2$ scattering. We also leave for future work the question of how one could do this at next-to-leading order. (See, however, Ref.~\cite{NagySoper}.) More ambitiously, we would like to extend the whole formalism, including the splitting functions, to next-to-leading order. Current work by others on soft-collinear effective theory may be helpful here \cite{SCET}.

\paragraph{Masses.} We gave included quark masses in our momentum mappings and splitting functions. However, there are a number of issues associated with masses that we do not address. Suppose that we start with a hard scattering at a scale $Q_0$ that is much larger than the mass of the bottom quark, $m_\Lb$.\footnote{Here we assume that the bottom quark is the heaviest quark counted as initial state parton. One could include production of top quarks, but would not include the top quark as a constituent of the proton unless $Q_0 \gg m_{\rm t}$.} Suppose additionally that we want to continue evolution down to a scale $\mu$ that is less than $m_{\rm b}$. Then we need a suitable variable flavor number scheme. At the leading order used in this paper, this is easy enough, but at higher orders of perturbation theory there are some subtle issues. Collins \cite{JCCheavyquarks} has addressed some of these issues as they arise in deeply inelastic scattering. One should, however, note that for hadron-hadron collisions, the power suppressed terms that are omitted when the cross section is written in a factored form as a hard scattering function convoluted with parton distributions are not of order $\Lambda^2_{\rm QCD}/Q_0^2$ but rather of order $m_\Lb^2/Q_0^2$ \cite{FrenkelTaylor}. This is not a problem since, in the applications we have in mind, $m_\Lb^2/Q_0^2 \ll 1$. We lack a theorem to tell us what to do if we want to keep the hard scattering but lower the resolution scale to $\mu^2 < m_\Lb^2$ so as to examine the final state in more detail.

\paragraph{Foundations.}
The formalism presented here is based on ideas of factorization, both at the amplitude level and at the cross section level, where summation over partonic states that are unresolved at a given scale $\mu^2$ is essential. As the discussion of masses makes clear, more work is needed to make these ideas sufficiently precise to justify the formalism.


\acknowledgments{We are grateful to J.~Collins, M.~Seymour, P.~Skands, and Z.~Tr\'ocs\'anyi for helpful conversations. This work was supported in part the United States Department of Energy and by the Swiss National Science Foundation (SNF) through grant no.\ 200020-109162 and by the Hungarian Scientific Research Fund grants OTKA T-60432.
}

\appendix

\section{Limit on momentum fraction after splitting}
\label{sec:etalimit}

In Sec.~(\ref{sec:ISsplittingkinematics}) we defined the kinematics of initial state splitting. For a collinear splitting from an initial state parton, say parton ``a,'', we have $\hat \eta_\La > \eta_\La$. It is not exactly evident that this holds in away from the collinear limit, especially with masses. Here we show that this holds under the kinematic conditions (\ref{eq:Q0condition}) and (\ref{eq:etalimits}).

Let
\begin{equation}
\begin{split}
\hat g(\hat\eta_\La) ={}& (\hat p_\La + p_\Lb)^2
\\
={}& \hat\eta_a \eta_\Lb s + m^2(\hat f_\La) + m^2(f_\Lb) 
+ \frac{m^2(\hat f_\La)\, m^2(f_\Lb)}{\hat\eta_a \eta_\Lb s}
\;\;,
\\
g(\eta_\La) ={}& (p_\La + p_\Lb)^2
\\
={}& \eta_a \eta_\Lb s + m^2(f_\La) + m^2(f_\Lb) 
+ \frac{m^2(f_\La)\, m^2(f_\Lb)}{\eta_\La \eta_\Lb s}
\;\;.
\end{split}
\end{equation}
Note that $\hat g(\hat\eta_\La)$ is an increasing function of its argument in the allowed region of momentum fractions:
\begin{equation}
\frac{1}{\eta_b s}\ \frac{d\hat g(\hat\eta_\La)}{d\hat\eta_\La}
= 1 - \left(\frac{m(\hat f_\La)\, m(f_\Lb)}{\hat\eta_a \eta_\Lb s}\right)^2
> 0
\;\;,
\end{equation}
since $\hat\eta_a \eta_\Lb s > m_{\rm H}^2$ in our allowed kinematic region according to Eq.~(\ref{eq:etalimits}). Since $\hat g(\hat\eta_\La)$ is an increasing function, we just need to show that $\hat g(\hat\eta_\La) > \hat g(\eta_\La)$.

Recall that the kinematics requires 
\begin{equation}
\begin{split}
\hat g(\hat\eta_\La) ={}& (\hat K + p_{m+1})^2
\;\;,
\\
g(\eta_\La) ={}& K^2
\;\;,
\end{split}
\end{equation}
where $\hat K^2 = K^2$. Thus
\begin{equation}
\begin{split}
\label{eq:gdifference}
\hat g(\hat\eta_\La) - \hat g(\eta_\La) ={}& 
\hat g(\hat\eta_\La) -  g(\eta_\La) +  g(\eta_\La) - \hat g(\eta_\La)
\\
={}& (\hat K + p_{m+1})^2 - K^2 +  g(\eta_\La) - \hat g(\eta_\La)
\\
={}& 2 \hat K \cdot p_{m+1} + m^2(\hat f_{m+1}) 
\\ & 
+  \left(m^2(f_\La) -  m^2(\hat f_\La)\right)
\left[ 1 +
\frac{ m^2(f_\Lb)}{\eta_\La \eta_\Lb s}
\right]
\;\;.
\end{split}
\end{equation}
We need to show that the right hand side of Eq.~(\ref{eq:gdifference}) is positive. If $m(f_\La)^2 \ge m(\hat f_\La)^2$, this is evident. There is only one case in which $m^2(f_\La) < m^2(\hat f_\La)$. That is when we have a $q \to q g$ or $\bar q \to \bar q g$ splitting in which the quark (or antiquark) enters the final state and the gluon enters the hard scattering. Then $m^2(\hat f_{m+1}) = m^2(\hat f_{\La})$ and $m^2(f_\La) = 0$. In that case,
\begin{equation}
\begin{split}
\label{eq:gdifference2}
\hat g(\hat\eta_\La) - \hat g(\eta_\La) 
={}& 2 \hat K \cdot p_{m+1} -  m^2(\hat f_\La)\
\frac{ m^2(f_\Lb)}{\eta_\La \eta_\Lb s}
\\
>{}&
 2 \hat K \cdot p_{m+1} -  m^2(\hat f_\La)
\;\;,
\end{split}
\end{equation}
where we have used $\eta_a \eta_\Lb s > m^2(f_\Lb)$ from Eq.~(\ref{eq:etalimits}). The minimum value of $\hat K \cdot p_{m+1}$ occurs when $p_{m+1}$ is proportional to $\hat K$. Then $\hat K \cdot p_{m+1} = m(\hat f_\La) \sqrt {K^2}$, so 
\begin{equation}
\begin{split}
\label{eq:gdifference3}
\hat g(\hat\eta_\La) - \hat g(\eta_\La) 
 >{}&
 m(\hat f_\La)\
 \left(
 2 \sqrt {K^2} -  m(\hat f_\La)\right)
 > m(\hat f_\La) \
 \left(
 2 \sqrt {Q_0^2} -  m(\hat f_\La)\right)
\;\;.
\end{split}
\end{equation}
This is positive as long as the condition (\ref{eq:Q0condition}) holds.

We have seen that 
\begin{equation}
\hat g(\hat\eta_\La) > \hat g(\eta_\La)
\;\;.
\end{equation}
Since $\hat g(\hat\eta_\La)$ is an increasing function of $\hat \eta_\La$, this implies that $\hat\eta_\La > \eta_\La$.

\section{Counting factors for the density matrix}
\label{sec:symmetryfactors}

In this appendix, we organize the singular contributions to the density matrix starting with the quantum amplitudes defined to be symmetric in the labels of the final state partons (or antisymmetric in the case of identical fermions). Then we introduce the relabelings that define our labeling scheme for parton splittings. This produces the counting factors $S_l(\{\hat f\}_{m+1})$ defined in Eq.~(\ref{eq:Sldef}). The counting factors are related to our parton labeling choices, which in turn related to the singularities of the amplitude in the limit in which masses can be neglected. The counting factors do not depend on parton masses. In order to keep our notation simple, in this appendix we simply take all the parton masses to vanish.

Let $\ket{\ME(\{\hat p, \hat f\}_{m+1})}$ be the exact tree level matrix element for a final state of $m+1$ partons, defined to be symmetric under the interchange of the labels for any two of the final state partons, or antisymmetric if the two partons are identical fermions. The matrix element may have a singularity when any of the dot products of two parton momenta, $\hat p_i\cdot \hat p_j$, approaches zero. Let us define approximate matrix elements $\ket{\ME(\{\hat p, \hat f\}_{m+1};i,j)}$ that approximate the complete matrix element when $\hat p_i\cdot \hat p_j$, approaches zero. Here $i \in \{\La,\Lb\}$ and $j \in \{1,\dots,m+1\}$ or $i,j \in \{1,\dots,m+1\}$ with $i < j$. There is more than one way to do this. We can, for instance, use the method of Sec.~\ref{sec:QuantumSpltting}. The approximate matrix elements $\ket{\ME(\{\hat p, \hat f\}_{m+1};i,j)}$ thus defined should include a theta function 
\begin{equation}
\theta\Biggl(|\hat p_i\!\cdot\! \hat p_j| 
< \min_{\substack{\{k,l\} \ne \{i,j\}\\ \{k,l\}\ne \{j,i\}}}
|\hat p_k\!\cdot\! \hat p_l|\Biggr)
\end{equation}
so that $\ket{\ME(\{\hat p, \hat f\}_{m+1};i,j)}$ is not singular when $p_k\cdot p_l$ for some other pair of partons approaches zero. Given the approximate matrix elements, the complete matrix element can be written as
\begin{equation}
\begin{split}
\label{eq:Mcomplete0}
\ket{\ME(\{\hat p, \hat f\}_{m+1})} \sim{}&
\sum_{j=1}^{m+1} \ket{\ME(\{\hat p, \hat f\}_{m+1};\La,j)}
+ \sum_{j=1}^{m+1} \ket{\ME(\{\hat p, \hat f\}_{m+1};\Lb,j)}
\\ &
+
\sum_{\substack{i,j = 1\\i < j}}^{m+1}
\ket{\ME(\{\hat p, \hat f\}_{m+1};i,j)}
\;\;.
\end{split}
\end{equation}
The right hand side of Eq.~(\ref{eq:Mcomplete0}) approximates the complete matrix element in any of the singular limits. For our purposes, it is convenient to define
\begin{equation}
\label{eq:ijsymmetry}
\ket{\ME(\{\hat p, \hat f\}_{m+1};i,j)} =
\ket{\ME(\{\hat p, \hat f\}_{m+1};j,i)}
\end{equation}
for $i,j \in \{1,\dots,m+1\}$, $i > j$. Then we can symmetrize the third term in Eq.~(\ref{eq:Mcomplete0}),
\begin{equation}
\begin{split}
\label{eq:Mcomplete}
\ket{\ME(\{\hat p, \hat f\}_{m+1})} \sim{}&
\sum_{j=1}^m \ket{\ME(\{\hat p, \hat f\}_{m+1};\La,j)}
+ \sum_{j=1}^m \ket{\ME(\{\hat p, \hat f\}_{m+1};\Lb,j)}
\\ &
+
\frac{1}{2}
\sum_{\substack{i,j = 1\\i\ne j}}^{m+1}
\ket{\ME(\{\hat p, \hat f\}_{m+1};i,j)}
\;\;.
\end{split}
\end{equation}

Now we construct the density operator from
\begin{equation}
\rho_0(\{\hat p,\hat f\}_{m+1}) = \ket{\ME(\{\hat p, \hat f\}_{m+1})}
\bra{\ME(\{\hat p, \hat f\}_{m+1})}
\;\;.
\end{equation}
Imagine expanding both $\ket{\ME(\{\hat p, \hat f\}_{m+1})}$ and 
$\bra{\ME(\{\hat p, \hat f\}_{m+1})}$ according to Eq.~(\ref{eq:Mcomplete}). There are a number of terms, with the general form
\begin{equation}
{\it const.}\times
\ket{\ME(\{\hat p, \hat f\}_{m+1};i,j)}
\bra{\ME(\{\hat p, \hat f\}_{m+1};i',j')}
\;\;.
\end{equation}
Not all of these contributions have soft or collinear singularities strong enough to produce a logarithmic divergence if one were to integrate over the momenta $\{\hat p\}_{m+1}$. 

One contribution that does have a strong enough singularity comes when $i,j,i',j'\in \{1,\dots,m+1\}$ and $i' = i$, $j' = j$ or $i' = j$, $j' = i$. The sum of these contributions is 
\begin{equation}
\label{eq:finalsplit}
\frac{1}{2}\sum_{\substack{i,j = 1\\i\ne j}}^{m+1}
\ket{\ME(\{\hat p, \hat f\}_{m+1};i,j)}
\bra{\ME(\{\hat p, \hat f\}_{m+1};i,j)}
\;\;,
\end{equation}
where we have used Eq.~(\ref{eq:ijsymmetry}). 

Another contribution that has a strong enough singularity comes when $i = i' = \La$ and $j,j'\in \{1,\dots,m+1\}$ with $j' = j$. Similarly, we can have $i = i' = \Lb$ and $j,j'\in \{1,\dots,m+1\}$ with $j' = j$. The sum of these contributions is 
\begin{equation}
\begin{split}
\label{eq:initialsplit}
& \sum_{j=1}^{m+1}
\ket{\ME(\{\hat p, \hat f\}_{m+1};\La,j)}
\bra{\ME(\{\hat p, \hat f\}_{m+1};\La,j)}
\\& +
\sum_{j=1}^{m+1}
\ket{\ME(\{\hat p, \hat f\}_{m+1};\Lb,j)}
\bra{\ME(\{\hat p, \hat f\}_{m+1};\Lb,j)}
\;\;.
\end{split}
\end{equation}

Another contribution that does have a strong enough singularity comes when $i,j,i',j' \in \{1,\dots,m+1\}$ and $i' = i$, $j' \ne j$ or else $j' = j$, $i' \ne i$ or else $i' = j$, $j' \ne i$ or else $j' = i$, $i' \ne j$. These four cases are really the same, with different labeling. Adding the contributions, we have
\begin{equation}
\label{eq:finalinterference}
\sum_{\substack{i,j = 1\\i\ne j}}^{m+1}
\sum_{\substack{k = 1\\ k\ne i\,,\,k \ne j}}^{m+1}
\ket{\ME(\{\hat p, \hat f\}_{m+1};i,k)}
\bra{\ME(\{\hat p, \hat f\}_{m+1};j,k)}
\;\;.
\end{equation}
Here we have a leading singularity only if $\hat f_k = {\rm g}$. This is the interference between gluon emission from line $i$ and gluon emission from line $j$.

We can also have interference between gluon emission from an initial state line and gluon emission from a final state line or between gluon emission from one of the initial state lines and gluon emission from the other. When we add all of these cases together and add them to the contribution in Eq.~(\ref{eq:finalinterference}), we get
\begin{equation}
\label{eq:allinterference}
\sum_{\substack{i,j \in \{\La,\Lb,\dots,m+1\}\\i\ne j}}
\sum_{\substack{k = 1\\ k\ne i\,,\,k \ne j}}^{m+1}
\ket{\ME(\{\hat p, \hat f\}_{m+1};i,k)}
\bra{\ME(\{\hat p, \hat f\}_{m+1};j,k)}
\;\;.
\end{equation}

There are no more combinations of $i,j,i',j'$ that give leading singular contributions to $\rho$. Thus the sum of the leading singular contributions is obtained by adding the contributions (\ref{eq:finalsplit}), (\ref{eq:initialsplit}), and (\ref{eq:allinterference}).

This formulation is fine for constructing an inverse shower, starting from a state with many partons and combining partons to reach a hard scattering with fewer partons. In order to construct a shower starting from the hard scattering, it is convenient to adopt a labeling convention in which $\rho(\{\hat p,\hat f\}_{m+1})$ is not symmetric under interchanges of the parton labels. This is easy to do.

Consider the contribution (\ref{eq:finalsplit}). Using the $i \leftrightarrow j$ symmetry of the matrix elements, this is
\begin{equation}
\begin{split}
&\sum_{i = 1}^{m}
\ket{\ME(\{\hat p, \hat f\}_{m+1};i,m+1)}
\bra{\ME(\{\hat p, \hat f\}_{m+1};i,m+1)}
\\&+
\frac{1}{2}
\sum_{i = 1}^m
\sum_{\substack{j = 1\\j\ne i}}^{m}
\ket{\ME(\{\hat p, \hat f\}_{m+1};i,j)}
\bra{\ME(\{\hat p, \hat f\}_{m+1};i,j)}
\;\;.
\end{split}
\end{equation}
We can now deliberately break the relabeling symmetry in the second term by interchanging the labels $j$ and $m+1$. That is, we choose to label the daughter parton that here caries the label $j$ by $m+1$ instead. Note that this changes $\rho(\{\hat p,\hat f\}_{m+1})$ for any fixed value of $\{\hat p,\hat f\}_{m+1}$. However, the result of integrating $\rho(\{\hat p,\hat f\}_{m+1})$ against any measurement function (which must be symmetric under label interchanges) stays the same. After the interchange $j \leftrightarrow m+1$, the sum over $j$ in the second term simply becomes a factor $m-1$. This gives
\begin{equation}
\begin{split}
&\sum_{i = 1}^{m}
\ket{\ME(\{\hat p, \hat f\}_{m+1};i,m+1)}
\bra{\ME(\{\hat p, \hat f\}_{m+1};i,m+1)}
\\&+
\frac{m-1}{2}
\sum_{i = 1}^m
\ket{\ME(\{\hat p, \hat f\}_{m+1};i,m+1)}
\bra{\ME(\{\hat p, \hat f\}_{m+1};i,m+1)}
\;\;.
\end{split}
\end{equation}
This is
\begin{equation}
\frac{m+1}{2}
\sum_{i = 1}^{m}
\ket{\ME(\{\hat p, \hat f\}_{m+1};i,m+1)}
\bra{\ME(\{\hat p, \hat f\}_{m+1};i,m+1)}
;\;.
\end{equation}
We now consider the possibilities for flavors: $\{\hat f_i,\hat f_{m+1}\}$ could be $\{{\rm g},{\rm g}\}$, $\{{\rm g},q\}$, $\{{\rm g},\bar q\}$, $\{q,{\rm g}\}$, $\{\bar q,{\rm g}\}$, $\{q, \bar q\}$, and $\{\bar q,q\}$, where $q$ stands for a quark flavor, $\bar q$ stands for an antiquark flavor, and a $q$ together with a $\bar q$ stands for a quark flavor and its corresponding antiquark flavor. Thus we can insert a factor
\begin{equation}
\begin{split}
1 ={}& 
\theta\big(\{\hat f_i,\hat f_{m+1}\} = \{{\rm g},{\rm g}\}\big)
\\&
+ \theta\big(\{\hat f_i,\hat f_{m+1}\} = \{{\rm g},q\}\big)
+ \theta\big((\{\hat f_i,\hat f_{m+1}\} = \{{\rm g},\bar q\}\big)
\\ &
+ \theta\big(\{\hat f_i,\hat f_{m+1}\} = \{q,{\rm g}\}\big)
+ \theta\big(\{\hat f_i,\hat f_{m+1}\} = \{\bar q,{\rm g}\}\big)
\\ &
+ \theta\big(\{\hat f_i,\hat f_{m+1}\} = \{q,\bar q\}\big)
+ \theta\big(\{\hat f_i,\hat f_{m+1}\} = \{\bar q,q\}\big)
\;\;.
\end{split}
\end{equation}
In the coefficients of the $\{{\rm g},q\}$ and $\{{\rm g},\bar q\}$ theta functions, we can further define the labeling by interchanging the labels $i$ and $m+1$, so that $m+1$ is the label for the gluon. In the $\{\bar q,q\}$ term, we can further define the labeling by interchanging the labels $i$ and $m+1$, so that $m+1$ is the label for the antiquark. With these label choices, our contribution is
\begin{equation}
\begin{split}
\label{eq:finalsplit1}
& (m+1)
\sum_{l = 1}^{m}S_l(\{\hat f\}_{m+1})\
\ket{\ME(\{\hat p, \hat f\}_{m+1};l,m+1)}
\bra{\ME(\{\hat p, \hat f\}_{m+1};l,m+1)}
\;\;,
\end{split}
\end{equation}
where
\begin{equation}
\label{eq:Sldef1}
S_l(\{\hat f\}_{m+1}) = \left\{
\begin{array}{cl}
1/2\;, &  l \in \{1,\dots,m\},\ \hat f_l = \hat f_{m+1} = {\rm g}  \\
  1\;, &  l \in \{1,\dots,m\},\ \hat f_l \ne {\rm g}, \hat f_{m+1} = {\rm g}  \\
  0\;, &  l \in \{1,\dots,m\},\ \hat f_l = {\rm g}, \hat f_{m+1} \ne {\rm g}  \\
  1\;, &  l \in \{1,\dots,m\},\ \hat f_l = q, \hat f_{m+1} = \bar q  \\
  0\;, &  l \in \{1,\dots,m\},\ \hat f_l = \bar q, \hat f_{m+1} = q  \\
\end{array}
\right.
\;\;.
\end{equation}
We thus derive the factor $(m+1)$ and the factor 1/2 for a final state ${\rm g} \to {\rm g} + {\rm g}$ splitting. 

Consider now the contribution (\ref{eq:initialsplit}). We can break the relabeling symmetry by interchanging the labels $j$ and $m+1$. Then there are $m+1$ equal terms, giving
\begin{equation}
\begin{split}
\label{eq:initialsplit1}
& (m+1)
\sum_{l \in \{\La,\Lb\}}S_l(\{\hat f\}_{m+1})\
\ket{\ME(\{\hat p, \hat f\}_{m+1};l,m+1)}
\bra{\ME(\{\hat p, \hat f\}_{m+1};l,m+1)}
\;\;,
\end{split}
\end{equation}
where
\begin{equation}
\label{eq:Sldef2}
S_l(\{\hat f\}_{m+1}) = 1  \quad l \in \{\La,\Lb\}
\;\;.
\end{equation}

Consider, finally, the interference diagrams, Eq.~(\ref{eq:allinterference}). We separate this into several terms according to the values of $i$ and $j$,
\allowdisplaybreaks
\begin{align}
\label{eq:interference1}
& \sum_{\substack{i,j \in \{\La,\Lb\}\\i\ne j}}
\sum_{k = 1}^{m+1}
\ket{\ME(\{\hat p, \hat f\}_{m+1};i,k)}
\bra{\ME(\{\hat p, \hat f\}_{m+1};j,k)}
\notag\\&+
\sum_{i \in \{\La,\Lb\}}\sum_{j = 1}^m
\sum_{\substack{k = 1\\ k \ne j}}^{m+1}
\ket{\ME(\{\hat p, \hat f\}_{m+1};i,k)}
\bra{\ME(\{\hat p, \hat f\}_{m+1};j,k)}
\notag\\&+
\sum_{i = 1}^m \sum_{j \in \{\La,\Lb\}}
\sum_{\substack{k = 1\\ k \ne i}}^{m+1}
\ket{\ME(\{\hat p, \hat f\}_{m+1};i,k)}
\bra{\ME(\{\hat p, \hat f\}_{m+1};j,k)}
\notag\\&+
\sum_{i \in \{\La,\Lb\}}
\sum_{k=1}^{m}
\ket{\ME(\{\hat p, \hat f\}_{m+1};i,k)}
\bra{\ME(\{\hat p, \hat f\}_{m+1};m+1,k)}
\\&+
\sum_{j \in \{\La,\Lb\}}
\sum_{k=1}^{m}
\ket{\ME(\{\hat p, \hat f\}_{m+1};m+1,k)}
\bra{\ME(\{\hat p, \hat f\}_{m+1};j,k)}
\notag\\&+
\sum_{j = 1}^m
\sum_{\substack{k = 1\\ k \ne j}}^{m}
\ket{\ME(\{\hat p, \hat f\}_{m+1};m+1,k)}
\bra{\ME(\{\hat p, \hat f\}_{m+1};j,k)}
\notag\\&+
\sum_{i = 1}^m 
\sum_{\substack{k = 1\\ k \ne i}}^{m}
\ket{\ME(\{\hat p, \hat f\}_{m+1};i,k)}
\bra{\ME(\{\hat p, \hat f\}_{m+1};m+1,k)}
\notag\\&+
\sum_{\substack{i,j =1\\i\ne j}}^m
\sum_{\substack{k = 1\\ k\ne i\,,\,k \ne j}}^{m+1}
\ket{\ME(\{\hat p, \hat f\}_{m+1};i,k)}
\bra{\ME(\{\hat p, \hat f\}_{m+1};j,k)}
\notag\;\;.
\end{align}
We relabel the indices, treating each term separately. In each case, we interchange $k \leftrightarrow m+1$. In the first term, this gives $m+1$ equal terms from the sum over $k$. In the second and third terms, this gives $m$ equal terms from the sum over $k$. In the fourth through seventh terms, each term in the sum over $k$ remains as one term. Finally, in the eighth term, there are $m-1$ equal terms from the sum over $k$. After relabeling, we have
\begin{align}
\label{eq:interference2}
& (m+1)\sum_{\substack{i,j \in \{\La,\Lb\}\\i\ne j}}
\ket{\ME(\{\hat p, \hat f\}_{m+1};i,m+1)}
\bra{\ME(\{\hat p, \hat f\}_{m+1};j,m+1)}
\notag\\&+
m\sum_{i \in \{\La,\Lb\}}\sum_{j = 1}^m
\ket{\ME(\{\hat p, \hat f\}_{m+1};i,m+1)}
\bra{\ME(\{\hat p, \hat f\}_{m+1};j,m+1)}
\notag\\&+
m\sum_{i = 1}^m \sum_{j \in \{\La,\Lb\}}
\ket{\ME(\{\hat p, \hat f\}_{m+1};i,m+1)}
\bra{\ME(\{\hat p, \hat f\}_{m+1};j,m+1)}
\notag\\&+
\sum_{i \in \{\La,\Lb\}}
\sum_{k=1}^{m}
\ket{\ME(\{\hat p, \hat f\}_{m+1};i,m+1)}
\bra{\ME(\{\hat p, \hat f\}_{m+1};k,m+1)}
\\&+
\sum_{j \in \{\La,\Lb\}}
\sum_{k=1}^{m}
\ket{\ME(\{\hat p, \hat f\}_{m+1};k,m+1)}
\bra{\ME(\{\hat p, \hat f\}_{m+1};j,m+1)}
\notag\\&+
\sum_{j = 1}^m
\sum_{\substack{k = 1\\ k \ne j}}^{m}
\ket{\ME(\{\hat p, \hat f\}_{m+1};k,m+1)}
\bra{\ME(\{\hat p, \hat f\}_{m+1};j,m+1)}
\notag\\&+
\sum_{i = 1}^m 
\sum_{\substack{k = 1\\ k \ne i}}^{m}
\ket{\ME(\{\hat p, \hat f\}_{m+1};i,m+1)}
\bra{\ME(\{\hat p, \hat f\}_{m+1};k,m+1)}
\notag\\&+
(m-1)\sum_{\substack{i,j =1\\i\ne j}}^m
\ket{\ME(\{\hat p, \hat f\}_{m+1};i,m+1)}
\bra{\ME(\{\hat p, \hat f\}_{m+1};j,m+1)}
\notag\;\;.
\end{align}

These terms can be combined, after changing the names of some of the summation indices, and added to the contributions (\ref{eq:finalsplit1}) and (\ref{eq:initialsplit1}) to give the revised density operator, which we can call $\rho_1$,
\begin{equation}
\begin{split}
\label{eq:newrho}
\rho_1(&\{\hat p, \hat f\}_{m+1}) = 
\\
&(m+1)\!\!
\sum_{l \in\{\La,\Lb,1,\dots,m\}}S_l(\{\hat f\}_{m+1})\
\ket{\ME(\{\hat p, \hat f\}_{m+1};l,m+1)}
\bra{\ME(\{\hat p, \hat f\}_{m+1};l,m+1)}
\\&
+
(m+1)\!\!
\sum_{\substack{l,k \in\{\La,\Lb,1,\dots,m\}\\l\ne k}}^m
\ket{\ME(\{\hat p, \hat f\}_{m+1};l,m+1)}
\bra{\ME(\{\hat p, \hat f\}_{m+1};k,m+1)}
\;\;.
\end{split}
\end{equation}
The first line here contains the direct terms, Eqs.~(\ref{eq:finalsplit1}) and (\ref{eq:initialsplit1}), while the second line is the interference terms, Eq.~(\ref{eq:interference2}).


\end{document}